\documentclass[twocolumn,showpacs,preprintnumbers,amsmath,amssymb,prb]{revtex4}
\usepackage{graphicx}
\usepackage{dcolumn}
\usepackage[tight]{subfigure}
\usepackage{amsmath}
\usepackage{verbatim}
\usepackage[usenames,dvipsnames]{color}
\usepackage{bm} 
\usepackage{bbm}
\newcommand{\assign}{:=}
\newcommand{\mathd}{\mathrm{d}}

\newcommand{\substacktwo}[2]{
  \substack{ \mathrm{#1} \\ \mathrm{#2}}
}

\newcommand{\wideeq}[1]{\begin{widetext}
    #1
  \end{widetext}}

\newcommand{\Fig}[1]{Fig.~\ref{#1}}

\newcommand{\Eq}[1]{Eq.~(\ref{#1})}
\newcommand{\eq}[1]{(\ref{#1})}
\newcommand{\Sec}[1]{Sec.~\ref{#1}}
\renewcommand{\sec}[1]{\ref{#1}}
\newcommand{\App}[1]{App.~\ref{#1}}

\newcommand{\Cite}[1]{Ref.~\cite{#1}}

\newcommand{\op}[1]{\hat{#1}}
\renewcommand{\vec}[1]{\mathbf{#1}}

\newcommand{\tens}[1]{\bm{\mathcal{#1}}}

\newcommand{\bra}[1]{\langle #1|}
\newcommand{\ket}[1]{|#1 \rangle}
\newcommand{\braket}[2]{\langle  #1 | #2 \rangle} 
\newcommand{\brkt}[1]{\langle #1 \rangle}
\newcommand{\Tr}{\text{Tr}}
\newcommand{\tr}[1]{\text{Tr}_{#1}}
\newcommand{\hc}{\text{h.c.}}
\begin{document}
\title{
  Transport of spin-anisotropy without spin currents 
}

\author{Michael Hell$^{(1,2)}$}
\author{Sourin Das$^{(3,1,2)}$}
\author{Maarten R. Wegewijs$^{(1,2,4)}$}
\affiliation{
  (1) Peter Gr{\"u}nberg Institut,
      Forschungszentrum J{\"u}lich, 52425 J{\"u}lich,  Germany \\
  (2) JARA- Fundamentals of Future Information Technology\\
  (3) Department of Physics and Astrophysics,\\
      University of Delhi, Delhi 110 007, India  \\
  (4) Institute for Theory of Statistical Physics,
      RWTH Aachen, 52056 Aachen,  Germany
}
\date{\today}

\begin{abstract}
  We revisit the transport of spin-degrees of freedom across an electrically
  and thermally biased tunnel junction between two ferromagnets with
  non-collinear magnetizations. Besides the well-known charge and spin
  currents we show that a non-zero \emph{spin-quadrupole current} flows
  between the ferromagnets. This tensor-valued current describes the
  non-equilibrium transport of spin-anisotropy relating to both local and
  non-local multi-particle spin correlations of the circuit. {{This
  quadratic spin-anisotropy, quantified in terms of the spin-quadrupole
  moment, is fundamentally a two-electron quantity.}} In spin-valves with an
  embedded quantum dot such currents have been shown to result in a quadrupole
  accumulation that affects the measurable quantum dot spin and charge
  dynamics. The spin-valve model studied here allows fundamental questions
  about spin-quadrupole \emph{storage} and \emph{transport} to be
  worked out in detail, while ignoring the detection by a quantum dot. This
  physical understanding of this particular device is of importance for more
  complex devices where spin-quadrupole transport can be detected. We
  demonstrate that, as far as storage and transport are concerned, the spin
  anisotropy is only partly determined by the spin polarization. In fact, for
  a thermally biased spin-valve the charge- and spin-current may vanish, while
  a pure \emph{exchange} spin-quadrupole current remains, {{which
  appears as a fundamental consequence of Pauli's principle}}. We extend the
  real-time diagrammatic approach to efficiently calculate the average of
  multi-particle spin-observables, in particular the spin-quadrupole current.
  Although the paper addresses only leading order and spin-conserving
  tunneling we formulate the technique for arbitrary order in an arbitrary,
  spin-dependent tunnel coupling in a way that lends itself to extension to
  quantum-dot spin-valve structures.
\end{abstract}

 \maketitle

\section{Introduction}\label{sec:intro}

Spintronics combines the concepts of electronic transport and spin physics.
One of the earliest examples in solid state physics was the tunnel
magnetoresistance effect, discovered by Julliere in 1975 {\cite{Julliere75}}:
the charge current through two tunnel-coupled ferromagnets decreases when
their magnetizations are changed from a parallel to an antiparallel
configuration. The simple explanation of Julliere {\cite{Julliere75}}, based
on the {\emph{spin}}-dependence of the density of states for spin-$\uparrow$
and $\downarrow$, has been refined and extended by later works. Slonczewski
calculated \ the spin current through the FM-I-FM junction \
{\cite{Slonczewski89}}, which can be detected by a second tunnel junction
{\cite{Potok02}}. The spin current is responsible for an exchange coupling
between magnetizations of the two ferromagnets
{\cite{Braun05,Slonczewski89}}). An important early application of spin
currents is spin injection from ferromagnets into non-magnetic systems, (for a
review see {\cite{Schmidt05}}).

Since then, the frontiers of spintronics have been pushed more and more
towards the nanoscale, in particular by attaching macroscopic leads to small
quantum dots. To name only a few interesting effects in which the transport
heavily relies on the spin physics, we mention the Kondo effect
{\cite{Glazman88,Goldhaber98}}, Pauli blockade {\cite{Ono02}}, and various
types of spin-blockade effects {\cite{Weinmann94}}. The spintronic features
mentioned for the mesoscopic systems also have a counterpart in
{{microscopic}} quantum dot physics. For instance, spin injection into
quantum dots and spin currents have been measured {\cite{Chye02}}. Moreover,
for non-collinearly magnetized ferromagnets, the above mentioned exchange
effects translates into a dipolar exchange field {\cite{Braun04set}}, which
can even lift the spin-valve effect.

Besides these analogies, there are, however, profound differences when{{
microscopic systems}} such as quantum dots are involved. Due to the spatial
confinement of electrons, Coulomb electron-electron interaction becomes
all-important and correlations between electrons play a prominent role. Spin
correlations are built up due to the exchange spin-spin interaction, which
results from the concerted action of charging effects and the Pauli principle.
This couples the spin-dipole moments of the individual electrons to high-spin
states ($S \geqslant 1$). Such high-spin quantum systems have non-trivial
higher spin-moments beyond the average spin, such as the{\emph{
spin-quadrupole moment}} (SQM), which is usually the dominant part. In the
physical language of atomic and molecular magnetism, the SQM characterizes the
quadratic{\emph{ spin anisotropy}}. It quantifies the preference of pairs of
spins that make up the large moment $S \geqslant 1$ to be \emph{aligned}
along a specific {\emph{axis}} irrespective of their {\emph{orientation}}
along this axis (up, down). SQM is also relevant to transport: for example, a
spin anisotropy barrier can completely determine the signatures of the
conductance through molecule magnets {\cite{Gatteschi}} and magnetic adatoms
{\cite{Brune09}}. However, in these devices the spin anisotropy appears rather
as a property \ ``fixed'' \ to the atoms/molecule and not something that could
be moved around. This latter idea has been introduced by recent publications
{\cite{Sothmann10,Baumgaertel11,Misiorny13}}, which point out that
spin-quadrupole moment, like spin-dipole moment, can be {\emph{injected}} and
accumulated in a high-spin quantum dot attached to ferromagnets. Thus, spin
anisotropy has turned out to be a true {\emph{transport}} quantity in some
ways similar to spin-dipole moment. Thus, the transport picture of spin
degrees of freedom needs to be extended beyond that offered by charge and spin
currents. This is at the heart of this paper, which studies the
{\emph{storage}} and \emph{transport} of spin-quadrupole moment in
spintronic devices, merging concepts of spintronics and electron-spin
{\emph{correlations}} (for example present in single-molecule magnets). The
aim of this paper is to answer the following three fundamental questions
raised by the above cited studies:

i. How is SQM \emph{stored} in {\emph{macroscopic}} system, i. e.,
ferromagnets?

ii. How is SQM {\emph{transported}} {\emph{macroscopically}} between such
reservoirs?

iii. How can one define an SQM {\emph{current operator}} and what is the
physical interpretation of its average?

The answers are by no means obvious since SQM, unlike charge and spin, is a
\emph{two}-electron quantity. We therefore resort to the simplest possible
setting -- the Julliere model of two tunnel-coupled ferromagnets
{\emph{without}} an embedded quantum dot. The idea is to take one step
``back'' relative to the references
{\cite{Sothmann10,Baumgaertel11,Misiorny12a,Misiorny13}} and to learn as much
as possible from this simple spin-valve model about the concepts essential to
multi-spin transport.

We emphasize from the start that we thereby completely ignore the
complications of the measurable effects of SQM currents, which seem to occur
only when SQM can accumulate in a quantum dot. In the tunnel-junction
spin-valve the charge current as in
{\cite{Sothmann10,Baumgaertel11,Misiorny13}} does not measure the
spin-current, although it displays spin-dependent effects. Similarly, this
study shows that the charge and spin current do not measure the SQM current.
Thus, our results in no way invalidate results of previous studies of charge
and spin currents; in this {\emph{simple}} setup they simply coexist with the
SQM currents. As long as one is {{only interested in the charge current,}}
one can ignore SQM currents in this setup. We will therefore not suggest any
concrete ``meters'' of SQM effects in this paper. These were addressed
elsewhere {\cite{Baumgaertel11,Sothmann10,Misiorny13}} where for instance in
{\cite{Misiorny13}} the Kondo effect was shown to be sensitive to the
quadrupolar analogue of the spin-torque.

Still, the physical insights gained by this study provide a sound foundation
for the discussion of their counterparts in more complex, interacting
nanoscale devices, which allow for SQM detection. For this reason, we will
also address how SQM transport through the spin-valve may be controlled by
various non-linear driving parameters such as voltage, temperature gradients
and magnetic parameters. Finally, we note that all our results are obtained
within a modern version of the real-time transport formalism, which we have
extended to deal efficiently with multi-particle spin-degrees of freedom.

The paper is structured as follows: in {\Sec{sec:model}} we formulate the
spin-valve model and discuss the physical situations it applies to. We define
the one- and \emph{two}-particle densities of states that enter into the
results. In {\Sec{sec:spinMultipoleStorage}} we show that simple Stoner
ferromagnets provide reservoirs of uniaxial spin anisotropy in addition to
spin polarization. We introduce a spin-multipole network picture extending the
idea of a charge and spin transport network. For multi-electron quantities,
such as spin anisotropy, this picture is radically different since they
describe local and{\emph{ non-local}} correlations. In
{\Sec{sec:spinMultipoleNetworks}}. we will see how this naturally suggests the
general definition of spin quadrupole current operators. In \
{\Sec{sec:spinMultipoleCurrents}} the non-equilibrium averages of these
operators are presented for our spin-valve model. We discuss the decomposition
of the spin-quadrupole currents into a dissipative part (spin-quadrupole
injection/emission) and a coherent part (spin-quadrupole torque), similar to
the spin dipole current. \ The appendices contain -- besides details -- a
systematic account of some important technical developments of the real-time
transport theory that we employ.

\section{Spin-valve model}\label{sec:model}

We start with an overview of the main concepts and ideas, which are central to
our comprehensive analysis, aimed at answering the three guiding questions
posed in the introduction. The key to understanding the first question, i. e.,
how SQM is {\emph{stored}}, is to investigate the microscopic origin of SQM by
considering a system two coupled spin $1 \text{/} 2$. This provides a natural
link to atomic and molecular physics, which will be discussed in
{\sec{sec:atomic}}. Note that we deal here with the {\emph{spin}}-quadrupole
moment of a system consisting of {\emph{electrons}} and not with the
{\emph{electric nuclear}} quadrupole moment, which have been investigated in
great detail {\cite{Cottingham}}.

We moreover introduce the Hamiltonian for the spin-valve structure (see Sec.
\ref{sec:Hamiltonian}) consisting of two tunnel-coupled ferromagnets, allowing
for non-collinear magnetization directions. The ferromagnets are described
using a Stoner model. Importantly, the spin-dependent one-particle density of
states is {\emph{not}} sufficient to quantify spin-multipole properties of
ferromagnets. In {\sec{sec:DOSModel}}, we introduce a {\emph{two-particle
density of states}} (see {\Eq{eq:twoDOS}}), which is required for the
calculation of the average spin-quadrupole moment and its current (Secs. \
{\sec{sec:sqmT0}} and {\sec{sec:spinMultipoleCurrents}}). It can only be
calculated if the explicit spin-dependence of the dispersion relation is
available. For all concrete results presented in this paper, we employ a
single, wide, flat band approximation, whose validity will be discussed in
Sect. {\sec{sec:wide-flat-band}}. Throughout we set $\hbar = e = c = k_B = 1$.

\subsection{Spin-quadrupole Moment: From Atomic Physics to
Spintronics}\label{sec:atomic}

To address the storage of SQM, we consider two electrons occupying two
different orbitals with the combined system being in a spin-triplet state. The
single-particle spin vector operators of these electrons, $s^1_i$ and $s^2_i$,
add up the total spin operator $S_i = s^1_i + s_i^2$ ($i = x, y, z$). From the
operator components of the latter, the spin-quadrupole moment {\emph{tensor}}
operator $\tens{Q} = \sum_{i j} Q_{i j}  \vec{e}_i \vec{e}_j$ can be
constructed,
\begin{eqnarray}
  Q_{i j} & = & \frac{1}{2} \{ S_i, S_j \} - \frac{1}{3} \vec{S}^2 \delta_{i
  j},  \label{eq:Qdef}
\end{eqnarray}
where $i, j = x, y, z$. In the triplet states $\ket{T +} = \ket{\uparrow
\uparrow}$ or $\ket{T -} = \ket{\downarrow \downarrow}$, the average spin
dipole moment is non-zero: $\bra{T m} \vec{S} \ket{T m} = m \vec{e}_z$, for $m
= \pm$. The average spin-quadrupole moment has non-zero components as well
(see {\App{app:tripletCorr}}):
\begin{eqnarray}
  \bra{T \pm} \tens{Q} \ket{T \pm} & = & \tfrac{1}{3} \vec{e}_z \vec{e}_z -
  \tfrac{1}{6} \sum_{l \neq z} \vec{e}_l \vec{e}_l .  \label{eq:Qtripmatel1}
\end{eqnarray}
Since the largest element of this tensor, given by the component $\bra{T \pm}
Q_{z z} \ket{T \pm}$, is positive, the spins are likely to be
\emph{aligned} with the $z$-th axes in state $\ket{T \pm}$ -- irrespective
of their \emph{orientation}. Thus besides spin-polarization,
spin-quadrupole moment is ``\emph{stored}'' in this two-electron system.
One may object and ask whether the quadrupole moment is not completely
determined by the spin-dipole moment since the tensor (\ref{eq:Qtripmatel1})
could be entirely expressed in terms of $\bra{T \pm} \vec{S} \ket{T \pm}$.
However, in a \emph{quantum} system, even without two-particle
interactions, we have $\brkt{S_i S_j} \neq \brkt{S_i} \brkt{S_j}$ due to
exchange processes. As a result a system may be purely ``quadrupolarized'',
i.e. $\brkt{\tens{Q}} \neq 0$ while $\brkt{\vec{S}} = 0$. An example of this
is the triplet state $\ket{T 0} = \frac{1}{\sqrt{2}} \left( \ket{\uparrow
\downarrow} + \ket{\downarrow \uparrow} \right. )$ for which the
expectation values of all spin components vanish, $\bra{T 0} \vec{S} \ket{T 0}
= 0$, but
\begin{eqnarray}
  \bra{T 0} \tens{Q} \ket{T 0} & = & - \tfrac{2}{3} \vec{e}_z \vec{e}_z +
  \tfrac{1}{3}  \sum_{l \neq z} \vec{e}_l \vec{e}_l,  \label{eq:Qtripmatel0}
\end{eqnarray}
indicating that this is a ``planar'' spin state, in contrast to the axial spin
state (\ref{eq:Qtripmatel1}). In the context of quantum information, this
state is one of the triplet Bell states \ $\ket{B_z} = \ket{T 0}$. The other
two Bell states \ $\ket{B_x} = \frac{1}{\sqrt{2}} \left( \ket{\uparrow
\uparrow} - \ket{\downarrow \downarrow} \text{} \right) $ , $\ket{B_y}
= \frac{1}{\sqrt{2}} \left( \ket{\uparrow \uparrow} + \ket{\downarrow
\downarrow} \text{} \right)$ further illustrate that states of zero
spin-polarization ($\bra{B_k} \vec{S} \ket{B_k} = 0$ for each $k = x, y, z$)
can be distinguished by their {\emph{spin anisotropy}}: the latter is
quantified by the average of the spin-quadrupole tensor (see
{\App{app:tripletCorr}}), which reads
\begin{eqnarray}
  \bra{B_k} \tens{Q} \ket{B_k} & = & - \tfrac{2}{3} \vec{e}_k \vec{e}_k +
  \tfrac{1}{3}  \sum_{l \neq k} \vec{e}_l \vec{e}_l . 
\end{eqnarray}
Since the largest element of this tensor, $\bra{B_k} Q_{k k} \ket{B_k}$, is
negative in state $\ket{B_k}$, the spins lie in the plane perpendicular to the
$k$-th axes without any definite orientation. Such states appear as
eigenstates of {\emph{bi}}-axial spin Hamiltonians of type $H = - D S_z^2 + E
( S_x^2 - S_y^2)$, which are also well-known in molecular magnetism. In
general, the average of $\tens{Q}$ in any triplet superposition state is a
symmetric tensor, whose principal values lie in the interval \ $[ - 2 \text{/}
2, + 1 \text{/} 3]$. In fact, a triplet quantum state is completely specified
by giving the average of \emph{both} the spin-dipole and the
spin-quadrupole moment: formally, one can show that an arbitrary mixed-state
density operator in the triplet subspace can be decomposed into a bases of
spin dipole and quadrupole operators {\cite{Sothmann10,Baumgaertel11}}.
{{In this sense, the spin-quadrupole moment is thus a degree of freedom
independent of the spin-dipole moment in any system of more at least two
spins.}} {{Quadrupole moments are not limited to the spin degree of
freedom only. One may define {\emph{pseudo}}-spin dipole and -quadrupole
operator whenever one deals with a system of at least three levels. Such
systems arise, for instance, when combining spin and orbital degrees of
freedom. Such pseudo-quadrupole moments then express other types of
correlations, which are inevitably needed to fully characterize the state of
such systems. In this paper we are, however, only concerned with the
spin-quadrupole moment, which is most relevant for spintronics.}}

The above ideas can be extended to one of the basic, circuit element of
spintronics: a ferromagnetic many-electron system (see Sec. \ref{sec:sqmT0}).
The average of the macroscopic spin operator \ $S_i = \sum_a s_i^a$, where
$s_i^a$ \ is the $i$-the component of the spin of electron $a$, quantifies the
magnetization of the ferromagnet. Similar to the spin, the macroscopic
spin-quadrupole moment can also be decomposed into a sum of microscopic
contributions coming from electron pairs. By inserting $S_i = \sum_a s_i^a$
into {\Eq{eq:Qdef}}, we obtain
\begin{eqnarray}
  Q_{i j} & = & \sum_{a < b} q^{a b}_{i j}, \\
  q^{a b}_{i j} & = & s_i^a s^b_j - \frac{2}{3}  \left( \vec{s}^a \cdot
  \vec{s}^b \right) \delta_{i j} . 
\end{eqnarray}
The average spin-quadrupole moment thus quantifies spin {\emph{correlations}}
between all possible \emph{electron pairs}. It can be shown that
$\tens{Q}$ captures the {\emph{triplet}} correlations between the spins (see
{\App{app:tripletCorr}}). {{Other types of spin correlations become
important if spin singlet states are additionally considered. This does not
only concern spin-singlet correlations, but also correlations of
Dzyaloshinskii-Moriya type, related to antisymmetric tensors quadratic in the
spin, as found in \ {\Cite{Sothmann10}}. Furthermore, observables expressed by
higher powers in the spin operators describe spin-multipole correlations of
higher rank (e.g., spin octupoles, etc.). Although all of these are of
interest, we focus in this paper only on two-electron spin-triplet
correlations, which are exclusively determined by the spin-quadrupole moment,
and were found in the simplest possible situation {\cite{Baumgaertel11}} to be
the dominant spin-multipole moment coupling to the spin-dipole dynamics.}}

For a ferromagnet, we will see later that the spin-dipolarization induces a
spin-quadrupolarization similar to the simple example of two-electron triplet
states $\ket{T \pm}$, see the discussion of Eq. (\ref{eq:Qtripmatel1}). This
will become evident when we identify a classical or {\emph{direct}}
contribution, which is completely determined by the spin polarization. In
addition, there is a quantum or {\emph{exchange}} contribution to spin
anisotropy, which is independent of spin. The latter reveals the two-electron
nature of spin-quadrupole moment and comes as a consequence of the
Pauli-principle. We will see that this pure quantum anisotropy can be
understood as an tensor-valued ``Pauli-exlusion hole'' in the triplet
spin-correlations, accounting for correlations that are forbidden by the Pauli
principle (see Sec. \ref{sec:microscopic}). This in particular makes the
spin-quadrupole moment an independent degree of freedom that is ``stored'' in
a ferromagnet in addition to the charge and the spin-dipole moment. The
studies {\cite{Baumgaertel11,Sothmann10}} indicate that it is a quantity that
must be reckoned with in nanoscale spintronic systems with high
spin-polarizations.

We now turn to the second, central question announced in the introduction: how
can one \emph{transport} spin-quadrupole moment (see Secs.
\ref{sec:SQMcurrent} and \ref{sec:spinMultipoleCurrents})? If we tunnel-couple
two ferromagnets and apply a finite voltage bias, it is well-known that
besides a charge current a spin-current will flow {\cite{Slonczewski89}} since
electrons carry both charge and spin as an intrinsic degree of freedom. But
can there also be a {\emph{flow}} of spin anisotropy? At first sight, one may
answer ``no'' because single electrons do not have an intrinsic
spin-quadrupole moment. However, as an electron spin tunnels from one
ferromagnet to the other, it retains its correlations with other electrons. By
this, triplet correlations initially stored locally in one of the ferromagnets
turn into \emph{nonlocal} triplet correlations between electrons in
different ferromagnets. This leads, even by tunneling of single electrons, to
a nonzero {\emph{spin-anisotropy current}}. The aspect of non-locality of SQM
is another essential aspect of its two-particle nature. Even on a macroscopic
level, spin-anisotropy transport can therefore only be understood in a network
picture accounting for both local and{\emph{ non-local }}sources of SQM. Such
a spin-multipole network picture -- radically different from that for charge
and spin -- will be developed here. It illustrates that storage and transport
of SQM cannot be understood independently from each other.

These general considerations bring us to the third main question of our
paper, namely how to define the SQM current operator (see
{\sec{sec:SQMcurrent}}). It can then be identified with the rate of change in
SQM stored in these local and non-local sources. To develop a further
understanding of SQM transport, we need to calculate the average currents for
the simple spin-valve. We will discuss how to decompose the result into
various physically meaningful contributions. Besides a direct and an exchange
part we find in analogy to the spin-current dissipative and coherent
contributions. The interplay of these contributions causes the SQM current to
generate a \emph{biaxial spin-anisotropy} for non-collinear ferromagnets.
This transport of spin-anisotropy opens up the interesting possibility to
generate anisotropic magnetic systems starting with isotropic ones, in a way
similar to creating spin-polarized systems by spin-transport. To our
knowledge, this has not been discussed so far, even though the effects of
``static'' spin anisotropy {\emph{on}} transport have been studied extensively
in atomic / molecular magnetism {\cite{Gatteschi}} and spintronics.
{{Based on this it is expected that spin-quadrupole currents play a role
in many nanoscale spintronics devices with significant quantum
spin-correlations.}}

In our comprehensive study of the dependence of the SQM current on physical
parameters we will find a striking result, highlighting the above mentioned
different nature of SQM transport as compared to spin transport. We predict
the possibility of a \emph{pure spin-quadrupole current}, i.e., a
quadrupole current not accompanied by charge and spin current. This SQM
current is entirely due to quantum exchange processes and is driven by a
density {{gradient}} of ``Pauli exclusion holes'' across the junction. A
clear notion of the Pauli exclusion holes will be defined in Sec.
\ref{sec:microscopic}. We find that the spin-anisotropy flow direction can be
controlled by the direction of the thermal bias - a non-trivial result as a
deeper analysis of SQM storage will reveal. Transport of spin correlations is
thus possible and controllable without affecting net spin polarization or
charge distribution. This remarkable conclusion illustrates most clearly that
the spin quadrupole moment is really an independent \emph{transport
quantity} that should be incorporated into spintronics theories. It also
indicates possible, promising applications: injection of such an SQM current
may, for instance, modify or even generate spin-anisotropy in an embedded
system without changing its spin-polarization. This may perhaps allow for
novel ways of performing operation in multi-spin systems.

\subsection{Spin-valve Hamiltonian}\label{sec:Hamiltonian}

{
  \begin{figure}[h]
    \includegraphics[width=0.9\linewidth]{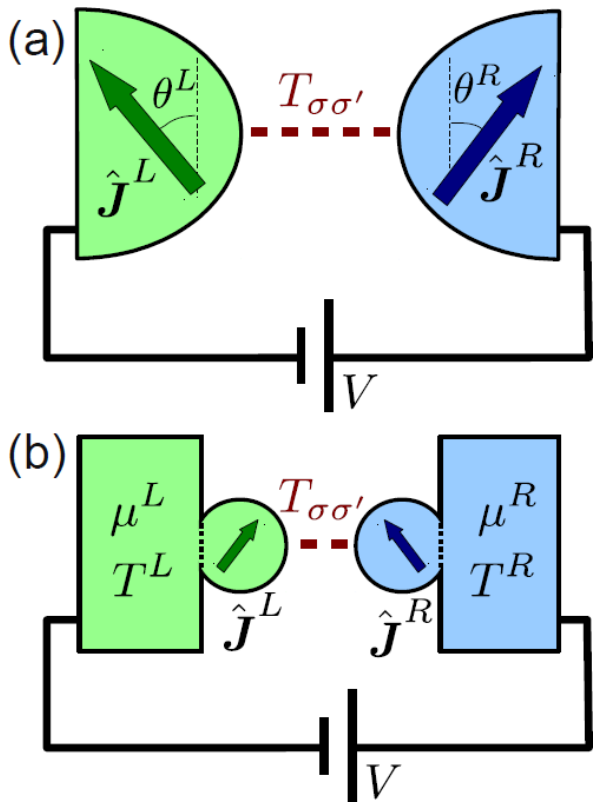}
    \caption{Spin-valve setup: a tunnel junction
between \ (a) \ macroscopic ferromagnets with Stoner fields \ $\vec{J}^r = J^r
\widehat{\vec{J}}^r$. (b) mesoscopic islands with local magnetic fields $B^r =
J^r \widehat{\vec{J}}^r$, each in equilibrium with energy- and particle-
reservoir. A combination of (a) and (b) is another possibility (not
shown).\label{fig:tunneljunction}}
  \end{figure}
}

We start from a quite general model Hamiltonian
\begin{eqnarray}
  H & = & H^L_0 + H^R_0 + H_T  \label{eq:Htot}
\end{eqnarray}
with the non-interacting Hamiltonians of subsystem $r = L, R$:
\begin{eqnarray}
  H_0^r & = & \sum_{n k \sigma} \varepsilon_{n k \sigma}^r c^{\dag}_{r n k
  \sigma} c_{r n k \sigma},  \label{eq:H0}
\end{eqnarray}
and the tunneling Hamiltonian with $T^{L R}_{\sigma \sigma'} = ( T^{R
L}_{\sigma' \sigma})^{\ast}$:
\begin{eqnarray}
  H_T &  & = \sum_{n n' k k' \sigma \sigma'} T^{L R}_{\sigma \sigma'} c_{L n k
  \sigma}^{\dag} c_{R n k' \sigma'_{}} + \hc  \label{eq:HT}
\end{eqnarray}
Here the field operators $c_{r n k \sigma}$ act on the single particle level
$k$ of band $n$ in subsystem $r = L, R$ with spin $\sigma = \uparrow,
\downarrow$. We assume that the single-electron spin $\vec{s}$ for every
orbital $( n, k)$ in the same ferromagnet can be quantized along a common
physical direction $\op{\vec{J}}^r$ (with $| \op{\vec{J}}^r | = 1$), i. e.,
\begin{eqnarray}
  &  & \left( \widehat{\vec{J}}^r \cdot \vec{s} \right) \ket{\sigma}_r =
  \sigma \ket{\sigma}_r,  \label{eq:eigenspin}
\end{eqnarray}
{{with $\ket{\sigma}_r = e^{- i \theta^r \op{\vec{m}} \cdot \vec{s} }
\ket{\sigma}_{\vec{e}_z}$ where we rotate by the angle $\theta^r$ between
$\widehat{\vec{J}}^r$ and $\vec{e}_z$ about the axis perpendicular to both
these vectors, defined by $\op{\vec{m}} = \vec{e}_z \times \op{\vec{J}^{}}^r
\text{/} | \vec{e}_z \times \op{\vec{J}}^r |$ .}} For simplicity, we assume
the tunneling amplitudes to be band- ($n$) and energy-($k$) independent;
moreover, the spin is conserved by the tunneling, $\left[ H_T, \vec{S} \right]
= 0$. Nevertheless, the tunneling amplitudes in {\Eq{eq:HT}}
\begin{eqnarray}
  T^{L R}_{\sigma \sigma'} & = & t { }  \text{}_L \braket{\sigma}{\sigma'}_R 
  \label{eq:tsigmacons}
\end{eqnarray}
are in general spin-dependent because the field operators $c_{L n k \sigma}$
and $c_{R n k \sigma}$ annihilate electrons with spins quantized along
non-collinear directions $\widehat{\vec{J}}^L \nparallel \widehat{\vec{J}}^R$.
The spin conservation in the tunneling is reflected by a
spin-{\emph{independence}} of the ``bare'' tunneling amplitude $t$. More on
this can be found in {\App{app:covrealtime}}, where we include spin-symmetry
breaking tunneling processes in our extension of the real-time transport
theory.

We model the two subsystems as reservoirs, each kept in a thermal equilibrium
state $\rho^r = e^{- ( H^r_0 - \mu^r N^r) / T^r} / Z_r$ where $Z^r = \Tr e^{-
( H^r_0 - \mu^r N^r) / T^r}$ is the grand-canonical partition function and
$N^r$ is the particle number operator of electrode $r$. Both electrodes are
have fixed electrochemical potentials $\mu^r $and temperatures $T^r$, whose
gradients drive the stationary state currents of interest. Note that even if
the tunneling is present, each electrode is held in equilibrium at each
instant of time.

\subsection{Two-Particle Density of States}\label{sec:DOSModel}

In {\Sec{sec:spinMultipoleStorage}} we will calculate the expectation values
of the charge and the spin multipoles involving sums over the mode index $k$.
We now indicate which quantities parametrize the spin information from the
ferromagnetic electrodes. As usual, we take the continuum limit and replace
sums over $k$ by a frequency integral. For one-particle quantities such as
charge and spin, one can expresses the results in terms of the spin-dependent
one-particle density of states (1DOS):
\begin{eqnarray}
  \nu^r_{\sigma} ( \omega) & = & \sum_{n, k} \delta_{} ( \varepsilon_{n k
  \sigma}^r - \omega) \\
  & = & \bar{\nu}^r ( \omega) ( 1 + \sigma n^r ( \omega)),  \label{eq:DOS}
\end{eqnarray}
where $\bar{\nu}^r ( \omega)$ is the {\emph{spin-averaged}} DOS
\begin{eqnarray}
  \bar{\nu}^r ( \omega) & = & \frac{\nu_{\uparrow}^r ( \omega) +
  \nu_{\downarrow}^r ( \omega)}{2} .  \label{eq:avdos}
\end{eqnarray}
All the spin-dependence of the 1DOS is contained in the {\emph{spin
polarization}} (of the 1DOS)
\begin{eqnarray}
  n^r ( \omega) & = & \frac{\nu_{\uparrow}^r ( \omega) - \nu_{\downarrow}^r (
  \omega)}{\nu_{\uparrow}^r ( \omega) + \nu_{\downarrow}^r ( \omega)} . 
  \label{eq:spinpol}
\end{eqnarray}
Importantly, we will find that the 1DOS {\eq{eq:DOS}}, although formulated for
a general one-particle energy spectrum $\varepsilon^r_{n k \sigma}$, is not
sufficient to quantify quantum transport of spin completely, in particular the
spin-spin correlations described by the spin-quadrupole moment. We will see
that the latter requires an additional, spin-dependent {\emph{two-particle
exchange DOS}} (2DOS):
\begin{eqnarray}
  \nu^r_{\sigma \sigma'} ( \omega, \omega') & = & \sum_{n , k} \delta
  ( \varepsilon_{n k \sigma}^r - \omega) _{} \delta ( \varepsilon_{n k
  \sigma'}^r - \omega') .  \label{eq:twoDOS}
\end{eqnarray}
The physical meaning of the 2DOS can be understood most easily by considering
two identical {\emph{copies}} of the same ferromagnet. The 2DOS is nonzero if
there is a pair of states for an electron with spin $\sigma$ at energy
$\omega$ in the first copy and an electron of spin $\sigma'$ at energy
$\omega'$ in the second copy, but within the{\emph{ same}} $k$-mode in the
{\emph{same}} band $n$. We emphasize that the latter restriction requires
additional modeling: one cannot make independent approximations for the 1DOS
and the 2DOS since they are not completely independent of each other: for
example, the spin-diagonal components of the 2DOS must satisfy the relation
$\nu_{\sigma \sigma}^r ( \omega, \omega') = \delta ( \omega - \omega')
\nu^r_{\sigma} ( \omega)$. Yet, the remaining components of the 2DOS,
$\nu_{\sigma \bar{\sigma}} ( \omega, \omega')$, where $\bar{\sigma} = -
\sigma$ denotes the opposite of $\sigma$, can {\emph{not}} be constructed from
1DOS. If the energies of electrons with spin $\sigma$ and $\bar{\sigma}$ are
related by a function $\varepsilon^r_{n k \bar{\sigma}} = g^r_{n \sigma
\bar{\sigma}} ( \varepsilon^r_{n k \sigma})$ (for example, if the dispersion
relation can be solved for $k$), then
\begin{eqnarray}
  \nu^r_{\sigma \sigma'} ( \omega, \omega') & = & \nu^r_{\sigma} ( \omega)
  \delta ( g^r_{n \sigma \sigma'} ( \omega) - \omega')   \label{eq:nursigma}
\end{eqnarray}
with $g^r_{n \sigma \sigma} ( \omega) = \omega$ trivially. Clearly, more than
the 1DOS is needed here. As a consequence, in general one has to start from
the spin dependent dispersion relation and calculate all required components
of the 1- and 2DOS consistently. {{Transport of two-particle transport
properties therefore probes more of the electronic structure of the
ferromagnets than the one-particle currents of charge and spin.}}

\subsection{Stoner Model and Flat Band
Approximation}\label{sec:wide-flat-band}

The central results of this paper,
{\Eq{eq:chargeCurrent}}-{\eq{eq:SQMCurrent}}, are valid for general case of
the above 1-DOS and 2DOS. However, since we focus on physically understanding
spin-anisotropy transport, rather than making material-specific predictions,
we keep all complications by band structure / dispersion relation features to
a minimum.

{\emph{Stoner model / external magnetic field.}} As explained above, we must
specify the spin-dependence of the dispersion relation for a consistent
treatment of the 1DOS and 2DOS. We model this by a rigid (i. e.,
energy-independent) splitting of absolute value $J^r_n$ between the spin
$\uparrow$ and spin $\downarrow$ states,
\begin{eqnarray}
  \varepsilon^r_{n k \sigma} & = & \varepsilon^r_{n k} - \sigma J^r_n \text{/}
  2,  \label{eq:Stoner}
\end{eqnarray}
which may be different in each band $n$. This model can be used to discuss
several situations, sketched in Fig. \ref{fig:tunneljunction} (a)-(b). In case
(a) macroscopic ferromagnets are treated within the Stoner model. In this case
$J^r_n$ can differ depending on the strength of the electron-electron
interaction in each band. In this case the {{restriction $J^r_n \gtrsim
T^r$ must imposed to avoid}} the breakdown of ferromagnetism (which is not
modeled by \ {\Eq{eq:H0}}). In case (b) we consider mesoscopic magnetic
islands, each in equilibrium with a reservoir. One may now let the $J^r_n$
model an external magnetic field, which may be different locally in each
electrode, i.e., we identify $J^r_n = B^r$. The main difference between (a)
and (b) is the relative importance of quantum exchange contributions in
two-particle spin quantities due to the smaller magnetic moment of the
reservoirs (see below). When considering nanoscopic islands charging and
non-equilibrium effects on the transport will of course be important, which
are neglected here. The main motivation for considering case (b) is that it
provides an interesting comparison with results for quantum dot spin-valves
where the latter effects are fully taken into account
{\cite{Sothmann10,Baumgaertel11}}. For readability we will discuss the results
throughout the paper in the language of case (a), ferromagnets with Stoner
splittings, unless explicitly stated otherwise.

{\emph{Flat Band Approximation.}} We secondly restrict ourselves to a single,
flat band (sketched in {\Fig{fig:DOS_1band}}) in each ferromagnet in the
limit of large bandwidth $2 D^r = 2 D$. The latter limit assumes that all
other energy scales $( T^r, J^r, \mu^r)$ are much smaller than the distance
$W$ of the band edge closest to all electro-chemical potentials, given
\begin{eqnarray}
  W & \assign & \underset{r, r', \sigma, p}{\min} \left( p D - \sigma
  \frac{J^r}{2} - \mu^{r'} \right),  \label{eq:distmuD}
\end{eqnarray}
which is positive since we assume all $\mu^r$ to lie inside the bands. We
refer to this in the following shortly as the {\emph{flat band
approximation}}, keeping in mind that we actually refer to a set of
assumptions. For all frequencies $\omega < W$, the spin-dependent DOS is given
by
\begin{eqnarray}
  \nu^r_{\uparrow} = \nu^r_{\downarrow} & = & \frac{N_o}{2 D}, 
\end{eqnarray}
where $N_o$ is the total number of orbitals in each subsystem and
\begin{eqnarray}
  \nu_{\sigma \sigma'}^r ( \omega, \omega') & = & \nu_{\sigma}^r ( \omega)
  \delta \left( \omega + \tfrac{\sigma - \sigma'}{2} J - \omega' \right), 
\end{eqnarray}
where we used Eq. (\ref{eq:Stoner}) to rewrite Eq. (\ref{eq:nursigma}). One
may criticize the simplicity of this approximation in that it does not account
for spin-polarization near the Fermi energy, but only for a Stoner shift,
which is only noticeable at the band edges. We will see in
{\Sec{sec:firstOrder}}, however, that this already captures plenty of
important aspects in SQM transport.

{
  \begin{figure}[h]
    \includegraphics[width=0.9\linewidth]{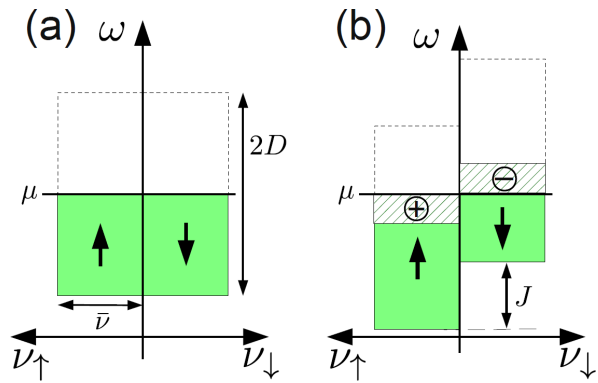}
    \caption{Spin-dependent density of states of a
single flat band for a non-magnetic system (left) and a Stoner ferromagnet
(right). The Stoner splitting redistributes a fraction of $J \text{/} 4 D$ of
the $N_o$ particles from spin-down to the spin-up states relative to the
non-magnetic case as indicated by the $+$ and $-$. We omit the electrode and
band index for simplicity.\label{fig:DOS_1band}}
\end{figure}
}

Clearly, the results for the average particle number, spin, SQM and their
currents have to be independent of the choice of both the coordinate system
and the spin-quantization axis (spin Hilbert space basis); they may only
depend on the physical vectors $\op{\vec{J}}^r$ and the scalar parameters
$\mu^r, T^r$ and $J^r_n$ and $t$ (below). A key technical result of the paper
is that we reformulate real-time diagrammatic transport theory such the
calculation is explicitly shows this covariance at \emph{every stage},
which also makes it much more efficient (see \ {\App{app:covrealtime}}).

Moreover, the usual modification that implements a spin-dependent DOS as
$\nu_{\sigma} = \bar{\nu}  ( 1 + \sigma n)$ with constant $\bar{\nu}$ and $n$
is only valid as long one deals with single-particle \emph{observables}
such as the spin (even when accounting for many-body effects). For these
calculations, all results can usually be expressed using the 1DOS. However,
when dealing with \emph{two-particle observables} relying on the 2DOS, it
is crucial to specify the dispersion relation as we discussed in
{\Sec{sec:DOSModel}}. The above spin-dependent but constant DOS physically
arises from mixing of different types orbitals in a tight-binding picture,
resulting in more than one band. These additional bands can often be ignored,
but this is no longer true for the 2DOS which is sensitive to these details.
To make this clear, we merely mention two possible valid alternative models
accounting consistently for a spin-polarization at the Fermi energy: (i) a
single curved band (see {\Fig{fig:DOS_2band}} (a)) and (ii) two bands with
different bandwidths and a large Stoner splitting so that {\emph{different}}
bands overlap at the Fermi energy (see {\Fig{fig:DOS_2band}} (b)). Since our
single, flat wide band approximation is already sufficient to illustrate
essential effects of spin-quadrupole storage and transport we will not pursue
these band-structure details further here.

{
  \begin{figure}[h]
    \includegraphics[width=0.9\linewidth]{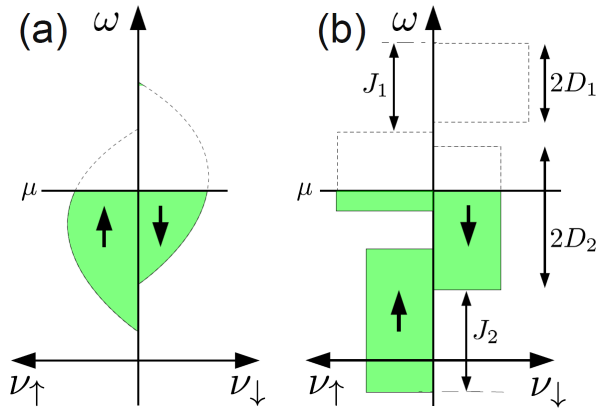}
    \caption{(a) Spin- and \emph{energy}-dependent
density of states for a \emph{one}-band Stoner ferromagnet (b)
Spin-dependent, \emph{constant} density of states for a
\emph{two}-band Stoner ferromagnet with different bandwidths $2 D_n$ and
Stoner splittings $J_n$ for band $n = 1, 2$.\label{fig:DOS_2band}}
\end{figure}}

\section{Spin-Multipole Storage}\label{sec:spinMultipoleStorage}

In this section, we show that a system of ferromagnets, each kept at
equilibrium, does not only store charge and spin polarization, but also
{\emph{stores spin anisotropy}}, quantified by the expectation value of the
spin-quadrupole moment operator {\eq{eq:Qdef}}.

In {\Sec{sec:T0}} we will first investigate the simplest case of a single
ferromagnet at zero temperature. {{We discuss how the average SQM tensor
relates to fluctuations in a macro-spin picture and relate this to the
microscopic triplet spin-spin correlations. We identify an {\emph{exchange}}
contribution, which accounts for a ``hole'' in the quantum two-particle
correlations of the spins due to the Pauli principle, giving rise to negative
or Pauli-forbidden anisotropy.}}

In {\Sec{sec:Tfinite}} we extend these considerations to finite temperatures
and multiple electrodes (without coupling them, i.e., $H_T = 0$), both of
which introduce new aspects. {{The case of two electrodes needs to be
carefully addressed in order to define an SQM current later on: we must
understand {\emph{from where}} and {\emph{to where}} SQM flows. It turns out
that the ferromagnetic electrodes \emph{cannot} simply be identified with
the sources of SQM and we formalize our considerations in a convenient general
spin-multipole network theory in {\Sec{sec:SQMnetwork}}.}}

\subsection{Single Electrode at $T = 0$}\label{sec:T0}

We first calculate and analyse the average particle number, spin-dipole moment
and spin-quadrupole moment of an isolated electrode in the simple limit of
zero temperature in the approximation of a Stoner-shifted flat band (see
{\sec{sec:wide-flat-band}}). In this subsection we will omit the electrode
index $r$ and band index $n$ and denote by $\brkt{\text{ \ }} = \bra{\psi_0}
\text{ \ } \ket{\psi_0}$ the $T = 0$ ground state average.

\subsubsection{Average Charge and Spin\label{sec:chargeSpinT0}}

We first review the average charge and spin-dipole moment for later comparison
of these one-particle quantities with the SQM, a two-particle
quantity. For zero temperature, {{all states with}} energy
$\varepsilon_{k \sigma} \leqslant \mu$ below the electro-chemical potential
$\mu$ are occupied and all levels with $\varepsilon_{k \sigma} > \mu$ are
empty (cf. {\Fig{fig:DOS_1band}}). Thus, the ground state average of particle
number operator
\begin{eqnarray}
  N & = & \sum_{k, \sigma} c^{\dag}_{k \sigma} c_{k \sigma}, 
  \label{eq:chargeOp}
\end{eqnarray}
corresponds to the sum of the green areas below the electro-chemical potential
in Fig. \ref{fig:DOS_1band}: with $\nu_{\sigma} = \bar{\nu}$ we find
\begin{eqnarray}
  \brkt{N} & = & \sum_{\sigma} \bar{\nu} \left( \mu + D + \frac{\sigma}{2} J
  \right) \\
  & = & N_o \left( 1 + \frac{\mu}{D} \right) . 
\end{eqnarray}
Here $N_o = 2 D \bar{\nu}$ is the number of orbitals in the bandwidth $2 D$.
The particle number is independent of the Stoner splitting $J$ in this simple
approximation.

The average of the spin operator,
\begin{eqnarray}
  \vec{S} & = & \sum_{k, \sigma} \vec{s}_{\sigma \sigma'} c^{\dag}_{k \sigma}
  c_{k \sigma'},  \label{eq:spinOp}
\end{eqnarray}
measures the spin-{\emph{dipolarization}} of the system, where $( s_i)_{\sigma
\sigma'} = ( \sigma_i)_{\sigma \sigma'} \text{/} 2$ and $\sigma_i$, \ $i = x
, y, z$ \ are the Pauli matrices. Choosing the coordinate system such
that $\vec{e}_z = \op{\vec{J}}$, we obtain for $T = 0$: $\brkt{S_x} =
\brkt{S_y} = 0$ and
\begin{eqnarray}
  \brkt{S_z} & = & \tfrac{1}{2} \sum_{\sigma} \sigma \bar{\nu}  \left[ \mu -
  \left( - D - \frac{\sigma}{2} J \right) \right] = \tfrac{1}{2} N_s . 
  \label{eq:spinavT0}
\end{eqnarray}
This equals the difference of the number of spin up and down electrons, i.e.,
the number of half-filled orbitals with polarized spins,
\begin{eqnarray}
  N_s & = & \bar{\nu} J = \frac{J}{2 D} N_o,  \label{eq:Ns}
\end{eqnarray}
and corresponds to the difference of the areas under two DOS curves below
$\mu$ in {\Fig{fig:DOS_1band}}.

\subsubsection{Average SQM and Spin Anisotropy\label{sec:spinanisotropy}}

The average SQM $\brkt{\tens{Q}} = \sum_{i j} \brkt{Q_{i j}} \vec{e}_i
\vec{e}_j$ is a real and symmetric tensor, which can therefore always be
diagonalized. With the above choice of the coordinate system with $\vec{e}_z =
\op{\vec{J}}$, $\brkt{Q_{i j}}$ is already diagonal by symmetry with respect
to rotations about $\op{\vec{J}}$. The average of the non-zero tensor operator
component
\begin{eqnarray}
  Q_{z z} & = & \tfrac{2}{3} S_z^2 - \tfrac{1}{3} ( S^2_x + S^2_y) 
  \label{eq:longminustrans}
\end{eqnarray}
now measures the \emph{spin anisotropy} with respect to the $z$-axis in
the ground state: \ $\brkt{Q_{z z}} > 0$ indicates that the spin is aligned
(but not necessarily oriented) with the easy $z$-axis, while $\brkt{Q_{z z}} <
0$ indicates an easy-plane configuration where the spin preferably lies in the
perpendicular $x y$-plane. If $\brkt{Q_{z z}}$ vanishes, neither alignment
longitudinal or transverse to the $z$-direction is favoured. This is the case,
e.g., for a spin-isotropic state for which $\brkt{S_x^2} = \brkt{S_y^2} =
\brkt{S_z^2}$; however, it can also be realized by states that are
{\emph{an}}isotropic in $x y$-plane, for which $\brkt{S_x^2} \neq
\brkt{S_y^2}$ while $\brkt{S_z^2} = \frac{1}{2} \left( \brkt{S_x^2} +
\brkt{S_y^2} \right)$. These two situations are thus distinguished by the
average of one other non-zero SQM tensor components $\brkt{Q_{x x}}$ or
$\brkt{Q_{y y}}$ (since $\sum_i \brkt{Q_{i i}} = 0$ these are not
independent).

{{We now investigate to what extent the average spin polarization in a
Stoner ferromagnet implies a uniaxial anisotropy.}} Classically, one expects
spin polarization to always imply some nonzero spin anisotropy, but the
converse need not be true as our example in {\Sec{sec:intro}} showed. We now
calculate the average SQM in two ways, first focusing an a collective
macrospin picture, common in atomic and molecular magnetism, and then
disentangling it into its microscopic contributions from electron pairs
relevant to spintronics.

\paragraph{Average Macrospin SQM}\label{sec:sqmT0}

The ground state of the ferromagnet is a maximally polarized pure spin state,
\ $\ket{\psi_0} = \ket{S, m = S} $(as sketched in {\Fig{fig:macrospin}}).

{
  \begin{figure}[h]
    \includegraphics[width=0.9\linewidth]{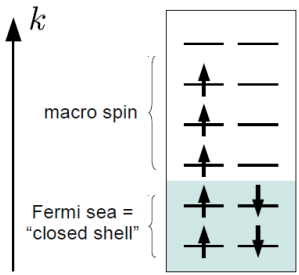}
    \caption{Schematics of the occupation of the
orbitals for an electrode at zero temperature: doubly occupied orbitals form a
zero spin state (Pauli principle) while all spins in the singly occupied
orbitals are parallel, maximizing the total spin (c. f.
text)\label{fig:macrospin}}
\end{figure}}

The value of the spin $S$ is determined from the half-filled orbitals with
$N_s$ polarized spins
\begin{equation}
  \begin{array}{lllll}
    S & = & \brkt{S_z} & \approx & \tfrac{1}{2} N_s
  \end{array} .
\end{equation}
Since $\ket{\psi_0}$ is a maximal spin eigenstate there are no quantum
fluctuations in the first, longitudinal part of {\Eq{eq:longminustrans}}:
$\brkt{S_z^2} = \brkt{S_z}^2$. The second, transverse contribution, however,
can be written as $S^2_x + S^2_y = S_- S_+ - i [ S_x, S_y] = S_- S_+ + S_z$
using $S_{\pm} = S_x \pm i S_y$. It has a non-vanishing part due to the
quantum spin commutation relations: since $S_+ \ket{\psi_0} = 0$, $\brkt{S^2_x
+ S^2_y} = \brkt{S_z}$. The $T = 0$ average {\Eq{eq:longminustrans}} is found
to be
\begin{eqnarray}
  \brkt{Q_{z z}} & = & \tfrac{2}{3} \brkt{S_z^2} - \tfrac{1}{3} \brkt{S^2_x +
  S^2_y}  \label{eq:Qrrzz}\\
  & = & \tfrac{2}{3} S^2 - \tfrac{1}{3} S,  \label{eq:qzzT0}
\end{eqnarray}
The spin-anisotropy, quantified by the average SQM, thus has competing
contributions: spin-polarization induces anisotropy in the $z$-direction
($\propto S^2$), but transverse spin fluctuations tend to suppress it
($\propto S$). The quantum fluctuations of the spin \ in the ground state
``resist'' perfect alignment of the spin, despite the maximal spin alignment.
In fact, {\Eq{eq:qzzT0}} also holds with $N_s = 1$, $S = 1 / 2$ in which case
the longitudinal term is completely cancelled by the transverse fluctuations:
a spin 1/2 is ``so quantum'' that it always has zero spin anisotropy due to
spin fluctuations, in fact, in {\emph{any}} state. Since the filled shells do
not contribute to the value of $S$, this suggests that {{$\brkt{Q_{z z}}$
at $T = 0$ only accounts for triplet correlations between the open shell
electrons with parallel spin. However, a full understanding of the transverse
fluctuations needs a further refinement of that picture.}}

\paragraph{Microscopic SQM Storage}\label{sec:SQMT0micro}

Above we linked the zero-temperature average SQM to the spin anisotropy stored
in a ferromagnet and related it to its average collective spin and its
transverse quantum fluctuations. We will investigate now how these quantum
fluctuations tend to smear out the spin, reducing the uniaxial anisotropy. For
this, we decompose the spin anisotropy into its microscopic contributions from
all particles: we start with the longitudinal contribution to $Q_{z z}$ in
{\Eq{eq:longminustrans}} and express the total spin operator $S_z = \sum_a
s_z^a$ as the sum of the single- electron spins:
\begin{eqnarray}
  &  & S_z^2 = \sum_a ( s_z^a)^2 + 2 \sum_{a < b} s_z^a s_z^b \\
  &  & = \sum_{k \sigma} \tfrac{1}{4} c^{\dag}_{k \sigma} c_{k \sigma} +
  \sum_{k k' \sigma \sigma'} \tfrac{\sigma \sigma'}{4} c^{\dag}_{k \sigma}
  c^{\dag}_{k' \sigma'} c_{k' \sigma'} c_{k \sigma} . 
\end{eqnarray}
Thus, $S_z^2 $ has both a {\emph{one}}- and a {\emph{two}}-electron part. When
averaging, the first term gives $\sum_{\sigma} \tfrac{1}{4} N_{\sigma}$ where
$N_{\sigma} = ( D + \sigma J) \text{/} 2$ is the number of orbitals occupied
with spin $\sigma$. For the two-particle part, we can first treat the
electrons as if they were classically distinguishable, yielding a contribution
whenever the {{states}} $( k, \sigma)$ and $( k', \sigma')$ are occupied.
This allows us to factorize the resulting expression $\left( \sum_{\sigma}
\tfrac{\sigma}{2} N_{\sigma} \right) \left( \sum_{\sigma' } \tfrac{\sigma'
}{2} N_{\sigma'} \right) = \tfrac{1}{4} N_s^2$ into the product of averages,
i. e., $\brkt{S_z}^2$ . We therefore call this a \emph{direct
(two-particle) contribution}. However, if $( k, \sigma) = ( k', \sigma')$, we
have to be careful: due to Pauli's principle, it is forbidden to annihilate
electrons in the same {{state}} twice, hence we have to {\emph{exclude}}
this possibility by a correction term $- \sum_{\sigma} \tfrac{\sigma
\sigma}{4} N_{\sigma} = - N_o$. We call this the {\emph{exchange
(two-particle) contribution}}, {{a denotation that will become more clear
in {\sec{sec:sqmAvEx}}.}} This yields altogether
\begin{eqnarray}
  \brkt{S^2_z} & = & \tfrac{1}{4} ( N_o + N_s^2 - N_o) = \brkt{S_z}^2, 
  \label{eq:longitudinal}
\end{eqnarray}
confirming the result trivially obtained in the macrospin picture (since
$\ket{\psi_0}$ is an eigenstate of $S_z$). The classical intuition is only
correct because of a non-trivial cancellation of a one-particle and
``quantum'' Pauli-exclusion term on a microscopic level. This importance of
this subtlety will become clear later (cf. Sec. \ref{sec:microscopic}).

We proceed with decomposing the transverse fluctuations into a one-particle
and a two-particle term,
\begin{equation}
  S_x^2 + S_y^2 = \sum_{i = x, y} \left[ \sum_a ( s_i^a)^2 + 2 \sum_{a < b}
  s_i^a s_i^b \right]
\end{equation}
\begin{equation}
  = \sum_{k \sigma} \tfrac{1}{2} c^{\dag}_{k \sigma} c_{k \sigma} + \sum_{k k'
  \sigma \sigma'} \tfrac{1 - \sigma \sigma'}{4} c^{\dag}_{k \bar{\sigma}}
  c^{\dag}_{k'  \bar{\sigma}'} c_{k' \sigma'} c_{k \sigma},
\end{equation}
and averaging over the ground state yields a non-vanishing one-particle term
$\tfrac{1}{2} \sum_{\sigma} N_{\sigma}$, describing transverse
{\emph{single}}-spin fluctuations. For the two-particle part, the direct term
vanishes as the individual spins are flipped in the modes $k$ and $k'$ so that
the ground state is not reproduced any more. This agrees with the fact that
the averages $\langle S_x \rangle = \langle S_y \rangle = 0$. However, we must
again treat the case $k = k'$ separately: when this mode is doubly occupied
and we have $\sigma' = \bar{\sigma}$, the sequence of the four field operators
together {\emph{exchanges}} the spins, reproducing the ground state. This
gives a correction $- N_{\downarrow}$, which is again due to Pauli's
principle: a configuration of two indistinguishable spins and the same
configuration with both spin exchanged cannot be told apart. This two-electron
exchange fluctuations together with the single-spin fluctuations make up for
the total transverse fluctuations of the macrospin,
\begin{eqnarray}
  & \begin{array}{lllll}
    \brkt{S^2_x + S^2_y} & = & \tfrac{1}{2} N_o - N_{\downarrow} & = &
    \brkt{S_z}
  \end{array} . &  \label{eq:transverse}
\end{eqnarray}
If we we next combine the longitudinal and the transverse term to obtain
$\brkt{Q_{z z}}$, we see that the one-particle contributions drop out:
\begin{eqnarray}
  \brkt{Q_{z z}} & = & \tfrac{1}{6} N_s^2 - \left( \tfrac{1}{6} N_o +
  \tfrac{1}{3} N_{\downarrow} \right) .  \label{eq:qzzsum}
\end{eqnarray}
As the SQM of a spin-1/2 vanishes (cf. the end of Sect. {\sec{sec:sqmT0}}),
the SQM exclusively measures true {\emph{two}}-spin correlations and does not
contain any single-spin information: The second bracket in {\Eq{eq:qzzsum}} is
a pure \emph{two-spin exchange }correction that accounts a kind of
``hole'' in the triplet correlations. The notion of this ``Pauli exclusion
hole'' will be explained prescisely in Sec. \ref{sec:microscopic}. It
physically arises from exchange contributions {\emph{both}} in $\brkt{S^2_z}$
and $\brkt{S^2_x + S^2_y}${\footnote{The exchange term is only by chance
proportional to $\brkt{S^2_x + S^2_y}$ at zero temperature. At finite
temperatures, this does not hold any more, showing that both terms are
physically quite distinct.}}. {\Eq{eq:qzzsum}} can be expressed as
\begin{eqnarray}
  \brkt{Q_{z z}} & = & \tfrac{1}{3} \times \tfrac{1}{2} N_s ( N_s - 1) . 
  \label{eq:QzzNs}
\end{eqnarray}
{{Thus, in the present case, the SQM counts the number of pairs of the
parallel spins in different half-filled orbitals.}} In accordance with the
macrospin picture, the doubly occupied orbitals can be simply ignored.
However, at finite temperatures, the Fermi edge becomes unsharp and modes
below the the electro-chemical potential $\mu$ also contribute to SQM. In
contrast to the macrospin picture, the present microscopic description already
included the entire Fermi sea into the description and can therefore be
extended to finite temperatures (see Sect. {\sec{sec:microscopic}}). For $T >
0$ we will also directly start from $\tens{Q}$ in second-quantized from, which
provides a clear way to demonstrate why SQM only senses spin-{\emph{triplet}}
correlations.

Importantly, these direct and exchange contribution to {\Eq{eq:qzzsum}} scale
differently with the number of polarized spins $N_s = \frac{J}{2 D} N_o$. For
a macroscopic ferromagnet, the exchange contribution to the SQM can be be
neglected due to the relative unimportance of excluding a single orbital among
many. In this case, SQM is entirely induced by spin-dipolarization. For $N_s
\rightarrow \infty$ the SQM per pair of polarized spins has only a finite
direct contribution of $1 / 3$ by {\Eq{eq:QzzNs}}, or alternatively, per
orbital $( J \text{/} 2 D)^2 \text{/} 3$. For mesoscopic ferromagnetic systems
with $N_s \sim 10 - 100$ polarized spins the exchange corrections start to
become relevant, and for magnetic molecular quantum dots in magnetic field
$N_s \sim 1 - 10$ and both terms can even be of comparable size. In both these
cases, the exclusion principle for a few quantum levels becomes relatively
important.

\subsection{Two Electrodes at $T > 0$}\label{sec:Tfinite}

We now extend the above analysis to two electrodes, which are, moreover, at
finite temperatures $T^L$ and $T^R$. This brings in two new aspects. First, in
{\Sec{sec:candsTfinite}}, we find that for finite temperatures that the
average SQM cannot be expressed anymore in the average spin as for $T = 0$.
The exchange SQM contribution is responsible for this difference, quantifying
pure quantum contributions to the anisotropy as we will see in
{\Sec{sec:sqmAvEx}}. This contribution involves a \ two-particle exchange DOS,
which is evaluated and discussed in {\Sec{sec:microscopic}}. This new quantity
is used to explain the notion of a ``Pauli exclusion hole'' in the triplet
spin correlations, which are encoded in the SQM. This provides the key to
understanding how quantum two-particle exchange processes allow for an SQM
current in the absence of spin-dipole current, the central result of the paper
in {\Sec{sec:spinFree}}.

The second new aspect, the subdivision of the system into smaller units,
touches upon the seemingly naive question of how to define an SQM current.
Clearly, an SQM current cannot quantify the ``amount'' of spin anisotropy that
flows {\emph{through}} a tunnel barrier as single tunneling electrons have
zero SQM: this idea only makes sense for a one-particle quantity such as
charge or spin. In contrast, SQM is a two-particle quantity, i. e., built up
by {\emph{pairs}} of electrons. As the electrons of a pair can stay at
different sides of the tunnel junction, SQM is not only stored
{\emph{locally}} in each ferromagnet, but also {\emph{non-locally}} between
the ferromagnets. The concept of storage of SQM thus needs to include
{\emph{nonlocal}} sources of SQM in addition to the local ones discussed so
far. In {\Sec{sec:SQMnetwork}} we develop a spin-multipole network theory to
aid the physical intuition and which will prove to be very helpful \ for the
discussion of SQM {\emph{transport}} later on and which has a wider range of
application than the model studied in this paper.

\subsubsection{Average Charge and Spin\label{sec:candsTfinite}}

In the following we calculate the average charge and spin-dipole moment in a
more technical way and in some more detail. We illustrate how to rewrite the
spin-dependent part of expectation values most elegantly in terms of
expressions independent of the choice of the coordinate system and of the spin
quantization axis. This serves as a good example of the manipulations we
present in {\App{app:covrealtime}} where we reformulate the real time
diagrammatic transport theory in an explicitly \emph{covariant} way.
Firstly, the one-particle operators {\eq{eq:chargeOp}} for the charge \ and
{\eq{eq:spinOp}} for the spin (now including the reservoir index $r$) are
jointly described by the four-component operator
\begin{eqnarray}
  R_{\mu}^r & = & \sum_{k, \sigma, \sigma'} ( r^r_{\mu})_{\sigma \sigma'}
  c^{\dag}_{r k \sigma} c_{r k \sigma'} .  \label{eq:chargespinOp}
\end{eqnarray}
Here $( r^r_{\mu})_{\sigma \sigma'} = \text{}_r \langle \sigma | r_{\mu} |
\sigma' \rangle_r$ denotes the matrix elements of the single-particle operator
$r_{\mu}$ for spin states quantized along $\op{\vec{J}}^r$. Using $r_0 =
\mathbbm{1}$ and $r_i = s_i$ ensures that $R_0 = N$ and $R_i = S_i$ for $i =
1, 2, 3$. We will from hereon distinguish whether the 0-component is included
or not by using Greek or Latin indices, respectively. Taking the average of
Eq. (\ref{eq:chargespinOp}) involves
\begin{eqnarray}
  \brkt{c_{r k \sigma}^{\dag} c_{r' k' \sigma'}} & = & f^r_+ (
  \varepsilon^r_{k \sigma}) \delta_{r r'} \delta_{k k'} \delta_{\sigma
  \sigma'} 
\end{eqnarray}
with the Fermi function
\begin{eqnarray}
  f^r_+ ( \omega) & = & \frac{1}{e^{( \omega - \mu^r) / T^r} + 1} . 
  \label{eq:occProb}
\end{eqnarray}
Recasting the sum over all $k$-modes as an integral over all energies by
inserting the DOS (see Eq. (\ref{eq:DOS})) yields
\begin{eqnarray}
  \brkt{R_{\mu}^r} & = & \sum_{\sigma , \sigma'} ( r^r_{\mu})_{\sigma
  \sigma'} \int \mathd \omega \delta_{\sigma \sigma'} \nu_{\sigma} f^r_+, 
  \label{eq:rrmu}
\end{eqnarray}
where we suppressed the $\omega$-dependence for brevity. Using Eq.
(\ref{eq:eigenspin}), i. e. $\left( \vec{J}^r \cdot \vec{s} \right) | \sigma
\rangle_r = \sigma | \sigma \rangle_r$, we may rewrite
\begin{eqnarray}
  \nu^r_{\sigma} ( \omega) \delta_{\sigma \sigma'} & = & \bar{\nu}^r ( \omega)
  \text{}_r \bra{\sigma} \check{\vec{n}}^r ( \omega) \cdot \check{\vec{r}}
  \ket{\sigma'}_r, 
\end{eqnarray}
introducing $\check{r}_0 = \mathbbm{1} \text{/} \sqrt{2}$ and $\check{\vec{r}}
= \sqrt{2} \vec{s}$ \ and the four-component vector $\widehat{\vec{n}}^r =
\sqrt{2} (1, \widehat{\vec{J}}^r n^r)$. The spin-dependent part of Eq.
(\ref{eq:rrmu}) can be recast as a trace in spin space:
\begin{eqnarray}
  \left\langle \vec{R}^r \right\rangle & = & \int \mathd \omega \bar{\nu}^r
  f^r_+  \Tr  \left[ \vec{r}  \left( \check{\vec{n}}^r \cdot \check{\vec{r}}
  \right) \right] .  \label{eq:covChargeSpin}
\end{eqnarray}
The trace is clearly {\emph{covariant}} in the general sense, i.e.,
form-invariant under changes of either the coordinate system and / or
quantization axis (it is not related to concepts from relativity; vectors with
four elements are just convenient). We obtain
\begin{eqnarray}
  \brkt{N^r} & = & \int d \omega 2 \bar{\nu}^r ( \omega) f^r_+ ( \omega), 
  \label{eq:avparticle}\\
  \brkt{\vec{S}^r} & = & \int d \omega 2 \bar{\nu}^r ( \omega) s^r ( \omega)
  \widehat{\vec{J}}^r,  \label{eq:avspin}
\end{eqnarray}
Analogous to the \emph{average occupation number} of a single level at
energy $\omega$ in {\eq{eq:avparticle}}, $f^r_+ ( \omega)$, we denote
\begin{eqnarray}
  s^r ( \omega) & = & f^r_+ ( \omega) \tfrac{1}{2} n^r ( \omega) 
  \label{eq:polProb}
\end{eqnarray}
in {\Eq{eq:avspin}} as the {\emph{average spin-polarization function}} of
electrons at frequency $\omega$, where $n^r ( \omega)$ is the
spin-polarization {\eq{eq:spinpol}}. Note that we only needs to use spin 1/2
operator algebra to calculate the average in {\Eq{eq:covChargeSpin}} in
coordinate-free form and the same can be done for all the less transport
calculations, see {\App{app:covrealtime}}.

\subsubsection{Network Picture: Non-Locality}\label{sec:SQMnetwork}

Eqs. (\ref{eq:avparticle}) and (\ref{eq:avspin}) show that each physical
electrode \ corresponds to a single source of charge and spin. We now
formalize the concept of particle and spin-dipole storage \ in terms of a
{\emph{network theory}}, which at first sight may seem superfluous. In fact,
it will prove to be helpful to compare this with the storage and transport of
spin-quadrupole moment.

The following considerations are formulated more compactly and hold more
generally for a composite system of any number of subsystems labeled by an
index $r$. Such a system may comprise of just two electrodes, each at
equilibrium, as discussed in this paper (then $r = L, R$), but it may also
include, e.g., strongly interacting quantum dots out of equilibrium as
discussed in {\cite{Sothmann10,Baumgaertel11,Misiorny13}} and in forthcoming
works. We first ask how the total charge and spin-dipole moment is distributed
over the subsystems. The answer is fairly intuitive for these one-particle
quantities: the total charge (spin) is the sum of the charge (spin) stored in
each electrode, i. e.,
\begin{eqnarray}
  R^{\mathrm{tot}}_{\mu} & = & \sum_r R^r_{\mu} .  \label{eq:oneParticleDec}
\end{eqnarray}
We can simply associate each subsystem shown in Fig. \ref{fig:network} (a)
with a {\emph{node}} of charge (spin) as depicted in Fig. \ref{fig:network}
(b). Note that decomposition {\eq{eq:oneParticleDec}} is even possible if
$R^{\mathrm{tot}}_{\mu}$ is {\emph{not}} conserved. (We postpone the discussion
of the {\emph{links}} in the network until we defined current operators in Sec
{\sec{sec:spinMultipoleNetworks}} where we complete the network theory.)

{
  \begin{figure}[h]
    \includegraphics[width=0.9\linewidth]{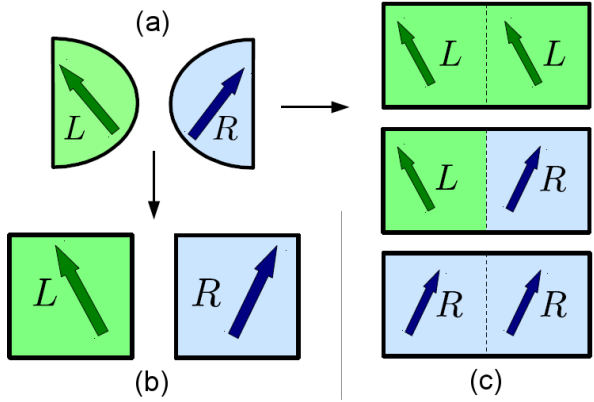}
    \caption{(a) Physical setup of two ferromagnets
and network picture for (b) the spin-dipole moment, a one-particle quantity
(like the charge) and for (c) the spin-quadrupole moment, a two-particle
quantity.\label{fig:network}}
\end{figure}}

This simple correspondence breaks down for SQM. When we ask how this
{\emph{two-particle}} quantity is distributed over composite system, the
answer is radically different. We start from the total SQM of the system,
written as
\begin{eqnarray}
  \tens{Q}^{\mathrm{tot}} & = & \vec{S}^{\mathrm{tot}} \odot \vec{S}^{\mathrm{tot}} 
  \label{eq:Qtot}
\end{eqnarray}
abbreviating the symmetric, traceless dyadic product of two vector operators
$\vec{a}$ and $\vec{b}$ as
\begin{eqnarray}
  \left( \vec{a} \odot \vec{b} \right)_{i j} & = & \tfrac{1}{2} ( a_i b_j +
  b_i a_j) - \tfrac{1}{3} \delta_{i j}  \vec{a} \cdot \vec{b} 
  \label{eq:symdyad}
\end{eqnarray}
We decompose $\tens{Q}^{\mathrm{tot}}$ by inserting $\vec{S}^{\mathrm{tot}} =
\sum_r \vec{S}^r$,
\begin{eqnarray}
  \tens{Q}^{\mathrm{tot}} & = & \sum_{\langle r r' \rangle} \tens{Q}^{r r'}, 
\end{eqnarray}
where $\langle r r' \rangle$ indicates that we sum only over all
{\emph{pairs}}
\begin{eqnarray}
  & \begin{array}{lllll}
    \tens{Q}^{r r'} & = & \tens{Q}^{r' r} & = & g^{r r'}  \vec{S}^r \odot
    \vec{S}^{r'},
  \end{array} &  \label{eq:Qalphaalphaprime}
\end{eqnarray}
and the factor $g^{r r} = 1$ and $g^{r r'} = 2$ ($r \neq r'$) accounts for the
double occurrence of each pair $r, r'$ with $r \neq r'$ in the expansion
{\eq{eq:Qtot}}. {\Eq{eq:Qalphaalphaprime}} is symmetric in $r$ and $r'$ since
\ $\vec{S}^r \odot \vec{S}^{r'} = \vec{S}^{r'} \odot \vec{S}^r$ and we can
write
\begin{eqnarray}
  \tens{Q}^{r r'} & = & \tfrac{1}{2} g^{r r'} ( \vec{S}^r \odot \vec{S}^{r'} +
  \vec{S}^{r'} \odot \vec{S}^r) .  \label{eq:Qaaprimesym}
\end{eqnarray}
Note that $\tens{Q}^{r r'}$ is a Hermitian operator and therefore an
observable because spin operators of different subsystems commute: $(
\vec{S}^r \odot \vec{S}^{r'})^{\dag} = \vec{S}^{r'} \odot \vec{S}^r =
\vec{S}^r \odot \vec{S}^{r'}$.

We now develop a network picture for the SQM by associating to each
\emph{pair} of subsystems $\langle r r' \rangle$ a single effective source
or node. For the \emph{two}-terminal spin-valve in Fig.
\ref{fig:network}(a) that we study, \emph{three} SQM-nodes appear in the
corresponding network picture of Fig. \ref{fig:network} (c). The total SQM is
stored in two \emph{local} nodes ($\tens{Q}^{L L} $, $\tens{Q}^{R
R}$) and in one \emph{non-local} node ($\tens{Q}^{L R} = \tens{Q}^{R L}
$). The (non)local nodes describe spin-triplet correlations between
pairs of electrons of the same (different) subsystem(s). {{This
non-locality of SQM storage is very important for the physical understanding
and definition of a SQM current operator.}} It is the injection of SQM
currents from these non-local nodes that drive the measurable local SQM
dynamics in embedded quantum dots, as found in Ref. {\cite{Baumgaertel11}}.
For the spin-valve considered here it now becomes clear how single electron
tunneling can transport SQM: first, local correlations, e. g., in the $\langle
L L \rangle$-node are turned into non-local correlations in the $\langle L R
\rangle$-node. The transfer of SQM is then completed by another
single-electron tunneling event that re-localizes the pair, but now in the
right electrode, contributing then to the $\langle R R \rangle$-node. This
picture will be refined once we defined SQM current operators in
{\Sec{sec:SQMcurrent}}.

\subsubsection{Direct and Exchange Contribution to Average
SQM}\label{sec:sqmAvEx}

{{We next inquire to which extent the stored SQM is independent of the
average spin-dipole moment, extending the discussion of
{\Sec{sec:SQMT0micro}}.}} The average of the SQM operator for node $\langle r
r' \rangle$ given by {\eq{eq:Qalphaalphaprime}}, can be decomposed it into a
direct and an exchange part using Wick's theorem for the averages of products
of field operators (see {\App{app:tripletCorr}} for details):
\begin{equation}
  \begin{array}{lll}
    \brkt{\tens{Q}^{r r'}} & = & \brkt{\tens{Q}^{r r'}}_{\mathrm{dir}} +
    \brkt{\tens{Q}^{r r'}}_{\mathrm{ex}}
  \end{array} .
\end{equation}
\emph{Direct SQM}. \ The first possible {\emph{direct}} contraction
combines field operators from the same spin operator in {\Eq{eq:Qrrzz}}. It
can therefore be factorized into the expectation values of the spin operators
given by {\Eq{eq:avspin}}:
\begin{eqnarray}
  \brkt{\tens{Q}^{r r'}}_{\mathrm{dir}} & = & \sum_{k k' \sigma \sigma'}
  \vec{s}^r_{\sigma \sigma} \odot \vec{s}^{r'}_{\sigma' \sigma'} f_+^r (
  \varepsilon^r_{k \sigma}) f^{r'}_+ ( \varepsilon^{r'}_{k' \sigma'}) 
  \label{eq:QDirFermi}\\
  & = & \brkt{\vec{S}^r} \odot \brkt{\vec{S}^{r'}} = q^{r r'}_{\text{dir}} 
  \widehat{\vec{J}}^r \odot \widehat{\vec{J}}^{r'}  \label{eq:redcontr}
\end{eqnarray}
with
\begin{eqnarray}
  q_{\text{dir}}^{r r'} & = & | \brkt{\vec{S}^r} | \, | \brkt{\vec{S}^{r'}} |.
  \label{eq:Qdir}
\end{eqnarray}
This direct SQM incorporates the cumulative effect of the energy resolved
spin-polarization $s^r ( \omega)$. It quantifies the uncorrelated contribution
of the quantum spins to the spin anisotropy: as intuitively expected, an
electrode with a{{ favoured spin {\emph{direction}} (polarization)
possesses a favoured spin {\emph{alignment}} (anisotropy)}}. For a macroscopic
system in equilibrium, the average SQM is dominated by the direct part, which
is completely determined by the average spin-dipole moment.

\emph{Exchange SQM}. For meso- and \ nanoscopic systems the last statement
ceases to be true due to the neglect of the Pauli's principle in the spin-spin
correlations. In the second {\emph{exchange}} contraction field operators of
different spin operators are contracted, giving a term
\begin{eqnarray}
  \brkt{\tens{Q}^{r r'}}_{\mathrm{ex}} & = & \delta^{r r'} \sum_{k \sigma
  \sigma'} \vec{s}^r_{\sigma \sigma'} \odot \vec{s}^{r'}_{\sigma' \sigma}
  f_+^r ( \varepsilon^r_{k \sigma}) f^r_+ ( \varepsilon^r_{k \sigma'}), 
  \label{eq:exCont}
\end{eqnarray}
which accounts for true {\emph{correlations}} in the sense of Spearman's rank
correlation coefficient{\footnote{{{Here, one has to treat the spin vector
as a stochastic variable when averaging over the grand-canonical ensemble.
}}Spearman's rank correlation coefficient $C_{A B}$ for two random variables
$A $, $B$ is defined by $C ( A, B) = \brkt{\left( A - \brkt{A} \right)
\left( B - \brkt{B} \right)}$. {\Eq{eq:correlation}} is a linear combination
of the $C ( S_i^r, S^{r'}_j)$ for different components of the spin operator so
that only triplet correlations are extracted. However, if all $C ( S_i^r,
S_j^{r'}) = 0$, this implies $\brkt{\tens{Q}^{r r'}}_{\mathrm{ex}} = 0$}}. This
becomes clear when rewriting {\Eq{eq:exCont}} using {\Eq{eq:redcontr}}:
\begin{eqnarray}
  \brkt{\tens{Q}^{r r'}}_{\mathrm{ex}} & = & \vec{} \brkt{\vec{S}^r -
  \brkt{\vec{S}^r}} \odot \vec{} \brkt{\vec{S}^{r'} - \brkt{\vec{S}^{r'}}} . 
  \label{eq:correlation}
\end{eqnarray}
Note that {\Eq{eq:exCont}} involves only {\emph{one}} sum over $k$.
{{Thus, the exchange term indeed scales linearly with the system size in
contrast to the direct term }}(see {\Eq{eq:redcontr}}) and can be neglected
for macroscopic systems (cf. last paragraph in Sect. {\sec{sec:SQMT0micro}}).
Here it is interesting to consider our Hamiltonian as a model for a mesoscopic
ferromagnet or a metallic island in a strong external magnetic field,
{\Fig{fig:tunneljunction}}(b). In this case the exchange contribution may even
become the dominant part in transport when the spin current vanishes: then the
spin-polarization $\brkt{\vec{S}^r}$ and therefore also $\brkt{\tens{Q}^{r
r'}}_{\mathrm{dir}}$ do not change, while $\brkt{\tens{Q}^{r r'}}_{\mathrm{ex}}$
does. When including a tunnel-coupling between the ferromagnets the transport
through the junction correlates spins of both systems and non-local exchange
SQM currents \emph{can} indeed arise. For this reason, we keep the
exchange term here and study it in some more detail.

\emph{Tensorial structure.} {\Eq{eq:correlation}} can be expressed as
\begin{eqnarray}
  \brkt{\tens{Q}}_{\mathrm{ex}} & = & - \delta^{r r'} q^{r r}_{\mathrm{ex}} 
  \widehat{\vec{J}}^r \odot \widehat{\vec{J}}^r  \label{eq:QexTensor}
\end{eqnarray}
with the positive quantity
\begin{eqnarray}
  q_{\mathrm{ex}}^{r r} & = & \tfrac{1}{4} \sum_k ( f_+^r ( \varepsilon^r_{k
  \uparrow}) - f_+^r ( \varepsilon^r_{k \downarrow}))^2 > 0. 
  \label{eq:qexksum}
\end{eqnarray}
(see {\App{sec:spinTraceTechnique}}). Clearly, only if $\varepsilon_{k
\uparrow} - \varepsilon_{k \downarrow} \ll T$ for all $k$, the exchange
contribution vanishes, i.e., for the Stoner model if $J \ll T$. However, if $J
< T$, each spin-polarized orbital $k$ gives a negative correction to the
direct spin anisotropy. We will refer to this as the {\emph{Pauli exclusion
hole}}, located in orbital $k$ with a ``distribution function'' $( f (
\varepsilon_{k \uparrow}) - f ( \varepsilon_{k \downarrow}))^2$. We give a
microcopic interpetation of this below in Sec. \ref{sec:microscopic}. A
gradient of these Pauli exclusion holes across the junction results drives an
exchange SQM current, which may even flow in the absence of a spin current,
see {\Sec{sec:spinFree}}.

We moreover note that the tensor $\brkt{\tens{Q}}_{\mathrm{ex}}$ has the same
principal axes as $\brkt{\tens{Q}}_{\text{dir}}$ (the reason for this is
discussed in the following section). Thus, the local SQM $\left\langle
\tens{Q}^{r r} \right\rangle \propto \widehat{\vec{J}}^r \odot
\widehat{\vec{J}}^r$, has a diagonal representation in any coordinate system
that includes the Stoner field direction, e.g., $\vec{e}_z =
\widehat{\vec{J}}^r$, with non-zero elements $\brkt{Q^{r r}_{x x}} =
\brkt{Q^{r r}_{y y}} = - \brkt{Q^{r r}_{z z}} \text{/} 2$. {{This shows
that the local SQM sources are {\emph{uniaxially}} anisotropic, and different
alignments in the plane perpendicular to $\op{\vec{J}}^r$ are not preferred}}.
Since the direct contribution exceeds the exchange contributions, we find as
expected$ \brkt{Q^{r r}_{z z}} > 0$, i.e., an easy-axis anisotropy favoring
the collinear orientation of the spins into the $z$-direction over any
orientation in the $x y$-plane, $\brkt{Q^{r r}_{x x}} = \brkt{Q^{r r}_{y y}} <
0$.

{{The non-local SQM $\brkt{\tens{Q}^{r r'}}$, $r \neq r'$, has three
non-degenerate principal values: it describes \emph{bi-axial
anisotropy}.}} It has a unique principal axes in which $\brkt{Q^{r r}_{z z}} >
\brkt{Q^{r r}_{y y}} > \brkt{Q^{r r}_{x x}}$, i.e., directions perpendicular
to the dominant easy axis ($z$) are distinguished, see
{\App{sec:appDiagonal}}.

\subsubsection{Microscopic Picture of SQM Storage}\label{sec:microscopic}

The physical meaning of the exchange SQM becomes transparent when revisiting
the microscopic picture of SQM storage. When calculating the direct SQM by
{\Eq{eq:QDirFermi}} one pretends to have to two distinct ferromagnets $r$ and
$r'$ and ``counts'' triplet correlations by adding all cross-correlations
between electrons occupying these distinguishable ferromagnets. This procedure
gives the full result for the non-local SQM (cf. {\Eq{eq:exCont}}): for $r
\neq r'$
\begin{eqnarray}
  \brkt{\tens{Q}^{r r'}} & = & \brkt{\tens{Q}^{r r'}}_{\mathrm{dir}} . 
\end{eqnarray}
This is correct as we we treat the two ferromagnets as distinguishable objects
by assumption (the total density operator is a direct product).

The direct, {\emph{local}} SQM $( r = r')$ also correctly ``counts'' the
local spin anisotropy as long as it concerns correlations of electrons from
different modes $k \neq k'$, which are also distinguishable (green lines in
{\Fig{fig:sqmmicroEx}}). However, this procedure fails for electrons occupying
the {\emph{same}} mode $k' = k$: a single mode (irrespective of whether being
singly or doubly occupied) does not contribute to the total SQM (see
{\App{app:tripletCorr}}). {{Thus, the local exchange SQM has to cancel the
contribution that the direct SQM \ {\eq{eq:QDirFermi}} \ incorrectly ascribes
to single modes (indicated by the red line in {\Fig{fig:sqmmicroEx}})}}

{
  \begin{figure}[h]
    \includegraphics[width=0.9\linewidth]{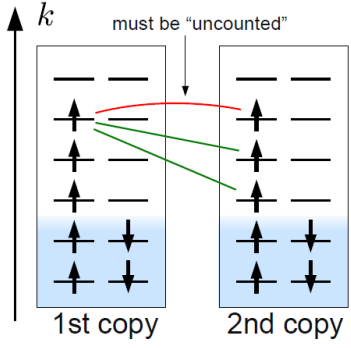}
    \caption{Microscopic contributions to the
local SQM $\brkt{\tens{Q}^{r r}}$. Two copies of the same ferromagnet are
considered and green lines indicate correlations between pairs of
distinguishable electrons in different orbitals (counted by the direct SQM).
The red line indicates the correlations between indistinguishable electrons in
the same orbital that the direct SQM counts {\emph{too much}}: according to
Pauli's principle two electrons cannot form a triplet state in the same
orbital. The exchange SQM contribution takes care of this and thus represents
a Pauli hole in the correlations, corresponding to the red line. When
considering only the 1st copy at finite temperature, the macrospin picture
discussed in Sect. {\sec{sec:sqmT0}} is recovered. For finite temperature, the
occupation probabilities are \ thermally smeared at the Fermi
edge.\label{fig:sqmmicroEx}}
\end{figure}}

For establishing an ``uncounting'' procedure to exclude the
single-m{\emph{}}ode SQM, one may again simply think of two identical, but
distinguishable copies of the same mode $k$ and calculate the direct SQM
generated from all these modes (see {\Fig{fig:sqmmicroEx}}). In this picture,
exchange SQM represents a ``spin-anisotropy hole'' ascribed to each mode and
therefore shows a formal analogy to a {\emph{one-particle quantity}}. This
analogy will reemerge when we consider the transport of SQM in Sect.
{\sec{sec:energyResolved}}.{{ To emphasize this multi-particle exchange
aspect, we will refer to this as a {\emph{Pauli exclusion hole}} in the spin
triplet correlations.}}

As a consequence, local exchange SQM must have the same tensorial structure
as the direct SQM, but with opposite magnitude, which is explicitly conveyed
by the negative sign in {\Eq{eq:QexTensor}}. Since $q_{\text{dir}}^{r r} > 0$
(by {\Eq{eq:Qdir}}), it follows also that $q_{\text{ex}}^{r r} > 0$ must hold.
This is confirmed explicitly by {\Eq{eq:qexksum}}, which shows that the
exchange SQM {{senses the spin alignment, a non-negative quantity that
accumulates when summing over all energies or $k$-modes, respectively. This
prohibits cancellations of signed contributions as they occur in the
spin-dipole moment.}} This means that spin-dipole moment may cancel whence
SQMs do {\emph{not}}. {\Eq{eq:qexksum}} also shows explicitly that exchange
corrections become negligible at high temperatures, i. e., if $T^r \gg
\varepsilon^r_{k \uparrow} - \varepsilon^r_{k \downarrow}$ for all $k$, as
expected.

\subsubsection{Energy-Resolved Exchange SQM Storage}\label{sec:energyResolved}

{{So far, it was helpfull to discuss the microscopic picture of SQM
storage in terms of contributions from \emph{orbitals} $k$. However, to
make progress in calculations we replace the $k$-sums by energy integrals. An
{\emph{energy-resolved}} picture of SQM storage will therefore be important
for understanding the key features of SQM transport compared to charge and
spin see {\Sec{sec:spinFree}}.}} For the rest of this chapter, we will only
discuss the local exchange SQM, i. e., $r' = r$, and therefore drop the
electrode index for brevity. Replacing the sum over $k$ in Eq.
(\ref{eq:qexksum}) by integrals over frequencies $\omega, \text{ } \omega'$
and inserting the {\emph{two-particle density of states}} {\eq{eq:twoDOS}}, we
can recast the exchange SQM into the form of {\Eq{eq:exCont}} after carrying
out the spin sum (see {\App{sec:appMultipole}}). The SQM exchange
{\emph{magnitude}} then reads as
\begin{eqnarray}
  q_{\mathrm{ex}} & = & \int \mathd \omega \bar{\nu} ( \omega) q_{\mathrm{ex}} (
  \omega)  \label{eq:Qtriplet}
\end{eqnarray}
The \emph{average exchange spin-quadrupolarization} for electrons at
frequency $\omega$,
\begin{eqnarray}
  q_{\mathrm{ex}} ( \omega) & = & f_+ ( \omega) a ( \omega), 
  \label{eq:quadProb}
\end{eqnarray}
with the {\emph{spin-anisotropy function}}
\begin{eqnarray}
  a ( \omega) & = & \sum_{\sigma} a_{\sigma} ( \omega), 
  \label{eq:tripletCorr}
\end{eqnarray}
and
\begin{eqnarray}
  \bar{\nu} ( \omega) a_{\sigma} ( \omega) & = & \int \mathd \omega' f_+ (
  \omega') \sum_{\sigma'} \tfrac{\sigma \sigma'}{4} \nu_{\sigma \sigma'} (
  \omega, \omega') .  \label{eq:asigma}
\end{eqnarray}
valid for general dispersion relations. Note that the integrand in
(\ref{eq:Qtriplet}) is {\emph{not}} a positive function, in contrast to each
term in Eq. (\ref{eq:qexksum}). For the discussion of the SQM currents, it
will be important to understand the meaning of the function $q_{\mathrm{ex}} (
\omega)$: it quantifies the cumulative exchange triplet correlation for
electrons occupying the same orbital. It is the formal analogue to the average
spin-polarization function $s ( \omega)$. To link the above result further to
the microscopic picture developed in Sec. {\sec{sec:microscopic}} and to
simplify the interpretation of the exchange SQM current in Sec.
{\sec{sec:firstOrder}}, we decomposed the spin-anisotropy function $a (
\omega)$ into its spin-dependent contributions $a_{\sigma} ( \omega)$: they
give the direct single-mode SQM, provided that an electron with spin $\sigma$
is present at frequency $\omega$ in the first copy while summing over the
contributions from the second copy in Fig. \ref{fig:sqmmicroEx} (cf.
{\App{app:spinAnisotropy}}). This reveals the formal anlogy between $a (
\omega)$ and average spin-polarization function in {\Eq{eq:polProb}},
{{given by \ $n ( \omega) \text{/} 2 \text{}$. The latter quantifies the
average spin at frequency $\omega$, provided we have full occupation is at
this frequency.}} However, in stark contrast to the latter, $a ( \omega)$ is
not solely a band structure property as it depends on a Fermi function due to
its two-particle origin. {{Note that the exchange SQM is entirely
described by $q_{\mathrm{ex}} ( \omega)$ and that the spin polarization $s (
\omega)$ does not enter, in contrast to the direct SQM. These two functions
have very different temperature and energy dependence, again making explicit
that the SQM is independent of the spin-polarization due to the presence of
exchange terms. }}

The functions $q_{\mathrm{ex}} ( \omega)$ and $a ( \omega)$ are of key
importance for the results of this paper and we will therefore explain their
basic physical meaning using the simple Stoner model $\varepsilon_{k \sigma} =
\varepsilon_k - \sigma J \text{/} 2$ and the flat band approximation (cf.
{\sec{sec:wide-flat-band}}). \ Then the spin-anisotropy function $a ( \omega)$
has the spin-resolved contributions
\begin{eqnarray}
  a_{\sigma} ( \omega) & = & \tfrac{1}{4} [ f_+ ( \omega) - f_+ ( \omega +
  \sigma J)] .  \label{eq:asigmaSimple}
\end{eqnarray}
In Fig. \ref{fig:anisotropyocc} we plot these two contributions and their sum
together with the average spin quadrupolarization $q_{\mathrm{ex}}$.

{
  \begin{figure}[h]
    \includegraphics[width=0.9\linewidth]{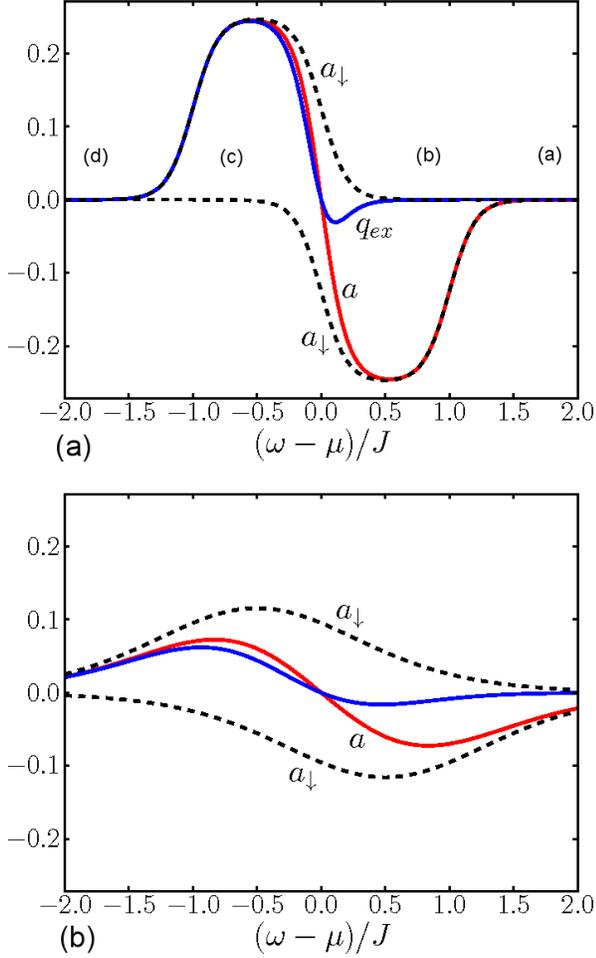}
    \caption{\label{fig:anisotropyocc}Average local
exchange quadrupolarization $q_{\mathrm{ex}} ( \omega)$ (blue), spin anisotropy
$a ( \omega)$ (red) and its two contributions $a_{\uparrow} ( \omega) > 0$
(broken line) and \ $a_{\downarrow} ( \omega) < 0$ (broken line) as function
of $( \omega - \mu) \text{/} J$ for two temperatures $T \text{/} J = 0.1$ \ in
(a) and $T \text{/} J = 0.5$ in (b). As $T$ approaches $J$ from below, the
weight of $a_{\uparrow} ( \omega)$ ($a_{\downarrow} ( \omega)$) considerably
shrinks (rises) and $q_{\mathrm{ex}} ( \omega)$ is strongly suppressed. See also
Sect. {\sec{sec:maxQu}}.}
\end{figure}}

We first discuss the shape of $a_{\sigma} ( \omega)$ for $\sigma = \uparrow,
\downarrow$ for low temperature $4 T \text{/} J < 1$ translating the arguments
of the microscopic picture of {\Fig{fig:sqmmicroEx}} into energy space in
{\Fig{fig:amicro}}. As mentioned, the function $a_{\sigma} ( \omega)$
characterizes the single-orbital SQM for a spin $\sigma$ occupying a mode at
energy $\omega$ and displays four regimes (in the following $\sim$ means `''up
to thermal smearing $T$''). These are marked (a)-(d) in
{\Fig{fig:anisotropyocc}} (a) and correspond to the regimes in
{\Fig{fig:amicro}}. We discuss them now in detail:

(a) $\omega \gtrsim \mu + J$ $\Rightarrow$ $a_{\uparrow} ( \omega) =
a_{\downarrow} ( \omega) = 0$: there are no occupied states at energy
$\omega$, so no exchange correction is needed.

(b) $\mu \lesssim \omega \lesssim \mu + J$ $\Rightarrow$ $a_{\downarrow} (
\omega) < 0, a_{\uparrow} ( \omega) = 0$: if a spin-$\uparrow$ is in the first
copy, the corresponding mode in the second copy has vanishing probability to
be occupied with electrons of {\emph{any}} spin since both $\varepsilon_{k
\uparrow} = \omega \gtrsim \mu$ and $\varepsilon_{k \downarrow} = \omega + J
\gtrsim \mu$. Thus, similar to regime (a), no exchange correction for
spin-$\uparrow$ electrons is needed and we obtain $a_{\uparrow} ( \omega) =
0$. In contrast, if a spin-$\downarrow$ \ is in the first copy, the
corresponding mode in the second copy is predominantly occupied with
spin-$\uparrow$ because $\varepsilon_{k \downarrow} = \omega \gtrsim \mu$, but
$\varepsilon_{k \uparrow} = \omega - J \lesssim \mu$. This contributes
{\emph{negatively}} to direct SQM, resulting in $a_{\downarrow} ( \omega) <
0$.

(c) $\mu - J \lesssim \omega \lesssim \mu$ $\Rightarrow$ $a_{\uparrow} (
\omega) > 0, a_{\downarrow} ( \omega) = 0$: if in this case a spin-$\uparrow$
is in the first copy, the corresponding mode in the second copy is also mostly
occupied with spin-$\uparrow$ since $\varepsilon_{k \uparrow} = \omega
\lesssim \mu$, but $\varepsilon_{k \downarrow} = \omega + J \gtrsim \mu$. This
gives a {\emph{positive}} correction to the direct SQM. In contrast,
$a_{\downarrow} ( \omega) = 0$ as $\varepsilon_{k \downarrow} = \omega$ and
$\varepsilon_{k \uparrow} = \omega - J$ refers to a mostly {\emph{doubly}}
occupied mode in the second copy, which has a vanishing direct SQM
contribution (cf. case (d)).

{{(d) $\omega \lesssim \mu - J$ $\Rightarrow$ $a_{\uparrow} ( \omega) =
a_{\downarrow} ( \omega) = 0$: the corresponding orbital deep inside the Fermi
sea is doubly occupied: $f ( \varepsilon_{k \uparrow}) = f ( \varepsilon_{k
\downarrow}) = 1$. By \ {\Eq{eq:qexksum}} the direct SQM due to both
spin-$\uparrow$ and spin-$\downarrow$-electrons cancel each other.}}

{
  \begin{figure}[h]
    \includegraphics[width=0.9\linewidth]{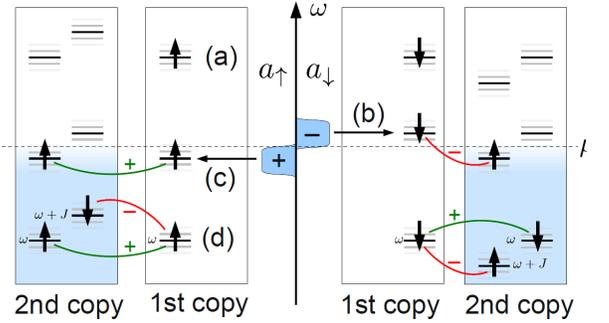}
    \caption{Microscopic picture of the
spin-quadrupolarization function $a_{\sigma} ( \omega)$, $\sigma = \uparrow,
\downarrow$ characterizing the \emph{energy} resolved spin-anisotropy
content of a ferromagnet (cf. {\Fig{fig:sqmmicroEx}}), see
text.\label{fig:amicro}}
\end{figure}}

Altogether, the anisotropy function $a ( \omega) = \sum_{\sigma} a_{\sigma} (
\omega)$ is exactly \emph{antisymmetric} with respect to the
electrochemical potential (see {\Fig{fig:anisotropyocc}} and
{\App{app:tripletCorr}})
\begin{eqnarray}
  a ( \mu + \omega) & = & - a ( \mu - \omega) . 
\end{eqnarray}
As mentioned at the outset, the average exchange quadrupolarization
$q_{\mathrm{ex}} ( \omega) = f_+ ( \omega) [ a_{\uparrow} ( \omega) +
a_{\downarrow} ( \omega)]$ has both positive and negative contributions;
however, $q_{\mathrm{ex}} > 0$ as the integrated $q_{\mathrm{ex}} ( \omega)$ in
{\Eq{eq:Qtriplet}} is always positive by {\Eq{eq:qexksum}}. At $T = 0$ only
positive correlations at $\omega < \mu$ count, and we recover the result
{\eq{eq:transverse}}. For $T > 0$, thermally excited spins $\downarrow$
negatively correlate with spins $\uparrow$ in the same orbital, thus reducing
$q_{\mathrm{ex}}$ (see {\Fig{fig:anisotropyocc}}). Eventually at $T \gg J$ this
cancellation reduces $q_{\mathrm{ex}}$ exactly to zero without ever becoming
negative. We now see explicitly that the exchange SQM only becomes
{\emph{thermally}} suppressed for $T \gg J$.

The average exchange quadrupolarization makes explicit that Pauli-forbidden
triplet correlations are stored by electrons in an energy window $\sim 2 J$
with opposite signs above and below the Fermi energy. Thus, the integrated
exchange quadrupolarization is {\emph{thermally}} suppressed for $T \gg J$
when the occupation probability \ is nearly constant across the energy scale
$J$.

\subsubsection{Parameter Dependence of Average Exchange SQM}\label{sec:maxQu}

In the flat band approximation (cf. Sect. {\sec{sec:wide-flat-band}}), \ the
integral {\eq{eq:Qtriplet}} can be carried out yielding (see
{\App{app:closedFormula}})
\begin{eqnarray}
  q_{\mathrm{ex}} & = & \left. \frac{\bar{\nu} T}{2} \left[ \frac{J}{2 T} \coth
  \left( \frac{J}{2 T} \right) \right. - 1 \right],  \label{eq:qrex}
\end{eqnarray}
which is positive since $x \coth ( x) > 1$, in agreement with the above
discussion. In the limit $J \text{/} T \rightarrow 0$, $q_{\mathrm{ex}}$
vanishes as it should (see above) and in the opposite limit of $T \text{/} J
\rightarrow 0$, $q_{\mathrm{ex}}$ scales linearly with $N_s$, the number of free
spins in the ferromagnet (cf. {\Eq{eq:Ns}}),
\begin{eqnarray}
  & \begin{array}{lllll}
    q_{\mathrm{ex}} |_{T = 0} & = & \tfrac{1}{4} \bar{\nu} J & = & \tfrac{1}{4}
    N_s
  \end{array}, &  \label{eq:qrex0}
\end{eqnarray}
in accordance with the $T = 0$ result {\eq{eq:qzzT0}}{\footnote{For
$\widehat{\vec{J}}^{} = \vec{e}_z$ we get $\brkt{\tens{Q}^{}}_{\mathrm{ex}} = -
q_{\mathrm{ex}}  \vec{e}_z \odot \vec{e}_z = - q_{\mathrm{ex}} [ \tfrac{2}{3}
\vec{e}_z \vec{e}_z - \tfrac{1}{3} ( \vec{e}_x \vec{e}_x + \vec{e}_y
\vec{e}_y)]$, that is, $\brkt{Q_{z z}}_{\mathrm{ex}} = - \tfrac{2}{3}
q_{\mathrm{ex}} = - \tfrac{2}{3} \frac{N_s}{4} = - \tfrac{1}{3} \brkt{S_z}_{T =
0}$, in agreement with {\eq{eq:qzzT0}}.}}. The average one-particle
spin-dipole moment $\brkt{S_z}_{T = 0} = \tfrac{1}{2} N_s$ thus basically
serves as a reference scale for two-particle $q_{\mathrm{ex}}$ (when multiplied
by $\hbar = 1$ in our units). The low $T < J$ behavior is most interesting
because in the regime $J < T$ the results do not apply to a Stoner
ferromagnet, for which the self-consistent magnetization $J$ would break down
(our model {\Fig{fig:tunneljunction}} (a) assumes fixed $J$). However,
considering our model as a description of mesoscopic islands in an external
magnetic field of strength $B = J$ ({\Fig{fig:tunneljunction}} (b), this range
may also be relevant. With this in mind we show in {\Fig{fig:sqmdensint}} the
pronounced temperature dependence of the exchange SQM (normalized to the spin
polarization) over the entire range for fixed $J$. This should be contrasted
with the average spin for which all $T$-dependence completely cancels out due
to the constancy of the assumed DOS. As already anticipated in Sec.
{\sec{sec:DOSModel}}, a two-particle quantity, the SQM, probes more of the
electronic structure of the ferromagnets than the one-particle charge and
spin-dipole moment do.

{
  \begin{figure}[h]
    \includegraphics[width=0.9\linewidth]{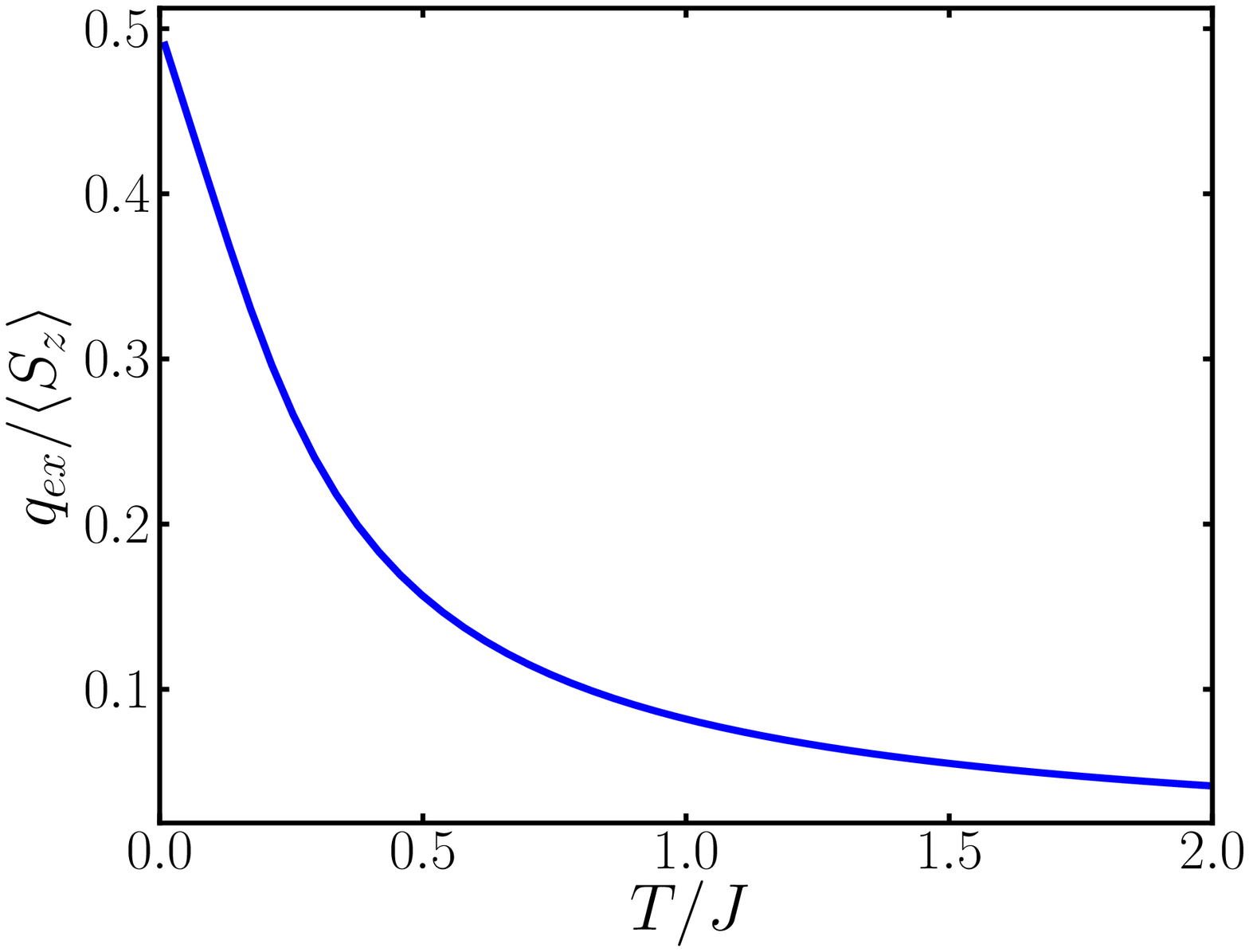}
    \caption{{\color{black} Magnitude of the
local exchange SQM, $q_{\mathrm{ex}}$, normalized to $\brkt{S_z} = N_s / 2$ as a
function of temperature $T \text{/} J$ for $\bar{\nu} = N_o \text{/} 2 D$, $D
= 25 J$, $\mu = 0$.}\label{fig:sqmdensint}}
\end{figure}}

In \ {\Fig{fig:spinPol}}, we plot \ the dependence of the exchange SQM
{\eq{eq:qrex}} on the Stoner splitting $J$, illustrating that while $\left|
\brkt{\vec{S}} \right|$ scales linearly with $J$,
$\brkt{\tens{Q}}_{\mathrm{ex}}$ initially increases quadratically and then
approaches a linear asymptote:
\begin{eqnarray}
  q_{\mathrm{ex}} & = & \tfrac{\bar{\nu} T}{4} \left\{ \begin{array}{ll}
    ( J \text{/} T)^2 / 6 & J \text{/} T \ll 1\\
    ( J \text{/} T) - 2 & J \text{/} T \gg 1
  \end{array} \right. .  \label{eq:linearSQM}
\end{eqnarray}
{
  \begin{figure}[h]
    \includegraphics[width=0.9\linewidth]{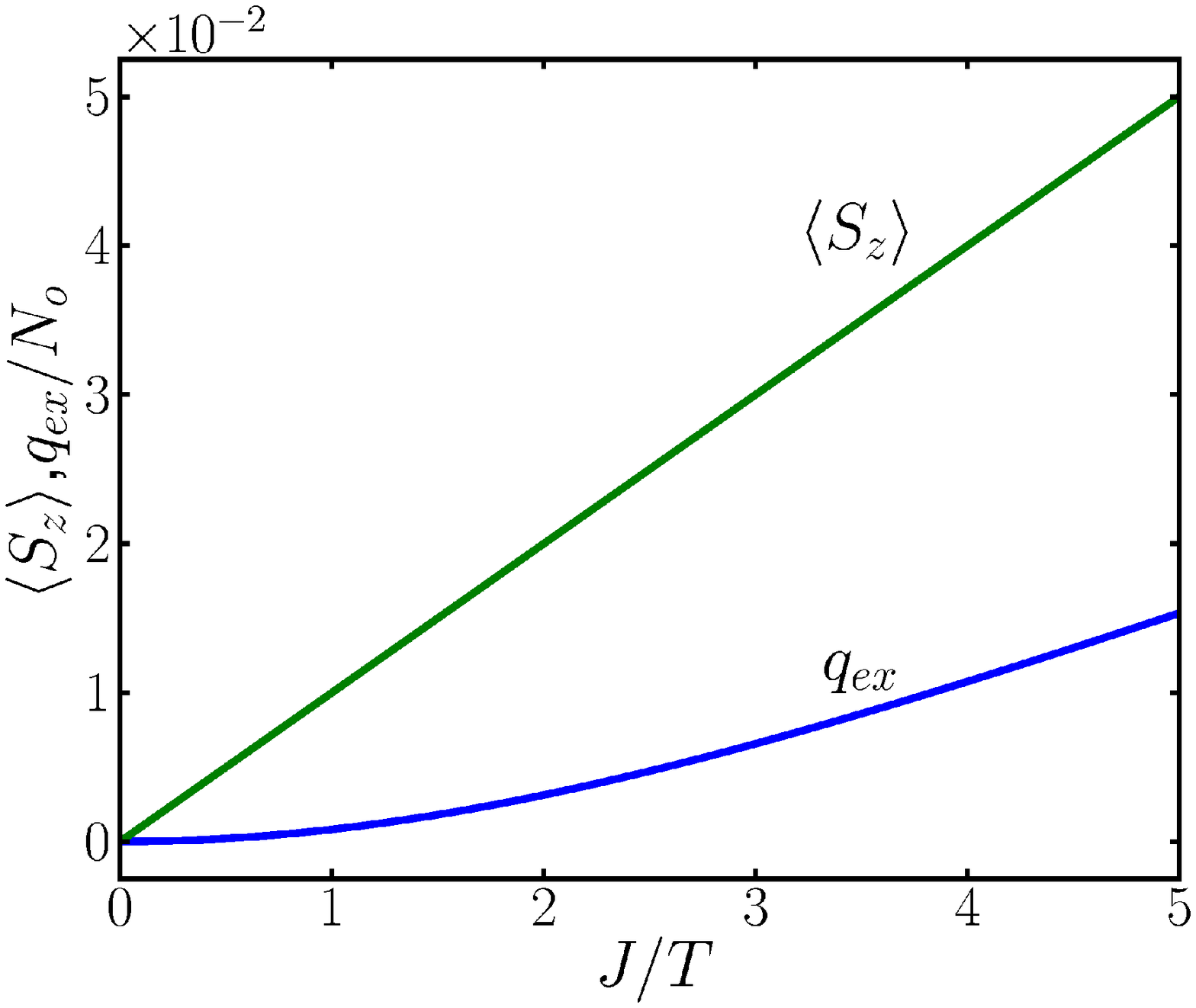}
    \caption{Magnitude of the spin
$\brkt{S_z} \approx N_s / 2 = N_o J / 2 D$ (green) ({\Eq{eq:spinavT0}}) and
the exchange SQM $q_{\mathrm{ex}}$ ({\Eq{eq:qrex}}) (blue) as a function of
Stoner splitting $J \text{/} T$ for $\bar{\nu} = N_o \text{/} ( 2 D)$, $D = 25
T$, $\mu = 0$.\label{fig:spinPol} }
\end{figure}}

\section{Spin-Multipole Current Operators}\label{sec:spinMultipoleNetworks}

We now turn to the third central question posed in the introduction - the
proper definition of the SQM current operator. \ In the previous section, we
answered the important question {\emph{from}} where and {\emph{to}} where SQM
can be transported in terms of {\emph{nodes}} in a spin-multipole network, cf.
{\sec{sec:SQMnetwork}}. We now investigate the {\emph{links}} between the
nodes in this network, which correspond to the SQM current operators as noted
at the end of Sec. {\sec{sec:Tfinite}}. Their definition and physical
interpretation requires some care because (i) like the spin, the total SQM is
not conserved in a device comprising Stoner ferromagnets and (ii) unlike the
spin, this two-particle quantity does not flow directly between local nodes,
but is buffered in non-local nodes.

To tackle point (i), we first revisit the one-particle charge and spin
current operator. The spin-dipole current does not have the intuitive
definition similar to the charge current (total outgoing current = rate of
loss of charge) since the spin is {\emph{not}} conserved internally in the
ferromagnets. Starting from continuity equations in integral form, the spin
currents rather have to be defined as the change in spin induced by the
{\emph{tunneling}}. {{In close analogy, we derive SQM current operators
obeying a continuity equation and current conservation laws.}}

Due to point (ii), we have also consider SQM current operators, accounting
for the flow of SQM between local and non-local nodes. These turn out to be of
central physical importance and reflect that on a microscopic level SQM is
carried by pairs of correlated spin dipoles. {{The flow of spin anisotropy
in an electronic system is thus inherently a two-particle process. We will see
that, as a result, the layout of the physical device and the network for SQM
transport are \emph{different}: the 2-terminal spin-valve requires a
serial 3-node SQM network.}} For more complex devices even the connectivity is
different {\cite{Hell13c}}.

\subsection{Charge and Spin-Dipole Current}\label{sec:NSnetwork}

The physical quantities of interest are the rates of change in local
quantities in the physical subsystems of the circuit due to transport
processes. For one-particle quantities such as charge and spin the physical
subsystems are in one-to-one correspondence with the nodes of the charge /
spin network, cf. {\Fig{fig:network}}. The time derivative operator
$\dot{R}^r_{\mu} ( t)$ giving the rate of change in the combined charge-spin
one-particle operator $R^r_{\mu}$ (cf. {\Eq{eq:chargespinOp}}),
$\tfrac{\mathd}{\mathd t} \langle R^r_{\mu} ( t) \rangle = \langle
\dot{R}^r_{\mu} ( t) \rangle$, is given by
\begin{eqnarray}
  \dot{R}^r_{\mu} ( t) & \assign & i [ H, R^r_{\mu}],  \label{eq:expTtime}
\end{eqnarray}
exploiting the von-Neumann equation $\dot{\rho} ( t) = - i [ H, \rho ( t)]$
and the cyclic invariance of the trace $\mathrm{tr} ( R^r_{\mu}  \dot{\rho} (
t)) = \mathrm{tr} ( i [ H, R^r_{\mu}] \rho ( t))$. We next decompose the total
system Hamiltonian $H$ into the part describing the decoupled subsystems,
$H_0$, and the tunneling $H_T = \sum_{\langle r s \rangle} H_T^{r s}$ with
$H_T^{r s}$ only accounting for tunneling processes between a pair of nodes
$r$ and $s$. This yields a continuity equation in integral form for operators,
\begin{eqnarray}
  \dot{R}^r_{\mu} ( t) & = & \dot{R}^r_{\mu} ( t) |_0 + \sum_{s \neq r}
  I_{R_{\mu}}^{r s},  \label{eq:RContinuity}
\end{eqnarray}
which decomposes the total rate of change in the charge (spin) operator in
node $r$ into two physically meaningful contributions: The first contribution
to {\Eq{eq:RContinuity}} is given by
\begin{eqnarray}
   R_{\mu}^r |_0 & = & i [ H_0, R_{\mu}^r] 
  \label{eq:internalchange}
\end{eqnarray}
and accounts for the time-evolution due to internal processes in node $r$. We
will depict this contribution in the network picture in \
{\Fig{fig:chargeSpinCurr}} by an external arrow attached to node $r$. The
second part $I_{R_{\mu}}^r = \sum_{s \neq r} I_{R_{\mu}}^{r s} = i [ H_T,
R_{\mu}^r]$ quantifies the rate of change induced by tunneling, i. e.,
this{\emph{ defines the current}} of observable $R_{\mu}^r$ {\emph{into}} node
$r$. In the form of {\Eq{eq:RContinuity}}, it has already decomposed into its
various contributions emanating from all other subsystems $s$:
\begin{eqnarray}
  I_{R_{\mu}}^r & = & \sum_{s \neq r} I_{R_{\mu}}^{r s},  \label{eq:currDec}
\end{eqnarray}
Whenever the {\emph{operator}} $I_{R_{\mu}}^{r s} \neq 0$, we depict this in
the network picture by an arrow inking the two nodes $r$ and $s$. Note that
still the \emph{average} current $\langle I_{R_{\mu}}^{r s} \rangle$ that
flows between the nodes may vanish, e. g., for a some special set of
parameters. So far, our considerations are quite general and also apply to
systems including quantum dots.

{
  \begin{figure}[h]
    \includegraphics[width=0.9\linewidth]{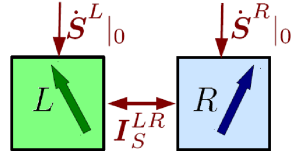}
    \caption{Network picture for
the spin with current \emph{operators} represented by links. The network
picture for charge is similar, but without external arrows pointing to the
nodes (indicated by $\dot{\vec{S}}^r |_0 $). The latter must be
introduced due to the internal violation of spin conservation in each
spin-node\label{fig:chargeSpinCurr}}
\end{figure}}

In our model (Eqs. {\eq{eq:Htot}}-{\eq{eq:HT}}), the particle number $R^r_0 =
N^r$ is conserved internally in each electrode individually and therefore
\begin{eqnarray}
  & \begin{array}{lllll}
    N^r |_0 & = & [ H_0, N^r] & = & 0
  \end{array}, & 
\end{eqnarray}
which is the 0-component of {\Eq{eq:internalchange}} and $\dot{N} =
\sum_{\alpha} I_N^{\alpha}$. \ These make up for the total change in charge.
The spin $R^r_i = S^r_i$ ($i \neq 0$), however, is \emph{not} conserved
internally in the ferromagnets:
\begin{equation}
  \begin{array}{lll}
    \dot{\vec{S}}^{\alpha} & = & \dot{\vec{S}}^r |_0 +
    \vec{I}^{\alpha}_{\vec{S}}
  \end{array},
\end{equation}
where $\vec{I}^{\alpha}_{\vec{S}}$ is the operator for spin-current into node
$\alpha$. Using {\Eq{eq:H0}}, one finds
\begin{equation}
  \begin{array}{lllll}
    \dot{\vec{S}}^r |_0 & = & \sum_n^{} J_n^r  \op{\vec{J}}^r \times
    \vec{S}^r_n & \neq & 0
  \end{array} \label{eq:Sdot0}
\end{equation}
for $\vec{S}^r_n = \sum_{k \sigma} \vec{s}^r_{\sigma \sigma'} c^{\dag}_{r n k
\sigma} c_{r n k \sigma}$ being the contribution from band $n$ to the spin of
electrode $r$. This describes a precession of $\vec{S}^r_n$ about the Stoner
field of electrode $r$.

To obtain an explicit starting point for the real-time calculation of the
average charge and spin current (see {\App{app:covrealtime}}), we use
{\Eq{eq:internalchange}} and {\Eq{eq:HT}} and recover the familiar form of the
charge and spin current operators,
\begin{eqnarray}
  I^{r s}_{R_{\mu}} & = & \sum_{n n' k k' \sigma \sigma'} ( - i r_{\mu} T)^{r
  s}_{\sigma \sigma'} c^{\dag}_{r n k \sigma} c_{s n' k' \sigma'} -
  \text{h.c.}, \,  \label{eq:TCurrent}
\end{eqnarray}
abbreviating the matrix product in spin-space $( r_{\mu} T)^{r s}_{\sigma
\sigma'} = \sum_{\tau} ( r_{\mu})_{\sigma \tau} T^{r s}_{\tau \sigma'}$. The
operator {\eq{eq:TCurrent}} describes the net current injected from node $s$
into $r$, accounting for tunneling processes from node $s$ to $r$ (the first
contribution in {\Eq{eq:TCurrent}}) and the reverse process (the second). Only
the sum of both terms is a Hermitian operator and therefore a physical
observable. Since both processes contribute with an opposite sign to the
current {\eq{eq:TCurrent}}, we obtain the antisymmetry relation
$I_{R_{\mu}}^{r s} = - I_{R_{\mu}}^{s r}$. This has an important physical
consequence: summing up all charge (spin) currents in the system yields the
zero operator:
\begin{eqnarray}
  \sum_r \sum_{s \neq r} I^{r s}_{R_{\mu}} & = & 0. 
  \label{eq:chargeSpinCurrentConservation}
\end{eqnarray}
This charge (spin) \emph{current} conservation law expresses that charge
(spin) is conserved by \emph{tunneling}, that is $[H_T,
R^{\mathrm{tot}}_{\mu}] = 0$. Since the total spin is not conserved under the
full time evolution (due to the \emph{internal} evolution
$\dot{\vec{S}}^{\mathrm{tot}} |_0 \neq 0$), there is no analogue of
{\Eq{eq:chargeSpinCurrentConservation}} for the total time derivative
$\dot{\vec{S}}^r$. We emphasize furthermore that this conservation law holds
on an \emph{operator} level and not only for expectation values.

\subsection{Spin-quadrupole Current}\label{sec:SQMcurrent}

We now try to proceed analogously for the SQM network in {\Fig{fig:network}}
of {\Sec{sec:Tfinite}}. Generally, we are interested in finding the rate of
change in the spin anisotropy stored in local nodes. To this end, we need to
consider the change in SQM, $\dot{\tens{Q}}^{r r'}$, in both the local nodes
($r = r'$) and the non-local nodes ($r \neq r'$). Taking the time derivative
of {\Eq{eq:Qaaprimesym}} and using {\Eq{eq:RContinuity}} we obtain:
\begin{eqnarray}
  \dot{\tens{Q}}^{r r'} & = & \dot{\tens{Q}}^{r r'} |_0 + \sum_{\langle s s'
  \rangle \neq \langle r r' \rangle} \tens{I}_{\tens{Q}}^{r r', s s'}, 
  \label{eq:SQMKirchhoff}
\end{eqnarray}
where $\brkt{s s'}$ denotes the sum over \emph{pairs} of indices $s s'$
(i.e., ignoring their order). This is the continuity equation in integral form
for the change in SQM in node $\brkt{r r'}$. The first term is the change in
SQM due to the internal time evolution
\begin{eqnarray}
  \dot{\tens{Q}}^{r r'} |_0 & = & \tfrac{g^{r r'}}{2} \left[ \dot{\vec{S}}^r
  |_0 \odot \vec{S}^{r'} + \vec{S}^r \odot \dot{\vec{S}}^{r'} |_0 + ( r
  \leftrightarrow r') \right],  \label{eq:internalSQM}
\end{eqnarray}
which involves the non-zero internal time evolution $\dot{\vec{S}}^r |_0$
given by {\Eq{eq:Sdot0}}. Like the spin, the SQM is thus \emph{not}
conserved in any of the nodes in our model. The responsible Stoner fields also
effectively exert a ``torque'' on the SQM, thereby rotating the principal axes
of this tensor. Similar to the spin, we depict this in the network picture
(Fig. \ref{fig:SQMNetworkLinks}) by one-sided arrows pointing at this node.

{
  \begin{figure}[h]
    \includegraphics[width=0.9\linewidth]{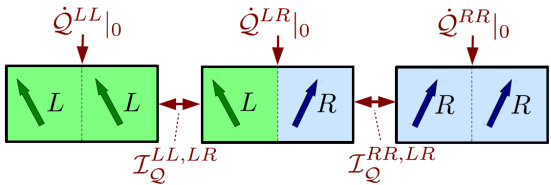}
    \caption{Network picture for SQM
including the links representing SQM current \emph{operators}. Similar to
spin, SQM is not conserved internally in the nodes giving rise to external SQM
currents. \label{fig:SQMNetworkLinks}}
\end{figure}}

The SQM current $\tens{I}_{\tens{Q}}^{r r'} = \sum_{\langle s s' \rangle \neq
\langle r r' \rangle} \tens{I}_{\tens{Q}}^{r r', s s'}$ is given by the
Hermitian tensor operator
\begin{eqnarray}
  \tens{I}_{\tens{Q}}^{r r'} & = & \tfrac{g^{r r'}}{2} \left[
  \vec{I}_{\vec{S}}^r \odot \vec{S}^{r'} + \vec{S}^r \odot
  \vec{I}_{\vec{S}}^{r'} + ( r \leftrightarrow r') \right] . 
  \label{eq:Iaaprime}
\end{eqnarray}
This is a central result of the paper. Since the spin and the spin current do
in general not commute as operators on Fock space{\footnote{$\left[ S^r_i,
I^r_{\vec{S} j} \right] \neq 0$ can be explicitly shown by inserting the
expressions for the spin operator (\ref{eq:spinOp}) and the spin current
operator (\ref{eq:TCurrent}), respectively, and applying the anti-commutation
relations of the field operators.}}, the individual terms in this expression
are not Hermitian and therefore not observables. The operator
{\eq{eq:Iaaprime}} reflects that in general the average SQM current is not
simply the product of spin and spin current since $\brkt{\vec{I}_{\vec{S} i}^r
\vec{S}^{r'}_j} \neq \brkt{\vec{I}_{\vec{S} i}^r} \brkt{\vec{S}_j^{r'}}$ due
to quantum mechanical exchange correlations, interactions, etc. {\color{blue}
}{{Therefore, SQM is not determined by spin-dipole moment: it requires a
separate description in spintronics transport theory.}} For the bilinear
tunnel coupling {\eq{eq:HT}}, the contribution to the net current from node
$\langle s s' \rangle$ into in node $\langle r r' \rangle$ is
\begin{eqnarray}
  \tens{I}_{\tens{Q}}^{r r', s s'} & = & \tfrac{g^{r r'}}{2} \left[
  \vec{I}_{\vec{S}}^{r s} \odot \vec{S}^{r'} \delta^{r' s'} + \vec{S}^r
  \delta^{r s} \odot \vec{I}_{\vec{S}}^{r' s'} \right] \nonumber\\
  ^{} &  &  + ( r \leftrightarrow s')] .  \label{eq:ISQM}
\end{eqnarray}
Notably, this SQM current is zero unless one of the indices $s, s'$ match the
indices $r, r'$. {{This puts an important restriction on the network
connectivity: the local SQM nodes are \emph{only} linked to non-local
nodes, and not to other local nodes. Changes of local spin-anisotropy,}}
\begin{eqnarray}
  \dot{\tens{Q}}^{r r} & = & \dot{\tens{Q}}^{r r} |_0 + \tens{I}_{\tens{Q}}^{r
  r},  \label{eq:Qaadot} \label{eq:SQMKirchhoffaa}
\end{eqnarray}
{{which are due to ransport thus only occur through changes in non-local
spin correlations:}}
\begin{eqnarray}
  & \begin{array}{lll}
    \tens{I}_{\tens{Q}}^{r r} & = & \sum_{s \neq r} \left(
    \vec{I}_{\vec{S}}^{r s} \odot \vec{S}^r + \vec{S}^r \odot
    \vec{I}_{\vec{S}}^{r s} \right),
  \end{array} &  \label{eq:Iaa}
\end{eqnarray}
where $\vec{I}_{\vec{S}}^{r s}$ is the spin-current operator from node $s$
into $r$.

{
  \begin{figure}[h]
    \includegraphics[width=0.9\linewidth]{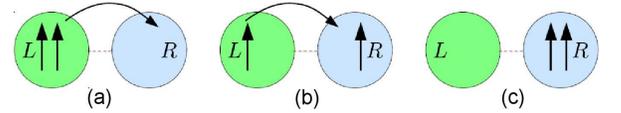}
    \caption{Illustration of the microscopic picture
of SQM transport. (a) Consider two electrons in electrode $L$ that contribute
to the SQM of the local node $\langle L L \rangle$. (b) By transferring one of
these electrons to subsystem $R$, the local spin-spin correlation is lost but
a new non-local correlations are established, increasing the SQM of node
$\langle L R \rangle$. {{Thus, emitting spin-polarized electrons non-local
correlations are set up in the circuit.}} (c) When the second electron follows
the first, local SQM correlations are created, but now in subsystem $R$ (node
$\langle R R \rangle$). Note that this picture is only meant to illustrate the
{\emph{non-locality}} aspect of SQM, but incorrectly portrays the spins as
distinguishable objects.\label{fig:SQMTransport}}
\end{figure}}

All the considerations so far in this section were are general. For the simple
spin-valve we consider n this paper the general theory above implies that SQM
cannot be directly transferred from the local node $\langle L L \rangle$ to
the local node $\langle R R \rangle$; it is rather first ``buffered'' in the
intermediate, non-local node $\langle L R \rangle$. This restriction on the
SQM network connectivity is related to the real-space picture of \ SQM
transport sketched in {\Fig{fig:SQMTransport}}. This picture unveils why SQM
transport is possible even in the single electron transport limit (leading
order in $H_T$), as discussed for spin-values with embedded
spin-\emph{isotropic} quantum dots
{\cite{Sothmann10,Baumgaertel11,Misiorny13}}. In this case we have on the
right hand side of {\Eq{eq:Qaadot}} $\dot{\tens{Q}}^{r r} |_0 = 0$ where $r$
now labels the quantum dot embedded in this spin-valve. Then SQM currents are
responsible for the change in the local QD spin-anisotropy of a subsystem $r$,
i. e. $\dot{\tens{Q}}^{r r} = \tens{I}_{\tens{Q}}^{r r}$ where
$\tens{I}_{\tens{Q}}^{r r}$ of {\Eq{eq:Iaa}} was already obtained in
{\Cite{Baumgaertel11}} for a spin-valve with an embedded quantum dot, however,
without using the new network picture. The calculations in
{\cite{Baumgaertel11}} demonstrate that in such more complicated devices the
SQM current $\tens{I}_{\tens{Q}}^{r r}$, averaged over the non-equilibrium
state, depends non-trivially on the average accumulated charge and spin of the
dot. {{The SQM current thus couples to measurable charge and spin currents
and should in general be considered for the description of spin and charge
dynamics.}}

Analogous to the charge and spin current the SQM current (\ref{eq:ISQM}) is
antisymmetric with respect to the node indices, i.e., when the two pairs of
subsystem-indices $r r'$ and $s s'$ are interchanged: \
$\tens{I}_{\tens{Q}}^{r r', s s'} = - \tens{I}_{\tens{Q}}^{s s', r r'}$. (Note
that the order of indices denoting a pair does not matter, i.e., $r r' = r'
r$). As a consequence, the SQM currents sum to the zero tensor operator:
\begin{eqnarray}
  \sum_{\langle r r' \rangle} \tens{I}_{\tens{Q}}^{r r'} & = & \sum_{\langle r
  r' \rangle} \sum_{\langle s s' \rangle \neq \langle r r' \rangle}
  \tens{I}_{\tens{Q}}^{r r', s s'} = 0.  \label{eq:SQMCurrentConservation}
\end{eqnarray}
Similar to spin, this \emph{SQM current conservation law} expresses the
conservation of total circuit SQM {\eq{eq:Qtot}} in the {\emph{tunneling}}.
This is a direct consequence of the total spin-dipole conservation by
tunneling: $\left[ H_T, \tens{Q}^{\mathrm{tot}} \right] = \vec{S}^{\mathrm{tot}}
\odot \left[ H_T, \vec{S}^{\mathrm{tot}} \right] + \left[ H_T,
\vec{S}^{\mathrm{tot}} \right] \odot \vec{S}^{\mathrm{tot}} = 0$.

Finally, we note that the restriction on the network connectivity found above
derives from the particular form of our tunneling Hamiltonian {\eq{eq:HT}},
which is bilinear in the field operators. The network thus describes the
connectivity on the \emph{operator} level. The topology of this network is
different when $H_T$ is, e.g., an effective exchange coupling that is quartic
in the fields. Since such a coupling is usually derived from the bilinear
tunneling model {\eq{eq:HT}} studied here, we will not dwell on this
further.{\footnote{Although the restrictions on the network connectivity
derives form the bilinearity of the tunneling Hamiltonian {\eq{eq:HT}}, this
does not mean that SQM cannot be exchanged directly between local nodes: when
calculating the \emph{averages}, including coherent processes of higher
order in this the bilinear coupling is indeed found to occur, see
{\cite{Hell13c}}.}}

\section{Average Spin-Multipole Currents}\label{sec:spinMultipoleCurrents}

In this section we complete the discussion of the third main question posed in
the introduction: {{we present explicit results for the average
spin-quadrupole current calculated to first order in the tunnel coupling
$\Gamma$ of the two ferromagnets and compare it to the average charge and spin
current.}} The calculations of these are compactly presented in in
{\App{app:FirstOrder}}, applying a covariant reformulation of the real-time
diagrammatic technique (for a systematic, self-contained technical
presentation see {\App{app:covrealtime}}). {{Covariance is used here in
the sense that all expressions are form-invariant under a change of the
spatial coordinate system and the spin quantization axis.}} An advantage of
this technique is that it can be extended to spin-valves with embedded quantum
dots {\cite{Hell13c}}.

In {\Sec{sec:firstOrder}} we discuss the results for a general multi-band
dispersion relation $\varepsilon^r_{n k \sigma}$, applying the insights
obtained from the spin-multipole network theory developed in
{\sec{sec:SQMnetwork}} and {\sec{sec:spinMultipoleNetworks}} and the
microscopic picture explained in {\sec{sec:microscopic}}. We decompose the SQM
current into physically meaningful contributions: direct vs. exchange (Pauli
exclusion hole aspects) and dissipative vs. coherent (quantum fluctuation
aspects). An intimate connection between storage (see
{\sec{sec:spinMultipoleStorage}}) and transport of charge, spin and SQM is
then established by comparing their {\emph{energy-resolved}} contributions
(see Sec. \ref{sec:energyResolved}).

In Sec. {\sec{sec:result1band}} we specialize to the flat band approximation
(cf. {\sec{sec:wide-flat-band}}) to obtain tangible analytical and numerical
results. Even in this simple limit -- where the dissipative spin-current
vanishes due to the energy independent DOS -- the average SQM current tensor
has a non-trivial parameter dependence. This concerns both its magnitude
(principal values) and its alignment (principal vectors), which are
substantially controllable by magnetic and electric parameters for
non-collinear Stoner vectors.

In Sec. {\sec{sec:spinFree}} we demonstrate that a {\emph{pure}} SQM current,
i. e., not accompanied by a spin current, is in principle possible. This
spin-anisotropy transfer is entirely carried by Pauli exclusion holes, giving
rise to a non-vanishing exchange SQM current. For a {\emph{temperature}} bias
between ferromagnets with collinear Stoner vectors, we even show that a pure
SQM currents even persists in the absence of a charge current. Electrodes with
non-trivial spin structure can thus ``talk'' to each other in ways not
described by charge and spin currents

\subsection{Charge, spin and SQM current}\label{sec:firstOrder}

The average charge, spin and spin-quadrupole current associated with the left
electrode read
\begin{eqnarray}
  \brkt{I^L_N} & = & I_{N, 0}^L + I^L_{N, F} \left( \widehat{\vec{J}}^L \cdot
  \widehat{\vec{J}}^R \right),  \label{eq:chargeCurrent}\\
  \brkt{\vec{I}^L_{\vec{S}}} & = & E_{\vec{S}}^L \widehat{\vec{J}}^L +
  A^L_{\vec{S}} \widehat{\vec{J}}^R + T^L_{\vec{S}} \left( \widehat{\vec{J}}^L
  \times \widehat{\vec{J}}^R \right),  \label{eq:spinCurrent}\\
  \brkt{\tens{I}^{L L}_{\tens{Q}}} & = & 2 \left\langle \vec{S}^L
  \right\rangle \odot \brkt{\vec{I}^L_{\vec{S}}} - \nonumber\\
  &  &  \widehat{\vec{J}}^L \odot \left[ E_{\mathrm{ex}}^L \widehat{\vec{J}}^L
  + A^L_{\mathrm{ex}}  \widehat{\vec{J}}^R + T^L_{\mathrm{ex}} \text{ } 
  \widehat{\vec{J}}^L \times \widehat{\vec{J}}^R \right] 
  \label{eq:SQMCurrent}
\end{eqnarray}
These were calculated in App. {\sec{app:FirstOrder}} to the first order in the
energy-resolved tunneling rate
\begin{eqnarray}
  \Gamma ( \omega) & = & 2 \pi t^2  \bar{\nu}^L ( \omega) \bar{\nu}^R (
  \omega), 
\end{eqnarray}
where $\bar{\nu}^r ( \omega)$ is given by {\Eq{eq:avdos}}. Here $\odot$
denotes the symmetric, traceless tensor product {\eq{eq:symdyad}}. The charge
current coefficients are (we will not write the $\omega$ dependence unless
confusion may arise)
\begin{eqnarray}
  I_{N, 0} & = & \int \mathd \omega 2 \Gamma \Delta,  \label{eq:nonmcharge}\\
  I^L_{N, F} & = & \int \mathd \omega 2 \Gamma \Delta n^L n^R, 
  \label{eq:ferrocharge}
\end{eqnarray}
where the spin-polarization function $n^r ( \omega)$ is given by
{\Eq{eq:spinpol}}. The well-known bias function,
\begin{eqnarray}
  \Delta ( \omega) & = & f^R_+ ( \omega) - f^L_+ ( \omega), 
  \label{eq:biasFunction}
\end{eqnarray}
is only non-zero in the bias window, $\mu^L \gtrsim \omega \gtrsim \mu^R$, up
to thermal smearing. The occurrence of the factor $\Delta ( \omega)$ signals
that a term arises from \emph{dissipative processes} in which the energy
of initial and final state have to be the same. The spin current coefficients
are
\begin{eqnarray}
  E^L_{\vec{S}} & = & \int \mathd \omega \Gamma \Delta n^L, 
  \label{eq:spinEmission}\\
  A^L_{\vec{S}} & = & \int \mathd \omega \Gamma \Delta n^R, 
  \label{eq:spinInjection}\\
  T^L_{\vec{S}} & = & \int \mathd \omega \Gamma \left( \beta^L 
  \frac{n^R}{\bar{\nu}^L} + \beta^R  \frac{n^L}{\nu^R} \right), 
  \label{eq:spinTorque}
\end{eqnarray}
the function $\beta^r ( \omega)$ incorporates the effect of the
spin-polarization of the DOS, $n^r ( \omega)$, through the principal value
integral,
\begin{eqnarray}
  \beta^r ( \omega) & = & \mathrm{Re} \int \frac{d \omega'}{\pi} 
  \frac{\bar{\nu}^r ( \omega') n^r ( \omega') f^r_+ ( \omega')}{\omega -
  \omega' + i 0},  \label{eq:betar}
\end{eqnarray}
integrating over all virtual state energies $\omega'$. Here and below such
functions, not limited by energy conservation as they involve virtual
intermediate states, appear in contributions from \emph{coherent
processes}. Finally, the exchange SQM emission, absorption and torque
coefficients
\begin{eqnarray}
  E^L_{\mathrm{ex}} & = & 2 \int \mathd \omega \Gamma \Delta a^L, 
  \label{eq:SQMEmission}\\
  A^L_{\mathrm{ex}} & = & 2 \int \mathd \omega \Gamma \Delta n^R  \tilde{a}^L, 
  \label{eq:SQMInjection}\\
  T^L_{\mathrm{ex}} & = & 2 \int \mathd \omega \Gamma ( \alpha^R f^L_+ -
  \beta^R) \frac{a^L}{\bar{\nu}^R},  \label{eq:SQMTorque}
\end{eqnarray}
depend on the local spin-anisotropy function $a^L$, cf. {\eq{eq:tripletCorr}},
and an additional {\emph{even}} spin-anisotropy function,
\begin{eqnarray}
  \tilde{a}^L ( \omega) & = & \int \mathd \omega' f^r_+ ( \omega')
  \sum_{\sigma, \sigma'} \frac{\sigma'}{4}  \frac{\nu^r_{ 
  \sigma \sigma'} ( \omega, \omega')}{\bar{\nu}^r ( \omega)} \\
  & = & \sum_{\sigma} \sigma a_{\sigma}^r ( \omega),  \label{eq:qLR}
\end{eqnarray}
where $\nu^r_{\sigma \sigma'} ( \omega, \omega')$ is the 2DOS
{\eq{eq:twoDOS}}. \ Finally, the torque coefficient $T^L_{\tens{Q}}$ involves
an additional function similar to {\Eq{eq:betar}} but without the distribution
function $f_+^R ( \omega)$ under the principal value integral:
\begin{eqnarray}
  \alpha^R ( \omega) & = & \mathrm{Re} \int \frac{d \omega'}{\pi}  \frac{n^R (
  \omega') \bar{\nu}^R ( \omega')}{\omega - \omega' + i 0} . 
  \label{eq:alphar}
\end{eqnarray}
The reader should note that the coefficients of Eqs.
{\eq{eq:SQMEmission}}-{\eq{eq:SQMTorque}} are defined such that a minus sign
appears in {\Eq{eq:SQMCurrent}}, which we introduced in agreement with the
sign convention for exchange SQM in {\Eq{eq:QexTensor}}, that is,
$\brkt{\tens{Q}^{L L}} = - q_{\mathrm{ex}}^L  \widehat{\vec{J}}^L \odot
\widehat{\vec{J}}^L$. Moreover, one obtains the expressions for
$\brkt{\tens{I}^{R R}_{\tens{Q}}}$ by formally substituting $L \leftrightarrow
R$ in {\Eq{eq:SQMCurrent}}, and $\brkt{\tens{I}^{L R}_{\tens{Q}}} = -
\brkt{\tens{I}^{L L}_{\tens{Q}}} - \brkt{\tens{I}^{R R}_{\tens{Q}}}$ follows
from the SQM current conservation law {\eq{eq:SQMCurrentConservation}} (see
also below). One checks that the results are invariant under scalar gauge
transformations (global energy shifts). Finally, we note that since positive
currents are defined as entering a node, positive (negative) absorption
coefficients correspond to injection (ejection) of particles and vice versa
for emission coefficients.

The SQM current {\Eq{eq:SQMCurrent}} is the central result of this paper. It
depends explicitly (but not exclusively) on the spin current
{\eq{eq:spinCurrent}}. Therefore, its physical interpretation is aided by
first giving a pertinent review of the different contributions to the charge
and spin-currents {\eq{eq:chargeCurrent}} and {\eq{eq:spinCurrent}},
respectively.

\subsubsection{Charge Current}

{\Eq{eq:chargeCurrent}} is a well-known and experimentally tested result for
the charge current, which accounts for single electron tunneling processes
between the left and right electrode. It has only dissipative contributions
(i.e., involving $\Delta ( \omega)$): the first part $I_{N, 0}$ in \ Eq.
(\ref{eq:chargeCurrent}) only depends in the average DOS $\bar{\nu}^r$ whereas
the second, spin-dependent correction that depends on the DOS
spin-polarizations $n^r$ through $I^L_{N, F}$ and on the angle between the
Stoner vectors through $\widehat{\vec{J}}^L \cdot \widehat{\vec{J}}^R = \cos
\theta$. The reduction going from $\theta = 0 \rightarrow \pi$ is the
celebrated spin-valve or tunnel magneto resistance (TMR) effect.{\footnote{The
current contribution of electrons at energy $\omega$ is changed by the factor
$( 1 + \cos ( \theta) n^L ( \omega) n^R ( \omega))$, i.e., when
\emph{both} spin-polarizations are nonzero. For parallel Stoner vectors
the fraction of electrons with spin $\sigma$ in electrode $r$ with energy
$\omega$ is changed from 1/2 to $( 1 + \sigma n^r ( \omega)) / 2$ as compared
to the non-magnetic case. These electrons can only access a fraction $( 1 +
\sigma n^{\bar{r}} ( \omega)) / 2$ of the states in the other electrode
$\bar{r}$ instead of 1/2 as in the non-magnetic case. The reason for this is
conservation of spin by the tunneling (cf. {\Eq{eq:tsigmacons}}), so all
states with opposite spin in the other electrode are forbidden final states.
In total, this changes the current by a factor $\sum_{\sigma} ( 1 + \sigma
n^r) ( 1 + \sigma n^{\bar{r}}) / 2 = 1 + n^L n^R > 1$. In the case of
non-collinear Stoner vectors electrons can access both spin channels of the
other electrode due to non-zero spin-overlap factors in {\Eq{eq:tsigmacons}},
giving an overall additional reduction factor $\cos ( \theta)$. }}

\subsubsection{Spin Current}\label{sec:spincurrent}

The spin current {\eq{eq:spinCurrent}} was obtained by Braun et. al and we
review here its two type of contributions.

\paragraph{Dissipative spin emission $\sim \op{\vec{J}}^L$ and absorption
$\sim \op{\vec{J}}^R$}The dissipative spin-dipole current (first two terms in
{\Eq{eq:spinCurrent}}, containing $\Delta ( \omega)$) is analogous to the
particle current: \ an electron emitted from the left node transports a
spin-dipole moment $\op{\vec{J}}^L / 2$ and the electrons absorbed from the
right node transports $\op{\vec{J}}^R / 2$. The expression for particle
current {\Eq{eq:nonmcharge}} simply has to be supplemented by a factor $n^r (
\omega) \op{\vec{J}}^r / 2$ to obtain the terms for spin emission
{\eq{eq:spinEmission}} and absorption {\eq{eq:spinInjection}}.

\paragraph{Coherent spin torque $\sim \op{\vec{J}}^L \times \op{\vec{J}}^R$}

The coherent spin current (last term in {\Eq{eq:spinCurrent}}) has no such
analogy to the particle current and corresponds to a spin-torque. This
corresponds to spin flips induced by{\color{blue}  }virtual fluctuations
between the left and the right electrode restricted by energy conservation
only in the final state, but not in the intermediate state. This is reflected
by its dependence on the principal value integral $\beta^r ( \omega)$ (cf.
discussion of {\Eq{eq:betar}}): an electron with spin $\sigma$ occupying a
level at energy $\omega$ in the left electrode can fluctuate to all empty
levels with energy $\omega'$ in the right electrode with an amplitude $\propto
1 / ( \omega - \omega')$. While in the right electrode, the electron spin
$\propto \op{\vec{J}}^L$ is not collinear to the Stoner field $\propto
\op{\vec{J}}^R$ in the right electrode and precesses about it, explaining the
factor $\widehat{\vec{J}}^L \times \widehat{\vec{J}}^R$ in the coherent spin
current {\eq{eq:spinTorque}}. Note that a net \emph{spin-torque} on the
magnetization of the left electrode only occurs if the 1DOS of {\emph{both}}
electrodes are spin-polarized.

Finally, we note that at zero bias, the dissipative spin-current vanishes,
$E_{\vec{S}}^L = A^L_{\vec{S}} = 0$, but the coherent spin-current (torque
contribution) remains, $T^L_{\vec{S}} \neq 0$: non-collinear ferromagnets keep
interacting by virtual fluctuations, thereby exerting a torque on each other.

\subsubsection{SQM Current}\label{sec:SQMCurrent}

\paragraph{SQM current conservation}The most immediate property of the SQM
current expression {\eq{eq:SQMCurrent}} is its formal lack of symmetry with
respect to interchanging the electrodes $L \leftrightarrow R$. This differs
notably from charge and spin current, for which the original expression is
reproduced with a minus sign when interchanging $L \leftrightarrow R$. This
distinction is related to the striking characteristics of the SQM network
picture compared to the charge and spin network (see {\Sec{sec:SQMCurrent}}):
the currents of the latter, $\brkt{I^L_{R_{\mu}}}$, describe the net flow into
the $L$-node coming from from the $R$-node (see {\Fig{fig:chargeSpinCurr}}).
Interchanging $L$ and $R$ yields then the opposite current from $R$ to the
$L$-node, reflecting the current conservation law
{\eq{eq:chargeSpinCurrentConservation}}: $\brkt{I^L_{R_{\mu}}} +
\brkt{I^R_{R_{\mu}}} = 0$. In contrast, $\brkt{\tens{I}^{L L}_{\tens{Q}}}$
describes the net flow of SQM from the local $\langle L L \rangle$-node to the
nonlocal $\langle L R \rangle$-node (see {\Fig{fig:SQMNetworkLinks}}). If we
interchange $L \leftrightarrow R$ in {\Eq{eq:SQMCurrent}}, we obtain the
current $\brkt{\tens{I}^{R R}_{\tens{Q}}}$ emanating from the $\langle R R
\rangle$-node. Importantly, $\brkt{\tens{I}^{L L}_{\tens{Q}}} +
\brkt{\tens{I}^{R R}_{\tens{Q}}} = - \brkt{\tens{I}^{L R}_{\tens{Q}}} \neq 0$
in accordance with the SQM current conservation law
(\ref{eq:SQMCurrentConservation}). This again emphasizes the relevance of the
nonlocal node $\langle L R \rangle$, which ``buffers'' the SQM currents from
both local nodes.

\paragraph{Direct and Exchange SQM Current}{{The SQM current allows for
\emph{two} physically meaningful, different decompositions. The first
decomposition is given by Eq. (\ref{eq:SQMCurrent}), which breaks up the SQM
current into different two-particle contributions, the first, {\emph{direct}}
term Eq. (\ref{eq:SQMCurrent}) and the second, {\emph{exchange}} term. This
has no analogue in the one-particle charge and spin current.}}

The direct current $\brkt{\tens{I}^{L L}_{\tens{Q}}}_{\mathrm{dir}}$ quantifies
the tunneling-induced change in the direct part of the average SQM
{\eq{eq:redcontr}}, $\brkt{\tens{Q}^{L L}}_{\mathrm{dir}} \assign
\brkt{\vec{S}^L} \odot \brkt{\vec{S}^L}$, which ignores the Pauli exclusion
hole (cf. {\Sec{sec:spinMultipoleStorage}}). Indeed, using {\Eq{eq:Iaa}}, we
reproduce the first term in Eq. (\ref{eq:SQMCurrent}):
\begin{eqnarray}
  \brkt{\tens{I}^{L L}_{\tens{Q}}}_{\mathrm{dir}} & \assign &
  \brkt{\dot{\tens{Q}}^{L L} - \dot{\tens{Q}}^{L L} |_0}_{\mathrm{dir}} \\
  & = & 2 \brkt{\vec{S}^L} \odot \brkt{\vec{I}_{\vec{S}}^L} . 
  \label{eq:dirCurr}
\end{eqnarray}
{{Similar to the average SQM, the direct average SQM \emph{current} is
completely determined by a product of average spin-dipole properties, here the
spin $\brkt{\vec{S}^L}$ and the spin-current $\brkt{\vec{I}_{\vec{S}}^L}$,
given by {\Eq{eq:avspin}} and {\eq{eq:spinCurrent}}, respectively. This
equation substantiates the classical picture of transport SQM or
spin-anisotropy sketched in {\Fig{fig:SQMTransport}}: when single electrons
move, the triplet correlations between pairs of electron spins first
delocalize and then relocalize, resulting in a change of the local SQM, the
part described by $\brkt{\tens{Q}^{L L}}_{\mathrm{dir}}$.}}

{{The exchange SQM current $\brkt{\tens{I}^{L L}_{\tens{Q}}}_{\mathrm{ex}}$,
the second term in Eq. (\ref{eq:SQMCurrent}), \ accounts for the
tunnel-induced change in $\brkt{\tens{Q}^{L L}}_{\mathrm{ex}}$, i.e., a negative
quantum anisotropy due to the Pauli-exclusion holes in the triplet spin
correlations. It \emph{cannot} be expressed in the average spin-current.}}
The above classical picture of SQM transport thus needs correction: by
reducing SQM transport to ``spin times spin-current'' one over-estimates the
anisotropy flow, by counting Pauli-forbidden, local triplet correlations
(those coming from the same orbital) and accounting for their transformation
into non-local correlations when one of the two electrons tunnels out. The SQM
exchange current compensates for this: it is an effective back-flow of
non-local anisotropy into to the local nodes of the SQM network
({\Fig{fig:SQMNetworkLinks}}). {{We now reach a central conclusion of the
paper: whenever the average spin-current is made to vanish
$\brkt{\vec{I}_{\vec{S}}^L} = 0$, a non-zero SQM exchange current is generally
present since $\brkt{\vec{S}^L \odot \vec{I}_{\vec{S}}^L} \neq
\brkt{\vec{S}^L} \odot \brkt{\vec{I}_{\vec{S}}^L}$ due to the Pauli exclusion
holes. In {\Sec{sec:spinFree}} we explicitly verify that the cancellations of
the single-particle contributions that cause the spin-current to cancel have
no counterpart for the \emph{two-particle exchange} SQM current. This
indicates the possibility of pure SQM transport, that is, without spin
current.}}

The most prominent distinction between the direct and exchange SQM currents is
that they differ by a relative factor $| \brkt{\vec{S}^L} |$. To see this
explicitly, we express the SQM current {\eq{eq:SQMCurrent}} as a symmetric,
traceless tensor product of the unit vector $\op{\vec{J}}^L$ with a linear
combination of $\op{\vec{J}}^L$, $\op{\vec{J}}^R$ and $\op{\vec{J}}^L \times
\op{\vec{J}}^R$ by inserting the explicit spin current {\eq{eq:spinCurrent}}:
\begin{eqnarray}
  \brkt{\tens{I}^{L L}_{\tens{Q}}} & = & \widehat{\vec{J}}^L \odot \left[
  E^L_{\tens{Q}} \widehat{\vec{J}}^L + A^L_{\tens{Q}} \widehat{\vec{J}}^R
  \right] \nonumber\\
  &  & + T^L_{\tens{Q}} \widehat{\vec{J}}^L \odot ( \widehat{\vec{J}}^L
  \times \widehat{\vec{J}}^R) ,  \label{eq:SQMasym}
\end{eqnarray}
Each of the coefficients has a direct and exchange contribution, respectively:
\begin{eqnarray}
  E^L_{\tens{Q}} & = & 2 E^L_{\vec{S}} \left( \brkt{\vec{S}^L} \cdot
  \op{\vec{J}}^L \right) - E_{\mathrm{ex}}^L,  \label{eq:ELdef}\\
  A^L_{\tens{Q}} & = & 2 A^L_{\vec{S}} \left( \brkt{\vec{S}^L} \cdot
  \op{\vec{J}}^L \right) - A_{\mathrm{ex}}^L,  \label{eq:JLdef}\\
  T^L_{\tens{Q}} & = & 2 T^L_{\vec{S}} \left( \brkt{\vec{S}^L} \cdot
  \op{\vec{J}}^L \right) - T_{\mathrm{ex}}^L .  \label{eq:TLdef}
\end{eqnarray}
Since all coefficients {\Eq{eq:spinEmission}}-{\eq{eq:spinTorque}} and
{\Eq{eq:SQMEmission}}-{\eq{eq:SQMTorque}} appearing on the right hand sides
are in general of the same order, the ratio of the direct SQM current to the
exchange SQM scales linearly with the average spin $| \brkt{\vec{S}^L} | \sim
N_s = ( J / D) N_o$ as expected from our analysis below {\Eq{eq:correlation}}
in Sec. {\sec{sec:SQMnetwork}}. Consequently, for a \emph{macroscopic}
ferromagnet, the SQM current is dominated by its direct part {\eq{eq:dirCurr}}
and is thus \emph{induced} by the spin-current.Furthermore, since the SQM
current accounts for the change in the correlations between the spin of a
transported electron with all \emph{other} spins in the system, only the
{\emph{SQM current per electron}} is expected to be a meaningful quantity in
the thermodynamic limit{\footnote{In contrast, charge and spin current are
finite in the thermodynamic limit by themselves and {\emph{not}} per electron.
This has the same reason as for the averages: the average spin \emph{per
electron}, but only the average SQM per electron {\emph{pair}} have a finite
limit, cf. discussion of the exchange SQM {\eq{eq:exCont}}. }}. As soon as one
of the subsystems is of meso- or nanoscopic dimensions the relative factor $|
\brkt{\vec{S}^L} |$ matters (cf. {\Sec{sec:SQMT0micro}}) and the exchange SQM
current should be reckoned with. For a nanoscopic system, the full SQM current
was already studied in {\cite{Baumgaertel11}}, while also including the
relevant charging and non-equilibrium effects, which were neglected here. With
this in mind, we will in the following always first discuss the direct part,
dominating the SQM current for macroscopic ferromagnets
({\Fig{fig:tunneljunction}} (a)), and then separately consider the exchange
correction, relevant for mesoscopic ferromagnets, ({\Fig{fig:tunneljunction}}
(b)).

\paragraph{Dissipative and Coherent SQM Current}The second, alternative
decomposition of the SQM current is that into a dissipative and coherent part,
the first and second term of {\Eq{eq:SQMasym}}, respectively, similar to the
spin-dipole current. For non-collinear $\op{\vec{J}}^L$ and $\op{\vec{J}}^R$
these terms are linearly independent tensors and their coefficients have very
different parameter dependencies. The tensorial structure of the total SQM
current is determined by their non-trivial interplay. Its discussion requires
explicit results and is therefore postponed to {\Sec{sec:angleDep}} where we
use the flat band approximation.

The decomposition of the direct SQM current follows by {\Eq{eq:dirCurr}}
directly from the decomposition of the spin-current {\eq{eq:spinCurrent}} into
a dissipative emission, dissipative absorption and coherent torque part. Since
the exchange SQM current is a correction to the direct current accounting for
\ Pauli-forbidden triplet correlations, cf. explanation below \
{\Eq{eq:dirCurr}}, it must have the same decomposition into emission,
absorption and torque part with coefficients given by Eqs.
{\eq{eq:ELdef}}-{\eq{eq:TLdef}}, respectively.

\subparagraph{Dissipative SQM emission $\sim \widehat{\vec{J}}^L \odot
\widehat{\vec{J}}^L$ and absorption $\sim \widehat{\vec{J}}^L \odot
\widehat{\vec{J}}^R$.}

The SQM emission can be microscopically understood as the delocalization of
triplet spin correlations from node $\langle L L \rangle$ to node $\langle L R
\rangle$ (see {\Fig{fig:network}}). The SQM absorption describes the converse
relocalization of such correlations from node $\langle L R \rangle$ to node
$\langle L L \rangle$. This is reflected by the tensorial structure of these
contributions to the average SQM currents: they coincide with the average SQM
stored in the node, from where they are emitted ($\widehat{\vec{J}}^L \odot
\widehat{\vec{J}}^L \sim \brkt{\tens{Q}^{L L}}$) or absorbed
($\widehat{\vec{J}}^L \odot \widehat{\vec{J}}^R \sim \brkt{\tens{Q}^{L R}}$).
Importantly, there is no SQM absorption in $\brkt{\tens{I}^{L L}_{\tens{Q}}}$
that originates from the $\langle R R \rangle$-node ( $\widehat{\vec{J}}^R
\odot \widehat{\vec{J}}^R \sim \brkt{\tens{Q}^{R R}}$) as expected from the
connectivity in the SQM network picture of {\Sec{sec:SQMcurrent}}.

The relation between the storage and transport is also reflected by the
integrands of the exchange SQM current in Eqs.
{\eq{eq:SQMEmission}}-{\eq{eq:SQMInjection}}: these resemble the average local
spin-quadrupolarization $q^L_{\mathrm{ex}} ( \omega) = f^L_+ ( \omega) a^L (
\omega)$ in the expression {\eq{eq:Qtriplet}} for $\brkt{\tens{Q}^{L
L}}_{\mathrm{ex}}$. These can formally be obtained by a replacement $f_+^L (
\omega) \rightarrow \Delta ( \omega)$, i.e., similar to the relation between
charge current and the dissipative spin current in {\Sec{sec:spincurrent}}.
{{However, the symmetry of the function $a^L ( \omega)$ with respect to
$\omega$ is is very different from that of $n^L ( \omega)$ appearing the spin
current emission ({\Eq{eq:spinEmission}}-{\eq{eq:spinInjection}}). This fact
underlies a key result of the paper in {\Sec{sec:spinFree}}.}}

{{The microscopic picture of exchange SQM storage can be extended to
capture a precise, physical understanding of the exchange SQM {\emph{current}}
as follows: as explained in Sect. {\sec{sec:microscopic}}, the Pauli exclusion
holes can be ``counted'' as single-mode cross correlations between two
identical copies of the same ferromagnet (cf. {\Fig{fig:amicro}}). For the
exchange SQM current, we have to imagine that the electron in the first copy
undergoes a tunneling process (representing to the spin current in
{\Eq{eq:Iaa}}), and the second copy is left unchanged (representing the spin
in {\Eq{eq:Iaa}}){\footnote{We also have to account for the situation in which
the role of the first and the second copy are interchanged; however, when
summing over all contributions this gives the same result as in the first
case. This yields the additional factor of 2 in the SQM current.}}. For the
dissipative exchange SQM emission, this becomes directly clear from the
expression {\eq{eq:SQMEmission}}: the Fermi function in $\Delta ( \omega)$
refers to the electron tunneling at energy $\omega$ in the first copy and the
local anisotropy function $a^L$ contains the single-mode correlation of that
electron with the average (unchanged) second copy. This term therefore
describes the flow of Pauli exclusion holes arising from the spin emission
$\sim \op{\vec{J}}^L$. The exchange coefficient {\eq{eq:SQMInjection}}
corresponding to the spin absorption $\sim \op{\vec{J}}^R$, the the first
factor $\Delta n^R$ relates to the absorbed spin from the right electrode and
the second factor $\tilde{a}^L$ represents the correlations of that absorbed
spin with the local spins in the left electrode. Notably, the spin-dependent
contributions $a^L_{\sigma}$ to $a^L = \sum_{\sigma} a_{\sigma}^L$ are added
in $\tilde{a}^L = \sum_{\sigma} \sigma a^L_{\sigma} > 0$ such that they always
count as positive. Here the sign of how to count the Pauli-forbidden
anisotropy is related to the sign of $n^R$: if $n^R > 0$, mostly
spin-$\uparrow$ is absorbed and the missing anisotropy generated by this is
positive, while for $n^R < 0$ mostly spin-$\downarrow$ is absorbed and the
missing anisotropy is negative as explained in Sect. {\sec{sec:microscopic}}}}

\subparagraph{Coherent SQM torque $\sim \widehat{\vec{J}}^L \odot (
\op{\vec{J}}^L \times \widehat{\vec{J}}^R)$} {{The coherent contribution
to the SQM current basically originates from the spin torque.}} This follows
by considering the direct contribution that derives from the spin current, cf.
{\Eq{eq:dirCurr}}. It accounts for the change of the correlation between of
the spin of an electron fixed in the left electrode with the spin of an
electron that \emph{virtually fluctuates} into the right electrode
(spin-flip scattering). Since during this fluctuation the latter spin
precesses about the Stoner field and a net conversion of local into non-local
correlations results, i.e., there is an associated SQM current. The exchange
SQM torque coefficient $T_{\mathrm{ex}}^L$ excludes the single-mode
correlations: in the microscopic picture only the electron in the first copy
undergoes a virtual fluctuation (indicated by $\beta^R$ and $\alpha^R$ in
{\Eq{eq:SQMTorque}}) , while the second copy is left unchanged (indicated by
$a^{L L}$ in {\Eq{eq:SQMTorque}}){\footnote{The analogy between SQM and spin
torque becomes explicit when rewriting $T^L_{\vec{S}} = \int \mathd \omega
\Gamma \left( \tfrac{\alpha^R}{\bar{\nu}^R} f^L_+ -
\tfrac{\beta^R}{\bar{\nu}^R} \right) n^L$ by interchanging $\omega
\leftrightarrow \omega'$ in the first term contributing to the double integral
{\eq{eq:spinTorque}}. Then $T^L_{\mathrm{ex}}$ is obtained from $T^L_{\vec{S}}$
by replacing {{the spin-polarization by the quadrupolarization function}},
$n^L \rightarrow a^L$.}}. This is the effect of the spin torque on the local
Pauli-exclusion holes.

\subsection{Parameter Dependence}\label{sec:result1band}

Having discussed the general structure and physical meaning of the main
results {\eq{eq:chargeCurrent}}-{\eq{eq:SQMCurrent}}, we now simplify them as
far as possible by making the flat band approximation. {{Although this is
a crude approximation, it reveals a general key feature of the exchange SQM,
namely its sensitivity to the spin {\emph{alignment}}, a non-negative quantity
that accumulates when summing over energies / $k$-modes. This prohibits
cancellations in the exchange SQM current as they occur due to \ signed
contributions in the charge and spin-dipole current. (cf.
{\Sec{sec:microscopic}})}}. As noted in {\Sec{sec:wide-flat-band}} the 2DOS
appearing in the exchange expressions requires modelling of the electrodes
that goes beyond the 1DOS. However, in the flat-band approximation we only
need {\Eq{eq:tripletCorrSimple}}. We furthermore apply the bias voltage
symmetrically to the ferromagnets, $\mu^L = + V / 2$ and $\mu^R = - V / 2$
while considering arbitrary non-collinear Stoner vectors $\op{\vec{J}}^L$ and
$\op{\vec{J}}^R$. We assume all further parameters to be symmetric: $J^r = J$,
$T^r = T$, $D^r = D$ and $\nu^r_{\sigma} = \nu_{\sigma}$ for $\sigma =
\uparrow, \downarrow$ (except for a temperature gradient discussed in
{\Sec{sec:spinFree}}). In this approximation the densities of states are fixed
by the bandwidths, $\nu_{\sigma} = 1 / 2 D$, and the tunneling rate is set by
the spin-conserving tunnel amplitude $t$: $\Gamma = 2 \pi ( t / 2 D)^2$, cf.
{\Eq{eq:tsigmacons}}. Together with the leading order approximation in
$\Gamma$ this limits the applicability of the results to the regime $\Gamma
\ll T^{}, J, \ll W$ (cf. {\Eq{eq:distmuD}}). The central equations
{\eq{eq:chargeCurrent}}-{\eq{eq:SQMCurrent}} now simplify to
\begin{eqnarray}
  \brkt{I^L_N} & = & I_{N, 0}^L,  \label{eq:chargeCurrentSimple}\\
  \brkt{\vec{I}^L_{\vec{S}}} & = & T^L_{\vec{S}} \left( \widehat{\vec{J}}^L
  \times \widehat{\vec{J}}^R \right),  \label{eq:spinCurrentSimple}
\end{eqnarray}
{\wideeq{\begin{eqnarray}
  \brkt{\tens{I}^{L L}_{\tens{Q}}} & = & 2 \left( \brkt{\vec{S}^L} \cdot
  \widehat{\vec{J}}^L \right)  \widehat{\vec{J}}^L \odot
  \brkt{\vec{I}^L_{\vec{S}}} - \widehat{\vec{J}}^L \odot \left[ E_{\tens{Q}}^L
  \widehat{\vec{J}}^L + T^L_{\tens{Q}}  \left( \widehat{\vec{J}}^L \times
  \widehat{\vec{J}}^R \right) \right]  \label{eq:SQMCurrentSimple}\\
  & = & - E_{\mathrm{ex}}^L \widehat{\vec{J}}^L \odot \widehat{\vec{J}}^L +
  \left( 2 \left( \brkt{\vec{S}^L} \cdot \widehat{\vec{J}}^L \right)
  T^L_{\vec{S}} - T^L_{\mathrm{ex}} \right) \widehat{\vec{J}}^L \odot \left(
  \widehat{\vec{J}}^L \times \widehat{\vec{J}}^R \right) . 
  \label{eq:SQMCurrentSimple2}
\end{eqnarray}}}

In this approximation the DOS {\eq{eq:DOS}} is not spin-polarized in the bias
window: $n^r ( \omega) = 0$ for $\mu^L \lesssim \omega \lesssim \mu^R$.
Therefore the charge current \ {\Eq{eq:chargeCurrentSimple}} reduces to its
non-magnetic part, i.e., there is no spin-valve effect. By the same token, the
dissipative part of the spin current {\eq{eq:spinCurrentSimple}} vanishes due
to the cancellation of particle and hole contributions. Thus only the coherent
spin torque part remains, whose coefficient we now estimate as follows:
inserting {\Eq{eq:betar}} into {\Eq{eq:spinTorque}}, we obtain
\begin{eqnarray}
  T^L_{\vec{S}} & = & 2 t^2 \mathrm{Re} \int \mathd \omega \int \mathd \omega' 
  \prod_r ( \bar{\nu}^r ( \omega) n^r ( \omega)) \nonumber\\
  &  & \times \frac{f_+^R ( \omega') - f_+^L ( \omega)}{\omega - \omega' + i
  0} .  \label{eq:spinTorqueAppr}
\end{eqnarray}
For our DOS approximation, we have $\bar{\nu}^L ( \omega) n^L ( \omega)
\bar{\nu}^R ( \omega') n^R ( \omega') = \mathrm{sgn} ( \omega \omega') / ( 4
D)^2$ if $| \omega | - D \leqslant J / 2$ and $| \omega' | - D \leqslant J /
2$ and zero otherwise. At these energies the Fermi functions are 0 or 1 and if
their difference in {\Eq{eq:spinTorqueAppr}} is nonzero, we can approximate $1
/ | \omega - \omega' | \approx 1 / ( 2 D)$, yielding
\begin{eqnarray}
  T^L_{\vec{S}} & \approx & - \frac{\Gamma}{4 \pi} \frac{J^2}{D} . 
  \label{eq:estimateSpinTorque}
\end{eqnarray}
where $\Gamma = 2 \pi | t |^2 / ( 2 D)^2$. The resulting spin current is
equivalent to a spin-torque exerted by an magnetic field $\vec{B}^R \approx
\bar{\nu}^R J | t |^2 / D \widehat{\vec{J}}^R $on the spin on $\vec{S}^L$
(insert {\Eq{eq:spinavT0}} for $\left| \vec{S}^L \right|$ into
{\Eq{eq:spinCurrentSimple}}). Finally, in the SQM current
{\eq{eq:SQMCurrentSimple}}-{\eq{eq:SQMCurrentSimple2}} the absorption
coefficient $A_{\tens{Q}}^L$ \ -- and with it, the non-local anisotropy
function $a^{L R}_{}$ -- drops out in this approximation because of the
vanishing of the spin-polarization, $n^r ( \omega) = 0$ in the bias window.

\subsubsection{Dissipative SQM Flow Direction}\label{sec:direction}

We first discuss the direction of the dissipative spin-anisotropy emission,
$\brkt{\tens{I}^{L L}_{\tens{Q}}} = - 2 E_{\tens{Q}}^L \widehat{\vec{J}}^L
\odot \widehat{\vec{J}}^L$, which entirely arises from the exchange term in
{\Eq{eq:SQMCurrentSimple}} (the direct part vanishes in absence of dissipative
spin current). Remarkably, the emission always results in a {\emph{loss}} of
local exchange spin-anisotropy $\brkt{\tens{Q}^{L L}}_{\mathrm{ex}} = - q^{L
L}_{\mathrm{ex}} \op{\vec{J}^{}}^L \odot \op{\vec{J}}^L$ irrespective of the
voltage bias direction, because $E_{\tens{Q}}^L$ is always {\emph{negative}}
(unless zero). By interchanging the left and right electrode index in all
expressions, we \ observe that the spin-anisotropy of the $\langle R R
\rangle$-node decreases as well. We conclude from the SQM current conservation
law that non-local spin-triplet correlations are built up irrespective of the
bias direction. This is in accordance with the physical intuition of SQM
{\emph{transport}} that we have developed using our network picture the
tunneling of electrons across the junction delocalizes spin-triplet
correlations. However, such a pure delocalization is special to a
voltage-biased tunnel junction. When we discuss the situation of thermal bias
later, we will see that an effective transfer of spin-anisotropy between the
local nodes is still possible.

These results can also be clearly understood in terms of the microscopic
picture of SQM {\emph{storage}} introduced in Sect.
{\sec{sec:energyResolved}}. {{To see this}}, we first note that the
exchange SQM emission {\eq{eq:SQMEmission}} is obtained from the average (see
Eqs. {\eq{eq:Qtriplet}} -{\eq{eq:quadProb}}) by replacing the Fermi function
$f^L_+$ by the bias function $\Delta$. If $\mu^L > \mu^R$, electrons
{\emph{leave}} the left electrode, which is indicated by $\Delta ( \omega) >
0$ at energies $\mu^R \lesssim \omega \lesssim \mu^L$. This destroys the
positive local exchange SQM content at this energy (given by $a^L ( \omega) >
0$, see also {\Fig{fig:anisotropyocc}}). Conversely, for the opposite bias
$\mu^L < \mu^R$, we find $\Delta ( \omega) < 0$ for $\mu^L \lesssim \omega
\lesssim \mu^R$ since the tunneling electrons {\emph{enter}} the left
electrode. Electrons at these frequencies provide {\emph{negative}} exchange
SQM (as $a^L ( \omega) < 0$). In both cases this results in a
{\emph{negative}} change in $q^L_{\mathrm{ex}}$.

\subsubsection{Scalar Parameter Dependence}\label{sec:scalarDep}

We next discuss the dependence of the direct and exchange SQM current on the
{\emph{scalar}} parameters $J$, $V = \mu^L - \mu^R$ and $T$. For the direct
SQM current,
\begin{eqnarray}
  \brkt{\tens{I}^{L L}_{\tens{Q}}}_{\mathrm{dir}} & = & 2 \left(
  \brkt{\vec{S}^L} \cdot \widehat{\vec{J}}^L \right) T^L_{\vec{S}} 
  \widehat{\vec{J}}^L \odot \left( \widehat{\vec{J}}^L \times
  \widehat{\vec{J}}^R \right), 
\end{eqnarray}
this is simple because $T^L_{\vec{S}}$ is nearly independent of $V$ and $T$
and increases as $J^3$ (use Eqs. {\eq{eq:estimateSpinTorque}} and
{\eq{eq:spinavT0}}), as shown in {\Fig{fig:dirJDep}}.

{
  \begin{figure}[h]
    \includegraphics[width=0.9\linewidth]{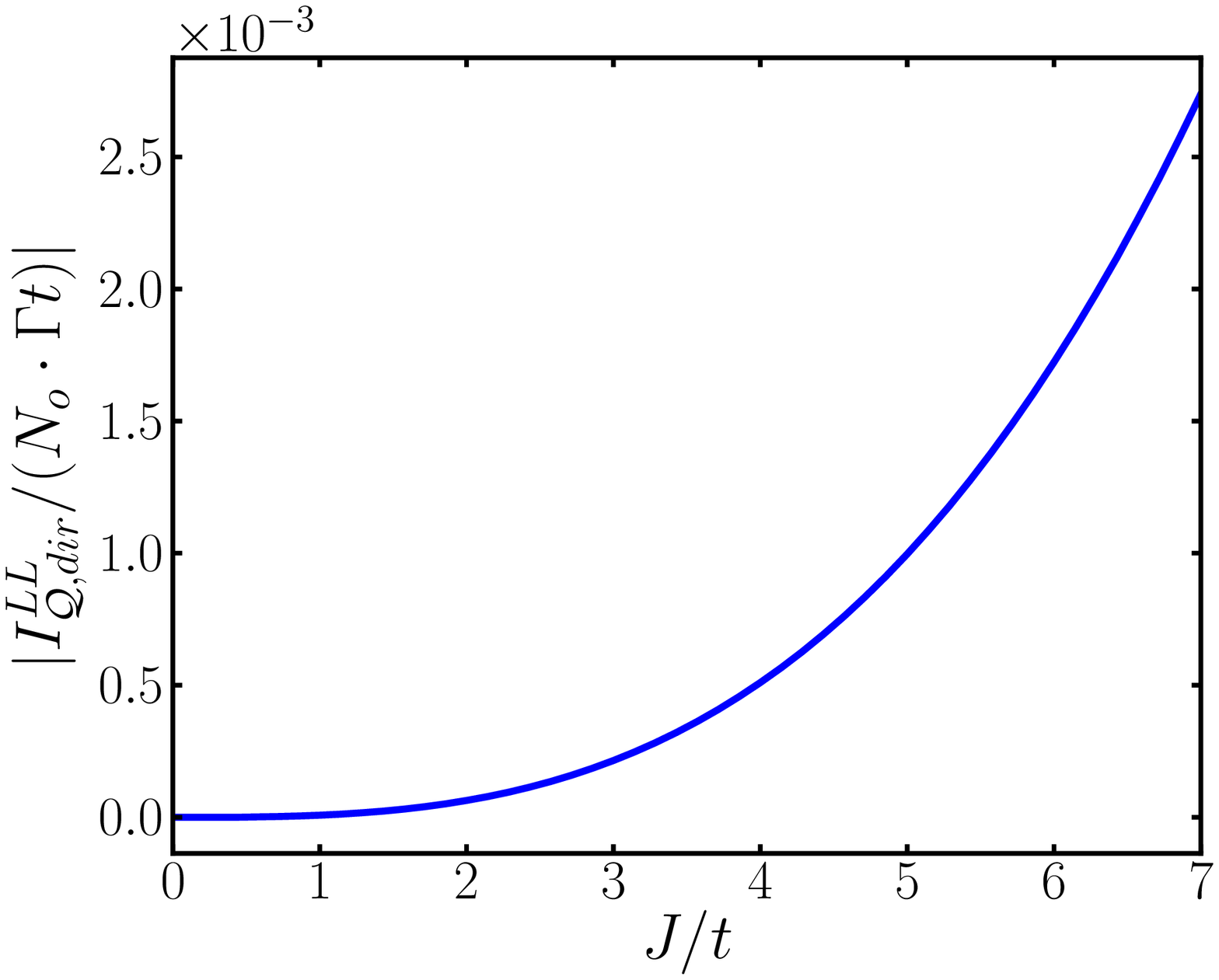}
    \caption{\label{fig:dirJDep}Dependence of $I^{L
L}_{\tens{Q}, \text{dir} \text{}} = 2 \brkt{\vec{S}^L} \cdot \op{\vec{J}}^L
T^L_{\vec{S}}$ on the Stoner splitting $J / t$, where $N_o$ is the fixed
number of orbital states of the left subsystem and $J = 5 t $, $D = 25
t$, $T = t / 2$.}\end{figure}}

The exchange SQM current, in contrast, shows a stronger dependence on the
scalar parameters. To see this, we first roughly estimate how its emission and
torque coefficients scale with $V$ and $J$ for low temperatures $T \lesssim V,
J$. For $T = 0$ the integrand of {\eq{eq:SQMEmission}} is the product of the
anisotropy function $a^{L L} ( \omega)$, which has a support of width $2 J$,
and the bias function $\Delta ( \omega)$ with a support of width $V$, and the
smaller one of these energy scales limits the SQM emission:
\begin{eqnarray}
  E_{\mathrm{ex}}^L & \approx & - \frac{\Gamma}{2} \min ( | V |, | J |), 
  \label{eq:emissionScale}
\end{eqnarray}
where $\Gamma = 2 \pi | t |^2 / ( 2 D)^2$. In contrast, the SQM torque
{\eq{eq:SQMTorque}} scales in the same way as the spin torque:
\begin{eqnarray}
  T_{\mathrm{ex}}^L & \approx & - \frac{\Gamma}{\pi}  \frac{J^2}{D}, 
  \label{eq:torqueEst} \label{eq:torqueScale}
\end{eqnarray}
where we also set $T = 0$ and proceeded analogous to the estimation of the
spin torque (cf. {\Eq{eq:estimateSpinTorque}}). The additional suppression
factor $J / D$ in {\Eq{eq:torqueEst}} relative to {\Eq{eq:emissionScale}} for
$V < J$ reflects that the SQM torque originates from coherent virtual
fluctuations to states near the band edges where the spin-polarization is
nonzero in an energy window proportional to the Stoner splitting $J$. This
gives rise to two regimes, in which the coherent exchange term
{\eq{eq:torqueScale}} is larger (smaller) than the dissipative exchange term
{\eq{eq:emissionScale}} for $V \lessgtr V^{\ast}$, where $E_{\mathrm{ex}}^L \sim
T^L_{\mathrm{ex}}$ occurs for{\footnote{{\Eq{eq:Vstar}} holds for $V^{\ast} <
J$, i. e., $J < D \text{/} \pi$, which is valid for the flat-band
approximation (cf. {\sec{sec:wide-flat-band}}). Otherwise
{\Eq{eq:emissionScale}} in already limited by $| J |$.}}
\begin{eqnarray}
  V^{\ast} & = & \frac{J^2}{\pi D} .  \label{eq:Vstar}
\end{eqnarray}

{
  \begin{figure}[h]
    \includegraphics[width=0.9\linewidth]{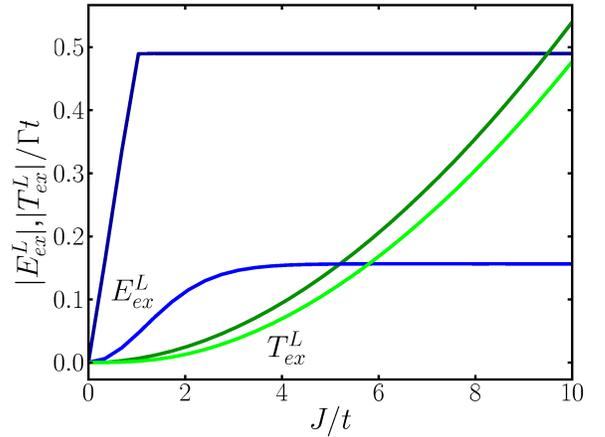}
    \caption{\label{fig:JDep}Dependence of the
magnitude of SQM emission $| E^L_{\mathrm{ex}} | \text{/} \Gamma t$ (blue) and
torque $| T_{\mathrm{ex}}^L | \text{/} \Gamma t$ (green) on the Stoner splitting
\ $J / t$ for $T = 0$ (dark colors) and $T = 0.5 t$ (light colors). The
residual parameters are $V = t$, $D = 25 t$. For $T = 0$, the crossover occurs
at $J \sim J^{\ast} = \sqrt{\pi D V \text{}} \approx 9 t$. The initial
non-linearity of the SQM emission coefficient for the finite temperature
(preceded by a linear regime), and the smaller saturation value compared to
the $T = 0$ case are due to the thermal smearing.}
\end{figure}}

\paragraph{Stoner-field dependence}In {\Fig{fig:JDep}} we show a numerical
calculation of the precise shapes {\eq{eq:SQMEmission}} and
{\eq{eq:SQMTorque}} of the exchange emission coefficient $E_{\mathrm{ex}}^L$ and
the torque coefficient $T_{\mathrm{ex}}^L$, confirming the estimates
{\eq{eq:emissionScale}} and {\eq{eq:torqueScale}}: they show that the torque
increases quadratically and the emission saturates to a constant on the scale
of bias $V$ (which is smaller than the value predicted by
{\Eq{eq:emissionScale}} due to finite temperature).

{
  \begin{figure}[h]
    \includegraphics[width=0.9\linewidth]{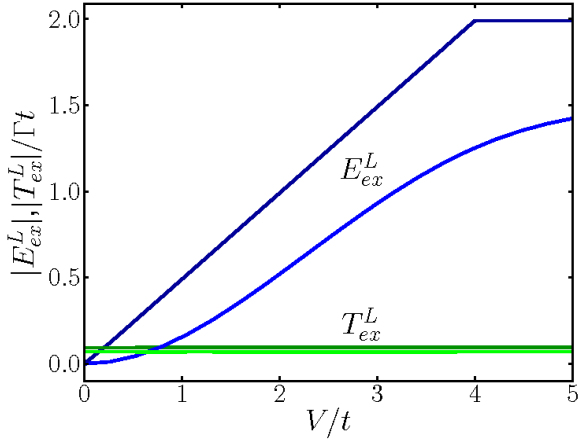}
    \caption{\label{fig:VDep}Dependence of the
magnitude of SQM emission $| E_{\mathrm{ex}}^L | / \Gamma t$ (blue) and torque
$| T_{\mathrm{ex}}^L | / \Gamma t$ (green) on the bias voltage $V / t$ for $T =
0$ (dark colors) and $T = 0.5 t$ (light colors). The residual parameters are
$J = 5 t $, $D = 25 t$. For both temperatures, the torque is constant
at $| T^L_{\mathrm{ex}} | \approx 0.15$ according to estimation
{\eq{eq:torqueEst}}. The small deviation of this value and the saturation
level of $| E_{\mathrm{ex}}^L | / \Gamma t$ for finite temperature compared to
$T = 0$ is due to the thermal smearing. Therefore, the rough estimate
$V^{\ast} \approx J^2 / \pi D \approx 1 / 3$ for the crossing point at
$E_{\mathrm{ex}}^L = T^L_{\mathrm{ex}}$ is exactly fulfilled only for $T
   = 0$.}
\end{figure}}

\paragraph{Bias dependence}{\Fig{fig:VDep}} shows the same crossover, but now
as function of the bias $V$ for fixed $J$ and $T$. The torque is constant and
given approximately by {\eq{eq:torqueEst}} and the emission saturates at the
value set by {\eq{eq:emissionScale}} when $V$ approaches the scale of $J$.
{{This \ voltage dependence allows for magnetic and electric control over
the orientation of the exchange SQM current \emph{tensor}, discussed in
the next {\Sec{sec:angleDep}}.}} The saturation at $V \sim J$ is an
interesting, new feature of the dissipative exchange SQM current, not present
in the charge or spin current. It provides access to the Stoner
\emph{shift} of the DOS, cf. {\Fig{fig:anisotropyocc}}, even without
spin-polarization of the DOS in the bias window. Similar to the spin current a
finite coherent SQM term remains at zero bias, even though the dissipative SQM
current vanishes.

{
  \begin{figure}[h]
    \includegraphics[width=0.9\linewidth]{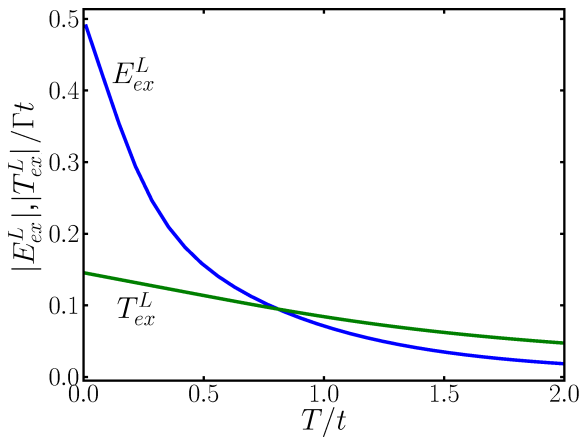}
    \caption{\label{fig:TDep}Dependence of the
magnitude of exchange SQM emission $| E_{\mathrm{ex}}^L | / \Gamma t$ (blue) and
torque $| T^L_{\mathrm{ex}} | / \Gamma t$ (green) on temperature $T / t$ for $V
= t $, $J = 5 t$ $D = 25 t$. }
\end{figure}}

\paragraph{Temperature dependence}In {\Fig{fig:TDep}} we show the temperature
dependence of the exchange SQM emission and torque coefficients, keeping $V$
and $J$ fixed. Both coefficients decrease monotonously with temperature, but
with very different characteristic dependencies on $T$. The reason is that the
emission integral {\eq{eq:SQMEmission}} incorporates Fermi functions, which
have a much stronger exponential dependence with $T^{- 1}$, whereas torque
integral {\eq{eq:SQMTorque}} comprises the renormalization function $\beta^R$,
which depends much weaker, namely algebraically on $T^{- 1}$. Moreover,
{\Fig{fig:TDep}} shows that exchange SQM emission is strongly suppressed when
$T$ approaches the voltage $V ( < J)$ when the bias function $\Delta = f^R_+ -
f_+^L$ is largely broadened over an energy range of $\sim 4 T \sim J$ (for the
parameters of {\Fig{fig:TDep}}), so that positive and negative contributions
of the spin-anisotropy function $a^L$ cancel each other. This is similar to
the discussion of the exchange SQM storage in Sec. {\sec{sec:energyResolved}}.

\subsubsection{Angle Dependence}\label{sec:angleDep}

The exchange SQM current, in contrast to the direct part, has a nontrivial
dependence {{on the angle}} between the two Stoner vectors
$\op{\vec{J}}^L$ and $\op{\vec{J}}^R$ due to the interplay of its dissipative
and coherent contributions (see Sec. {\sec{sec:angleDep}}). This requires a
more extensive analysis since we are dealing with a {\emph{tensor}}-valued
current. There are two relevant questions relating to the orientation of the
SQM tensors. {{The first question is whether the orientation of the local
SQM $\brkt{\tens{Q}^{L L}}$ is changed by the injected SQM current
$\brkt{\tens{I}_{\tens{Q}}^{L L}}$.}} One can show that if these
\emph{tensors} commute, $[ \brkt{\tens{I}_{\tens{Q}}^{L L}},
\brkt{\tens{Q}^{L L}}] = 0$, {{then}} the SQM current corresponds
{{only}} to a change in the principal values of the local SQM
{{\emph{without}}} changing its principal axes (see
{\App{sec:approtation}}). Now {\Eq{eq:SQMCurrent}} shows that for
non-collinear $\widehat{\vec{J}}^L$ and $\widehat{\vec{J}}^R$ \ the average
SQM current $\brkt{\tens{I}_{\tens{Q}}^{L L}}$ is a superposition of three
linearly independent, symmetric and traceless tensors, $\widehat{\vec{J}}^L
\odot \widehat{\vec{J}}^L$ (emission {\eq{eq:SQMEmission}}) ,
$\widehat{\vec{J}}^L \odot \widehat{\vec{J}}^R$(absorption
{\eq{eq:SQMInjection}}) and $\widehat{\vec{J}}^L \odot ( \op{\vec{J}}^L \times
\widehat{\vec{J}}^R)$ (torque {\eq{eq:SQMTorque}}). The latter two tensors do
not commute with $\brkt{\tens{Q}^{L L}} \propto \widehat{\vec{J}}^L \odot
\widehat{\vec{J}}^L$ for non-collinear $\widehat{\vec{J}}^L$ and
$\widehat{\vec{J}}^R$ and vanish only for collinear $\widehat{\vec{J}}^L$ and
$\widehat{\vec{J}}^R$. This holds for both the direct and exchange
contributions. Therefore the injected average SQM current
$\brkt{\tens{I}_{\tens{Q}}^{L L}}$ will tend to change the principal axes of
the average local SQM $\brkt{\tens{Q}^{L L}}$, besides changing its principal
values.

The second question is whether the direct and exchange SQM currents tend to
induce the same rotation of the principal axis of the SQM, which is equivalent
to these tensors commuting, $[ \brkt{\tens{I}_{\tens{Q}}^{L L}}_{\mathrm{dir}},
\brkt{\tens{I}_{\tens{Q}}^{L L}}_{\mathrm{ex}}] = 0$. To show that this is not
the case, we now first explicitly find the (different) principal axes and
values of $\brkt{\tens{I}^{L L}_{\tens{Q}}}_{\mathrm{dir}}$ and
$\brkt{\tens{I}^{L L}_{\tens{Q}}}_{\mathrm{ex}}$. This will furthermore allow us
to plot and discuss these average SQM currents in a clear way.

\paragraph{Principal axes and values}The following analysis holds for any
dispersion $\varepsilon^r_{n k \sigma}$. We first diagonalize the full average
SQM current tensor {\Eq{eq:SQMCurrent}} and then show how the direct and
exchange part can be obtained from the result. Since the former is real and
symmetric tensor it can always be diagonalized:
\begin{eqnarray}
  \brkt{\tens{I}^{L L}_{\tens{Q}}} & = & \sum_{\lambda = \pm, 0} I_{\lambda}
  \op{\vec{v}}_{\lambda} \op{\vec{v}}_{\lambda} . 
\end{eqnarray}
Here the $\op{\vec{v}}_{\lambda}$ denote the orthonormal {\emph{principal
axes}} (the hat indicating normalization) and the $I_{\lambda}$ denote
{\emph{principal SQM currents}}, which quantify the magnitude of the SQM
current. (We dropped the superscripts ``$L L$'' on $I_{\lambda}$ and
$\op{\vec{v}}_{\lambda}$ for brevity). In {\App{sec:appDiagonal}} we show that
for non-collinear Stoner vectors ($\cos ( \theta) = \widehat{\vec{J}}^L \cdot
\widehat{\vec{J}}^R \neq \pm 1$) the principal SQM currents are
\begin{eqnarray}
  I_0 & = & - \tfrac{1}{3} D_{\theta},  \label{eq:Izero}\\
  I_{\pm} & = & \tfrac{1}{6} D_{\theta} \pm \tfrac{1}{2} S_{\theta}, 
  \label{eq:Iplusminus}
\end{eqnarray}
which add up to 0 as they should (traceless tensor). The unnormalized
principal axes read
\begin{eqnarray}
  \vec{v}_0 & = & \widehat{\vec{J}}^L \times \left( A_{\tens{Q}}^L
  \widehat{\vec{J}}^R + T^L_{\tens{Q}} \widehat{\vec{J}}^L \times
  \widehat{\vec{J}}^R \right),  \label{eq:vzero}\\
  \vec{v}_{\pm} & = & ( D_{\theta} \pm S_{\theta})  \widehat{\vec{J}}^L +
  A^L_{\tens{Q}} \widehat{\vec{J}}^R + T^L_{\tens{Q}}  \widehat{\vec{J}}^L
  \times \widehat{\vec{J}}^R .  \label{eq:vplusminus}
\end{eqnarray}
In Eqs. {\eq{eq:Izero}}-{\eq{eq:vplusminus}}, we used the abbreviations
\begin{eqnarray}
  D_{\theta} & : = & \left( E^L_{\tens{Q}} + A^L_{\tens{Q}} \cos \theta
  \right),  \label{eq:Dtheta}\\
  S_{\theta} & : = & \sqrt{D_{\theta}^2 + \left( \left( T^L_{\tens{Q}}
  \right)^2 + \left( A^L_{\tens{Q}} \right)^2 \right) \sin^2 \theta} . 
  \label{eq:Stheta}
\end{eqnarray}
The principal SQM currents obey the inequalities
\begin{eqnarray}
  & I_- \leqslant I_0 \leqslant I_+ &  \label{eq:inequal}
\end{eqnarray}
since $S_{\theta} \geqslant D_{\theta}$. For collinear Stoner vectors ($\theta
= 0, \pi$), two eigenvalues are degenerate,
\begin{eqnarray}
  I_+ & = & \left( \tfrac{p}{6} + \tfrac{1}{2} \right) \left| E^L_{\tens{Q}} +
  A^L_{\tens{Q}} \right|, \\
  I_- & = & \left( \tfrac{p}{6} - \tfrac{1}{2} \right) \left| E^L_{\tens{Q}} +
  A^L_{\tens{Q}} \right|, \\
  I_0 & = & \tfrac{p}{3} \left| E^L_{\tens{Q}} + A^L_{\tens{Q}} \right| 
  \label{eq:eigen0pi}
\end{eqnarray}
with $p = \mathrm{sgn} ( D_0)$, $\op{\vec{v}}_+ = \op{\vec{J}}^L$ and any two
vectors in the plane perpendicular to $\op{\vec{J}}^L$ are principal axes of
the SQM current. {{We thus see that in general, the principal SQM currents
take three different values, i.e. a \emph{bi-axial spin-anisotropy} is
transported whenever the Stoner vectors are non-collinear. We emphasize that
\emph{by itself} the average spin-current \emph{vector} provides no
information about the non-collinearity of the spin-valve. The SQM current
tensor, in contrast, does: the transported anisotropy only becomes uniaxial
for collinear Stoner vectors ($\theta = 0, \pi$), in which case
$\op{\vec{J}}^L$ is the hard axis{\footnote{If $\theta \neq 0, \pi$, the SQM
current could only be uniaxial if $A^{L_{}}_{\tens{Q}} = 0$ and
$T^L_{\tens{Q}} = 0$ is fulfilled at the same time (otherwise the
$I_{\lambda}$ are clearly angle dependent according to Eqs. (\ref{eq:Dtheta})
and (\ref{eq:Stheta})). This is in general not expected as the SQM torque
senses the band structure over a wide energy range.}}}}. {{The possibility
of injecting biaxial anisotropy into molecular scale systems is of interest
since this type of anisotropy it is associated with interesting quantum-spin
tunneling effects {\cite{Gatteschi03rev}}.}}

We obtain the diagonal form of the direct and exchange SQM individually by
replacing the coefficients {\Eq{eq:ELdef}}-{\eq{eq:TLdef}} in the above
formulas by $E^L_{\tens{Q}} \rightarrow E^L_{\text{dir}} = \left( \vec{S}^L
\cdot \widehat{\vec{J}}^L \right) E^L_{\vec{S}}$ and $E^L_{\tens{Q}}
\rightarrow - E^L_{\mathrm{ex}}$, respectively, etc. Since these two sets of
coefficients are in general different functions of the various parameters, we
conclude that {{the direct and exchange SQM tensors have different
principal axes and do not commute, $[ \brkt{\tens{I}_{\tens{Q}}^{L
L}}_{\mathrm{dir}}, \brkt{\tens{I}_{\tens{Q}}^{L L}}_{\mathrm{ex}}] \neq 0$.}}
They therefore tend to induce the different rotations of the principal axis of
the local SQM. \ As a result the total principal SQM currents are {\emph{not}}
the sum of the direct and exchange principal SQM currents.

\paragraph{Flat band approximation}So far the considerations were general. We
now investigate the magnetic tuning of the exchange SQM orientation by the
Stoner vectors. For the flat band approximation,
{\Eq{eq:chargeCurrentSimple}}-{\eq{eq:SQMCurrentSimple}}, the principal SQM
currents and the principal axes are symmetric with respect to $\theta = \pi /
2$, i. e.,
\begin{eqnarray}
  I_{\lambda} \left( \tfrac{\pi}{2} - \alpha \right) & = & I_{\lambda} \left(
  \tfrac{\pi}{2} + \alpha \right), \\
  \op{\vec{v}}_{\lambda} \left( \tfrac{\pi}{2} - \alpha \right) & = &
  \op{\vec{v}}_{\lambda} \left( \tfrac{\pi}{2} + \alpha \right) . 
\end{eqnarray}
This is due to the vanishing SQM absorption coefficient, $A^L_{\tens{Q}}$, in
this limit. In {\Fig{fig:angledepEx}} we plot both the direct and the exchange
principal SQM currents as a function of the angle $\theta \in [ 0, \pi / 2]$.
We observe that both $I_0^{\mathrm{dir}}$ and $I_0^{\mathrm{ex}}$ are constant. In
both cases, the repulsion of both \ $I_{\pm}^{\mathrm{dir} / \mathrm{ex}}$ is
caused by the respective torque coefficient (see {\Eq{eq:Dtheta}}), which
increases with $J$ and decreases with $D$ (cf.
{\Eq{eq:estimateSpinTorque}}-{\eq{eq:torqueScale}}).

{
  \begin{figure}[h]
    \includegraphics[width=0.9\linewidth]{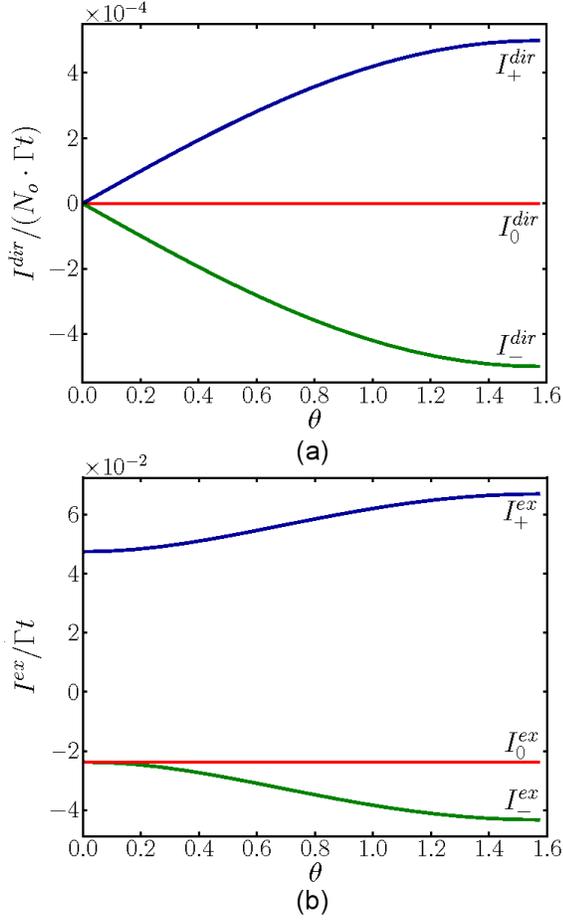}
    \caption{\label{fig:angledepEx}Angle
dependence of the principal SQM currents (a) $I_{\lambda}^{\text{dir}} / ( t
\cdot N_s)$ and (b) $I_{\lambda}^{\mathrm{ex}} / t$, respectively (red: $I_0$,
green: $I_-$, blue: $I_+$). The principal direct SQM currents only depend on
the nonzero spin torque $I_{\lambda}^{\mathrm{dir}} = \lambda 2| \vec{S}^L
||T^L_{\vec{S}} \sin \theta |$ $( \lambda = 0, \pm)$, whence the principal
exchange SQM currents depend both the nonzero exchange SQM emission and
torque. Parameters: $J = 5 T$, $D = 25 t  , V = T = t$,
$\Gamma = 2 \pi / 2500$. Note that the inequalities {\Eq{eq:inequal}} and
$\sum_{\lambda} I_{\lambda}^{\mathrm{dir}} = \sum_{\lambda}
I_{\lambda}^{\mathrm{ex}} = 0$ are fulfilled.}{}
\end{figure}}

To analyse the angle dependence of the principal axes
$\op{\vec{v}}_{\lambda}$, we construct a right handed coordinate system from
the non-collinear, non-orthogonal Stoner vectors: let $\vec{e}_z =
\widehat{\vec{J}}^L$, $\vec{e}_x = \widehat{ \vec{J}}^L \times
\widehat{\vec{J}}^R / \sin \theta$, $\vec{e}_y = \vec{e}_z \times \vec{e}_x$.
In Fig. \ref{fig:angleDep} we show the trajectories of the direct and exchange
principal axes as we rotate $\widehat{\vec{J}}^R = - \sin \theta \vec{e}_y +
\cos \theta \vec{e}_z$ through $\theta \in [ 0, \pi / 2]$. Then both
$\op{\vec{v}}_0^{\text{dir}} = \op{\vec{v}}^{\mathrm{ex}}_0 = \vec{e}_y$ are a
fixed directions in this coordinate system, i.e., independent of the angle
$\theta$. The other principal axes $\op{\vec{v}}_{\pm}^{\text{dir$/
\mathrm{ex}$}}$ lie in the $x z$-plane perpendicular to
$\op{\vec{v}}^{\text{dir$/$ex}}_0$ and are different for the direct and
exchange contribution.

The direct SQM current only consists of a torque contribution and therefore
its principal axes are independent of $\theta${\color{red} } in the above
coordinate system fixed by $\widehat{\vec{J}}^L$ and $\widehat{\vec{J}}^R$.
With $A^L_{\text{dir}} = E^L_{\text{dir}} = 0$ and $T^L_{\text{dir}} = 2 (
\vec{S}^L \cdot \widehat{\vec{J}}^L) T^L_{\vec{S}}$,
{\Eq{eq:vzero}}-{\eq{eq:vplusminus}} give $\op{\vec{v}}^{\text{dir}}_{\pm} =
\left( \pm \vec{e}_z - \vec{e}_x \right) \text{/} \sqrt{2}$ for $\theta \neq
0, \pi$ (when the direct SQM current is nonzero). In contrast, the exchange
SQM current shows a nontrivial competition of the torque and emission
contributions: in Fig. \ref{fig:angleDep} we plot the trajectories of its
principal axes in the plane perpendicular to $\op{\vec{v}}_0^{\mathrm{ex}} =
\vec{e}_y$ as $\theta$ is increased to from 0 to $\pi / 2$.

{{We emphasize that already in this simple model of flat-band
ferromagnets the spin-anisotropy flow has non-trivial tensorial structure: in
general the principal axes of the exchange SQM tensor are neither collinear to
$\widehat{\vec{J}}^L$, $\widehat{\vec{J}}^R$ nor collinear to
$\widehat{\vec{J}}^L \times \widehat{\vec{J}}^R$. Only for nearly collinear
configurations ($\theta \approx 0, \pi$), when the torque contribution is
negligible, do we have such a simple result: $\widehat{\vec{v}}_+^{\mathrm{ex}}
\approx \vec{e}_x = \widehat{\vec{J}}^L$ and $\widehat{\vec{v}}_- \approx
\vec{e}_z = \widehat{\vec{J}}^L \times \widehat{\vec{J}}^R \text{/} \sin (
\theta)$.}}

{
  \begin{figure}[h]
    \includegraphics[width=0.9\linewidth]{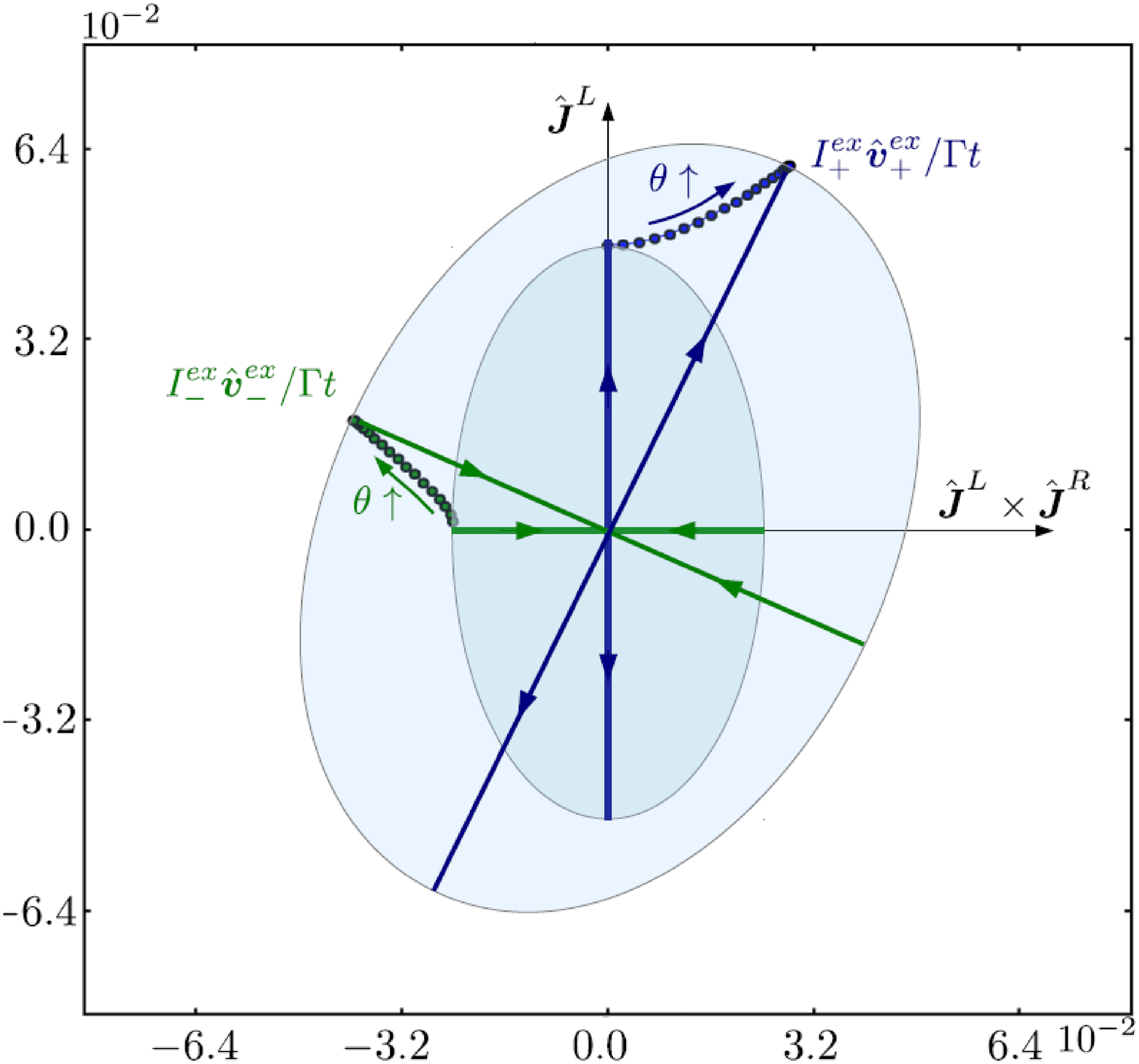}
    \caption{\label{fig:angleDep}Angle
dependence of the exchange SQM tensor $\brkt{\tens{I}_{\tens{Q}}^{L
L}}_{\mathrm{ex}}$. Each point corresponds an principal vector $I^{\mathrm{ex}}_+
\op{\vec{v}}^{\mathrm{ex}}_+ / t$ (blue) and $I^{\mathrm{ex}}_-
\op{\vec{v}}_-^{\mathrm{ex}} / t$ (green) of $\brkt{\tens{I}_{\tens{Q}}^{L
L}}_{\mathrm{ex}}$ for increasing angle $\theta$ in steps of $\pi / 36$ from $0$
to $\pi / 2$. Each pair of vectors $I^{\mathrm{ex}}_{\pm}
\op{\vec{v}}_{\pm}^{\mathrm{ex}} / t$ for one angle defines the semi axes of an
ellipse, drawn for the extremal values $\theta = 0$ and $\theta = \pi / 2$.
The axes of this ellipse indicate the principal axes of
$\brkt{\tens{I}_{\tens{Q}}^{L L}}_{\mathrm{ex}}$ and and the diameter $2 |
I_{\pm} / t |$ gives the principal SQM currents. The arrows of the semi axes
indicate the sign of the principal SQM current: it is positive (negative) if
the arrow points away from (towards) the origin. The full tensor can be
visualized by including the principal vector $I^{\mathrm{ex}}_0
\op{\vec{v}}^{\mathrm{ex}}_0 / t \propto \vec{e}_y$ (pointing into the plane),
completing the ellipse to an ellipsoid. $\op{\vec{J}}^R$ is rotated from
$\op{\vec{J}}^L$ $( \theta = 0)$ out of the plane when $\theta$ increases. The
residual parameters are $D = 25 T$, $J = 5 T$ , $V = T = t$, \ $\Gamma = 2 \pi
/ 2500$.}
\end{figure}}

A further striking property of the exchange principal axes is that they can
not only be tuned magnetically by the Stoner vectors but also to a large
extent {\emph{electrically}}. As discussed in {\Sec{sec:scalarDep}} there is a
crossover, shown in {\Fig{fig:VDep}}, from a torque-dominated (low bias) to an
emission-dominated exchange SQM (large bias). As a result, increasing the
voltage results in a change of the principal axes of the SQM current - an
effect that is comparable to that resulting from tuning the angle $\theta$
between the magnetizations in {\Fig{fig:angleDep}}.

{
  \begin{figure}[h]
    \includegraphics[width=0.9\linewidth]{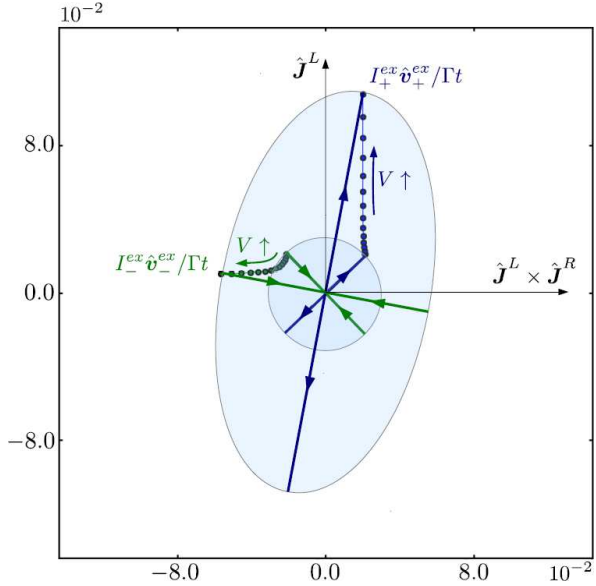}
    \caption{\label{fig:EVVDep}Voltage
dependence of the exchange SQM tensor. For explanation see
{\Fig{fig:angleDep}} where each points corresponds to a a different voltage $V
/ t$, increasing in steps of $1 / 10$ from $0$ to $1.5$. The remaining
parameters are $\theta = \pi / 4$, $D = 25 T$, $J^L = J^R = 5 t$ , $T = t$,
$\Gamma = 2 \pi / 2500$.}
\end{figure}}

The voltage scale at which the direct and exchange SQM current compete, i.e.,
$| E^L_{\mathrm{ex}} | \sim | T^L_{\mathrm{dir}} | = 2| \vec{S}^L |
T_{\vec{S}}^L$, can be estimated{\footnote{Assuming $N_s \gtrsim 10$, we
neglect the exchange SQM torque since $T_{\vec{S}}^L \sim T_{\mathrm{ex}}^L$
(cf. Eqs. {\eq{eq:estimateSpinTorque}} and {\eq{eq:torqueScale}}).}} using $|
\vec{S}^L | \sim J N_o / D \sim N_s$ (cf. {\Eq{eq:Ns}}) to
\begin{eqnarray}
  V & \sim & ( J^2 / D) N_s .  \label{eq:Vcross}
\end{eqnarray}
Since by {\Eq{eq:emissionScale}} $E^L_{\mathrm{ex}}$ has a voltage dependence
only for $V \lesssim J$, the electric tuneability is feasible only when the
crossover scale {\eq{eq:Vcross}} \ $\lesssim$ $J$: this is the case when the
number of polarized spin $N_s \lesssim D / J$, i. e. the number of orbitals is
limited to $N_o \lesssim ( D \text{/} J)^2$, which can still be fairly large
number. This is a first indication that for mesoscopic ferromagnets this
electrical tuneability of the SQM orientation may be possible and may be an
interesting topic for future analysis. We emphasize the crudeness of our model
here and the importance of investigating charging and non-equilibrium effects,
see Ref. {\cite{Hell13c}}.

\subsection{Transport of spin anisotropy without spin
current}\label{sec:spinFree}

Finally, we explore the possibility of a \emph{pure spin-anisotropy
current}, i. e. a non-vanishing SQM current in the absence of \ a spin
current, which was anticipated in Sec. {\sec{sec:SQMCurrent}}.

\subsubsection{Conditions for zero charge and spin current} We first discuss
the conditions for a vanishing spin current for a general band structure
$\varepsilon^r_{n k \sigma}$. We expect zero spin current only for
{\emph{collinear}} magnetizations. If the magnetizations are non-collinear,
the spin current has three non-collinear contributions and demanding that all
these vanish requires $E_{\vec{S}}^L = A_{\vec{S}}^L = T_{\vec{S}}^L = 0$.
This might be possible, but only for special band structures and parameter
values ($T^r, \mu^r$), but this is beyond the scope of this paper. For
collinear magnetizations, the spin torque automatically vanishes and the spin
current reads $\brkt{\vec{I}_{\vec{S}}} = \int \mathd \omega \Gamma \Delta
\left( n^L \op{\vec{J}}^L + n^R \op{\vec{J}}^R \right)$ with
$\widehat{\vec{J}}^L | | \widehat{\vec{J}}^R$. A generic situation with
cancelling spin current is then given for antiparallel magnetized
($\widehat{\vec{J}}^L = - \widehat{\vec{J}}^R$) ferromagnets with identical
spin-polarization of the 1DOS in the bias window, that is, $n^L ( \omega) =
n^R ( \omega)$, so that the bracket in the above integrand is zero. This
defines a parameter \emph{regime} for which $\brkt{\vec{I}_{\vec{S}}^{}} =
0$, as one may still apply any voltage or temperature bias. Again, there might
be exotic material combinations, for which the spin current even vanishes for
$\widehat{\vec{J}}^L = + \widehat{\vec{J}}^R$. For our crude flat band
approximation (cf. Sect. {\sec{sec:wide-flat-band}}), the dissipative spin
current is zero in any case, so that collinearity $\widehat{\vec{J}} =
\widehat{\vec{J}}^L = \pm \widehat{\vec{J}}^R$ of the Stoner vectors is
already sufficient for cancelling spin current.

We can even go one step further and envisage a situation, for which the charge
current vanishes as well: note that it still has a non-magnetic contribution
{\eq{eq:nonmcharge}} $\propto \int \mathd \omega \Delta$ ($\Gamma$ is constant
in the bias window and we consider $T^r \ll D$). For pure voltage bias $\mu^L
\neq \mu^R$, but $T^L = T^R$, the bias function $\Delta ( \omega)$ is
symmetric and positive and we will always have a charge current. However, for
a pure {\emph{temperature}} bias, $T^L \neq T^R$ and $\mu^L = \mu^R$, the bias
function $\Delta = f^R_+ - f^L_+$ is {\emph{anti}}symmetric and the charge
current contributions above and below the common electrochemical potential
cancel. Since $T^R > T^L$ the right ``hot'' electrode has a larger (smaller)
occupation probability for electrons with energy $\omega > \mu$ ($\omega <
\mu$) than the left electrode. Consequently, the particle current flowing from
left to right for electrons with energy $\omega > \mu$ exactly cancels the
charge current for electrons flowing from the right to the left at energies
$\omega < \mu$. Integrated over all frequencies this gives the zero net charge
current.

\subsubsection{Pure quadrupole current} Strikingly, in contrast to charge and
spin current, the SQM current remains {\emph{non-zero}} for a pure thermal
bias. It comes entirely from the exchange emission part:
\begin{eqnarray}
  \brkt{\tens{I}^{L L}_{\tens{Q}}} & = & - 2 \Gamma \int \mathd \omega \Delta
  a^L  \widehat{\vec{J}} \odot \widehat{\vec{J}} \neq 0,  \label{eq:IQtbias}
\end{eqnarray}
where $\Gamma$ can be pulled out of the integral since the DOS is constant at
energies $\omega$ for which $a^L ( \omega) \neq 0$. Since the spin-anisotropy
function $a^L ( \omega)$, {\Eq{eq:tripletCorr}}, is antisymmetric as well we
integrate an overall symmetric function and $\brkt{\tens{I}^{L L}_{\tens{Q}}}$
is non-zero. This is a central result of the paper. In {\Fig{fig:pureSQMTDep}}
we plot the total SQM current for a thermal bias with collinear Stoner vectors
as function of the temperature difference, as given by {\Eq{eq:IQtbias}}. In
{\Fig{fig:pureSQMJDep}} we plot the dependence on the Stoner field $J^L$.

{
  \begin{figure}[h]
    \includegraphics[width=0.9\linewidth]{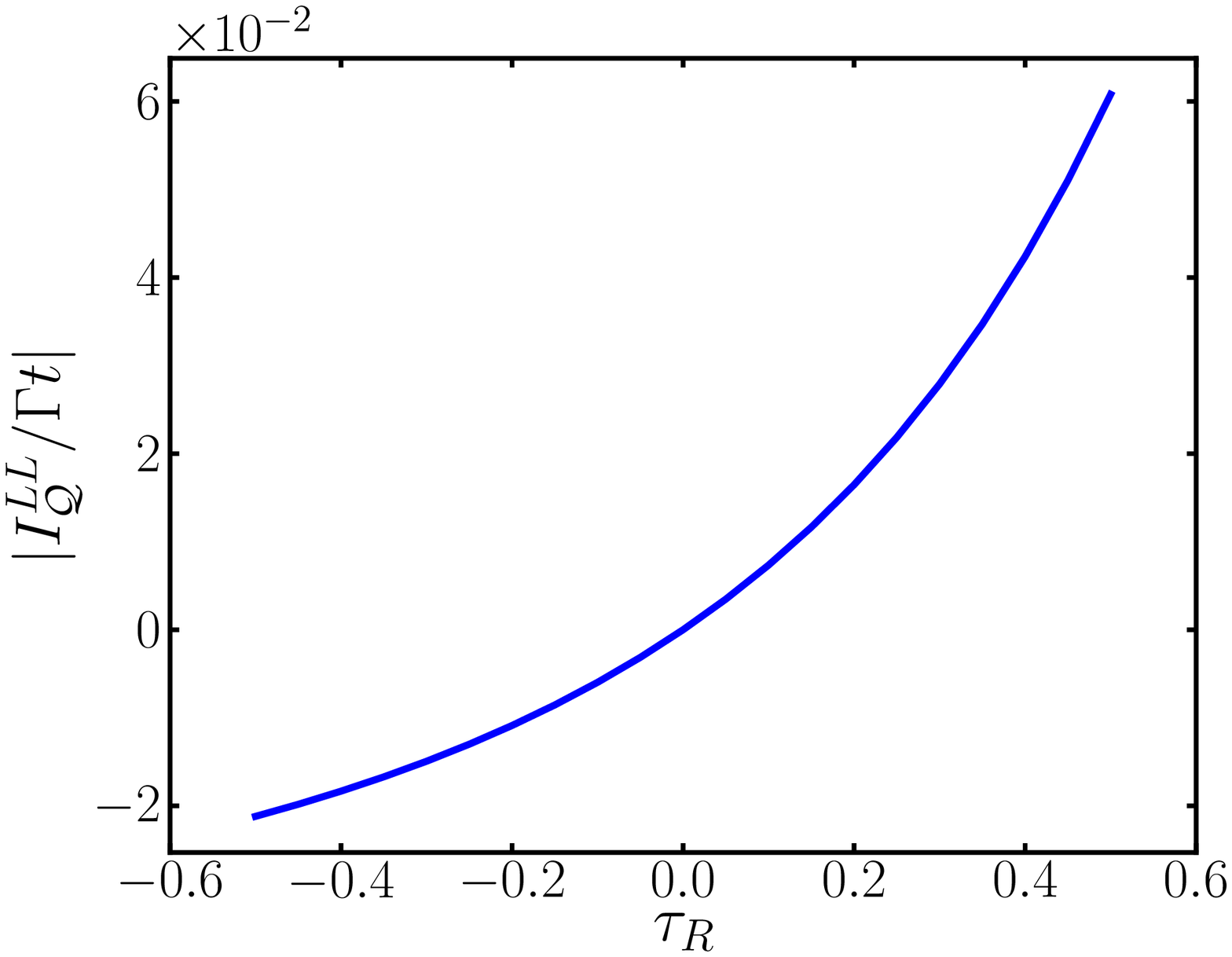}
    \caption{\label{fig:pureSQMTDep}$\left|
I_{\tens{Q}^{}}^{L L} \right| / t$ (from {\Eq{eq:IQtbias}} with
$\brkt{\tens{I}^{L L}_{\tens{Q}}} = - I_{\tens{Q}^{}}^{L L} \widehat{\vec{J}}
\odot \widehat{\vec{J}}$ as a function of the temperature bias ratio $\tau^R =
( T^R - T^L) / T^R$ for $T^L = t$, $J = 5 t$, $D = 25 t$ and $\Gamma = 2 \pi /
2500$.}
\end{figure}}

{
  \begin{figure}[h]
    \includegraphics[width=0.9\linewidth]{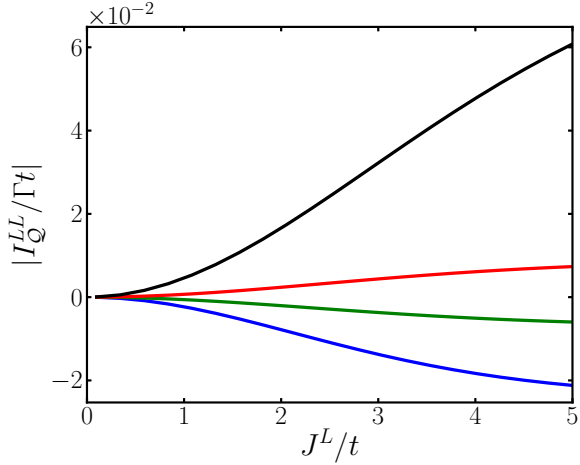}
    \caption{\label{fig:pureSQMJDep}Same as
{\Fig{fig:pureSQMTDep}}, but now showing $\left| I_{\tens{Q}^{}}^{L L} \right|
/ t$ as function of Stoner splitting $J^L / t$ for fixed thermal bias $\tau^R
= ( T^R - T^L) / T^R = - 0.5, - 0.1, 0.1 , 0.5$ (from topmost to
bottommost curve). The antisymmetry of the linear result
{\Eq{eq:pureSQMlinear}}, $I_{\tens{Q}}^{L L} ( \tau_R) \approx I_{\tens{Q}}^{L
L} ( - \tau_R)$, breaks down in the non-linear regime as shown in \ as
{\Fig{fig:pureSQMJDep}} for $\tau_R = \pm 0.5$.}
\end{figure}}

The linear response{\footnote{{\Eq{eq:pureSQMlinear}} is obtained by
substituting $x = ( \omega - \mu^L) / T^L$ in {\Eq{eq:IQtbias}} and expanding
the bias function $\Delta ( \omega) \approx - f' ( x) x \tau^R$ with $f ( x) =
( e^x + 1)^{- 1}$. {{This gives \ $I_{\tens{Q}}^{L L} \approx
\frac{\Gamma}{2} T^L \tau^R \int \mathd x f' ( x) x \left[ 2 f ( x) -
\sum_{\sigma} f ( x + \sigma j) \right]$ to which we apply {\Eq{eq:intFF}}
from {\App{app:closedFormula}}.}}}} in the temperature bias ratio $\tau^R = (
T^R - T^L) / T^R \ll 1$ varied for fixed $T^L$ gives for the SQM current
magnitude, defined here by $\brkt{\tens{I}^{L L}_{\tens{Q}}} = - I^{L
L}_{\tens{Q}} \widehat{\vec{J}} \odot \widehat{\vec{J}},$
\begin{eqnarray}
  I_{\tens{Q}}^{L L} & = & \frac{\Gamma}{2}  ( T^L - T^R) \times \nonumber\\
  &  & \frac{T^L}{T^R} \left[ 1 - \left( \frac{J^L / 2 T^L}{\sinh ( J^L / 2
  T^L)} \right)^2 \right] .  \label{eq:pureSQMlinear}
\end{eqnarray}
For fixed, different temperatures, the magnitude of the SQM current increases
monotonously as a function of the Stoner splitting as shown in
{\Fig{fig:pureSQMJDep}}. It eventually saturates for $J^L \approx 10 T^L$ at
the asymptotic value of $I_{\tens{Q}}^{L L} \approx ( \Gamma \text{/2}) \tau^R
T^L$.

A crude understanding of the above results is the following. Since the
magnitude of local exchange SQM decreases with temperature (Pauli exclusion
effects get washed out thermally), cf. {\Fig{fig:sqmdensint}} and
{\Eq{eq:qrex}}, the thermal gradient induces a ``gradient in the
correlations'' {{resulting in the SQM flow of Pauli exlusion holes from
the colder to the hotter reservoir, roughly speaking.}}

{{We emphasize, however, that this should not be interpreted as a direct
transfer of spin-correlations between the two local SQM nodes since they first
have to be converted into non-local spin-correlations: in the language of our
network picture, these are first buffered in the non-local intermediate
node.}} This becomes clearer in view of the SQM conservation law
{\eq{eq:SQMCurrentConservation}}, which reads of our device (cf.
{\Fig{fig:SQMNetworkLinks}}) after averaging $\brkt{\tens{I}^{L R}_{\tens{Q}}}
= - \brkt{\tens{I}^{L L}_{\tens{Q}}} - \brkt{\tens{I}^{R R}_{\tens{Q}}}$.
Interchanging the role of the left and the right electrode in
{\Eq{eq:pureSQMlinear}}, we see that the change in the local spin-anisotropy
of the $\langle L L \rangle$- and $\langle R R \rangle$-node have opposite
sign. Taking only the $O ( \Delta T)$- contribution, we may replace $T^L$ and
$T^R$, respectively, by the average temperature in the second line of
{\Eq{eq:pureSQMlinear}}: we then find that there is no {\emph{net}} creation
of non-local spin-correlations only to first order in in the thermal bias
$\Delta T$, i. e., $\brkt{\tens{I}^{L R}_{\tens{Q}}} = O ( \Delta T^2)$.

A more rigorous explanation of the thermally driven SQM current is based on a
microscopic point of view (cf. Secs. {\sec{sec:microscopic}} and
{\sec{sec:energyResolved}}). These considerations may be useful for proposals
for more complicated device setups that would allow for the detection a pure
SQM current (an issue that is not covered here). The exchange SQM in
{\eq{eq:IQtbias}} is quantified by the anisotropy function $a^L ( \omega)$
(cf. {\Eq{eq:tripletCorr}} and {\Fig{fig:anisotropyocc}}), which describes the
Pauli exclusion hole to which an electron at energy $\omega$ contributes. The
microscopic reason why $a^L ( \omega)$ changes sign was explained in detail in
{\Sec{sec:energyResolved}}: basically for $\omega < \mu$ ($\omega > \mu$) a
given electron at energy $\omega$ most likely sees a parallel (antiparallel)
spin at energy $\omega + J$ ($\omega - J$). Electrons with opposite energies
relative to $\mu$ thus contribute with an opposite sign to the Pauli exclusion
hole. Since the thermal bias transports electrons above and below the Fermi
edge into opposite directions, the contributions to the local average SQM
$\brkt{\tens{Q}^{L L}}$ thus {\emph{add up}}, explaining why {\Eq{eq:IQtbias}}
is finite. Notably, the thermal bias drives this flow of spin correlations
between the ferromagnets without any other one-particle quantity being net
transported. For example, the charge of each electron is independent of its
energy and therefore the contributions above and below the Fermi energy
cancel.

Importantly, in this case the direction of the spin-anisotropy flow can be
{\emph{controlled}} by the sign of the temperature gradient: for $T^L < T^R$,
the SQM current magnitude $I^{L L}_{\tens{Q}}$ is negative, i. e., local
{{planar}} spin-triplet correlations are net delocalized by the tunneling.
{{The left local node therefore {\emph{loses}} Pauli exclusion holes.}}
This can be understood following arguments similar to that of the purely
voltage-biased tunnel junction (see Sect. {\sec{sec:direction}}). However, in
stark contrast to pure voltage bias, the current magnitude $I^{L
L}_{\tens{Q}}$ becomes positive if $T^R < T^L$. This physically means that net
local {{planar}} spin-correlations are {\emph{created}} by tunneling. The
reason is that electrons are injected into the left electrode {\emph{below}}
the Fermi energy. As these electron obey Pauli's principle, they are
{\emph{forced}} to form new Pauli-exclusion holes. Furthermore, the electrons
are extracted only {\emph{above}} the Fermi energy and they carry away
{{positive (axial)}} spin correlations (leaving a negative contribution
behind). For the contrasting situation of a pure voltage bias, the
energy-resolved flow direction has to be opposite: electrons care only net
injected (extracted) at energies larger (lower) then the Fermi energy.

As a conclusion, the possibility to control the spin-anisotropy flow direction
by the thermal bias applied to the tunnel junction is a non-trivial result of
this paper. This fact and the prediction of a pure spin-quadrupole moment
current demonstrate most clearly that triplet-spin correlations form an
independent degree of freedom which is not only stored in a system of
ferromagnets, but can also be transported between them.

\section{Summary and Outlook}

In this paper we investigated fundamental questions about the spin anisotropy,
as quantified by the spin-quadrupole moment (SQM), which arise when it is
\emph{considered as a transport quantity}. In the physical language of
atomic and molecular magnetism, the SQM characterizes the quadratic spin
anisotropy, which is usually its dominant part. It quantifies the preference
of spins to be \emph{aligned} along a specific {\emph{axis}} irrespective
of their {\emph{orientation}} along it (up, down). {{We addressed three
central questions related to the quantum\emph{ transport }of
spin-anisotropy: (i) how can SQM be stored in and (ii) transported between two
ferromagnets in a spintronic circuit; (iii) how can one define an
\emph{SQM current tensor operator}, derive SQM continuity equations and
SQM current conservation laws; how does the non-equilibrium steady-state
average of the SQM current relate to the spin-current?}}

Our work was motivated by studies
{\cite{Sothmann10,Baumgaertel11,Misiorny13}} that indicated that the physical
picture of the transport of \emph{spin degrees of freedom} through
magnetic nanostructures needs to be extended. A refinement of this picture,
resulting from this paper, is as follows. Electrons are charged particles with
an intrinsic spin-dipole moment and vanishing higher spin moments. Therefore,
the motion of an {\emph{isolated}} electron is associated with a charge and
spin current only. However, in a multi-electron system the electron becomes
correlated with other electrons. Moving this electron therefore implies a
change of correlations. In particular, the transfer of spin-triplet
correlations between different subsystems is quantified by the spin-quadrupole
current. This complements the results of prior studies
{\cite{Sothmann10,Baumgaertel11,Misiorny13}}, which demonstrated that these
tensor-valued currents lead to an accumulation of SQM. The latter couples to
the accumulation of spin and charge and their measurable currents. {{In
this paper we ignored the complications of this accumulation, as well as
interaction and non-equilibrium effects that appear in nanoscale spintronic
devices. We exclusively focused on the description of transort of SQM between
macro- and mesoscopic circuit elements.}}

{{Answering question (i), we found that macroscopic ferromagnets, the
basic elements of spintronic devices, store a macroscopic SQM, which is
generated by their internal Stoner field.}} This {\emph{direct}}
spin-anisotropy is of easy-axis type and scales {\emph{quadratically}} with
the number of half-filled, spin-polarized orbitals $N_s$. This follows the
classical intuition that orientation of spins also implies their alignment.
{{However, for mesoscopic systems, an additional {\emph{quantum exchange}}
contribution to the SQM becomes relevant {\cite{Misiorny13}}, which scales
linearly with $N_s$.}} It quantifies the effect of Pauli-exclusion holes that
exist in the triplet \emph{two-particle} spin correlations, expressing the
simple fact that electrons in the same orbital do not form a triplet spin
state. This Pauli-forbidden spin-anisotropy is of the easy-plane type,
countering therefore the direct easy-axis anisotropy.

Importantly, the effect of the Pauli exclusion holes is cumulative, i. e.,
their contributions always add up and cannot cancel each other. This is in
stark contrast to the average spin-dipole moment, for which contributions from
electrons with opposite spin orientation can cancel each other. This shows
that the average SQM is a degree of freedom {\emph{independent}} of
one-particle quantities such as average charge and spin.

To answer question (ii) we developed a spin-multipole transport theory with
an associated network picture. For spin-dipole moment, each ferromagnet is
represented by a node of the network storing spin-dipole moment. However, due
to its two-particle nature in electronic systems, SQM is also stored as
non-local correlations between spins from different spin-polarized subsystems.
The network picture of SQM therefore incorporates also {\emph{non-local SQM
nodes}}. {{As a consequence, the \ SQM network differs from the physical
layout of the system of ferromagnets, both in the number of nodes and in their
connectivity.}} For the \emph{two}-terminal spin-valve that we studied in
this paper, this network thus consists of \emph{three} SQM nodes. This
network theory also applies also to spin-valves with embedded quantum dots
{\cite{Hell13c}}.

Based on this microscopic picture, we \ inferred the proper definition of the
{\emph{spin-quadrupole current tensor operators}}, answering question (iii).
{{By a continuity equation, the SQM currents generate the change of the
local anisotropy due to quantum transport processes.}} They furthermore obey a
current conservation law expressing the conservation of SQM in the tunneling.
For the two-terminal spin-valve it reads $\mathcal{I}^{L L}_{\mathcal{Q}} +
\mathcal{I}^{R R}_{\mathcal{Q}} + \mathcal{I}^{L R}_{\mathcal{Q}} = 0$.

Finally, we found by explicit calculation that the non-equilibrium
steady-state average of \emph{all} these these SQM currents is non-zero,
even for this elementary spintronic setup and analysed these in detail.
{{Similar to the average SQM, these average SQM currents have a
decomposition into classical and quantum \emph{two-particle}
contributions, similar to the average SQM itself.}} The \emph{direct SQM
current} is implied by a non-zero average spin and \emph{spin current}. It
reflects the classical intuition that ``orientation implies alignment''. In
addition to this, we found an quantum \emph{exchange SQM current}, which
is profoundly different from spin currents.

In analogy to the spin, we also distinguished dissipative and coherent
contributions to the SQM: the spin precession responsible for the spin-torque
term in the spin-current -- lifting the spin out of the plane of the Stoner
vectors -- has a counterpart in the SQM current. These spin-torque SQM terms
similarly result from coherent fluctuations by virtual tunneling into a
ferromagnet (i. e., spin-dependent scattering) which probe the spin-dependence
of the {{entire band structure}}. This effect is also responsible for the
exchange field {\cite{Koenig03}} in quantum dot spin-valves. The different
bias-voltage dependence of the dissipative and coherent terms allows for
electric control of both the magnitude and the \emph{orientation} of the
spin-anisotropy current \emph{tensor}.

{{Furthermore, for non-collinear ferromagnets, the spin-anisotropy current
was found to be a {\emph{bi-axial}} tensor. Its three distinct principal
values and axes reflect the lowered symmetry of a non-collinear setup, which
is not revealed by the spin current, which is just a {\emph{vector}}.}}
{{We showed that this dependence on the Stoner vectors allows for
substantial magnetic tuning of the SQM current tensor orientation.}} The
possibility of \emph{injecting biaxial anisotropy} into, e.g., molecular
magnets is of interest since the intrinsic, spin-orbit generated anisotropy of
this type is associated with quantum-spin tunneling effects
{\cite{Gatteschi03rev}}.

{{The striking central result of this paper, as announced by the paper
title, is a pure SQM current whenever the spin current vanishes by net
cancellation of one-particle contributions. This spin-anisotropy flow is \
driven by a gradient of Pauli exclusion holes in the triplet spin-spin
correlations, that is, a true quantum two-particle current.}} We illustrated
this general result for a temperature-biased junction connecting two
anti-parallel ferromagnets with a flat-band DOS. In this case, a pure SQM
current generates an \emph{uniaxial, easy plane} anisotropy, i.e., a
negative anisotropy that counteract an easy axis anisotropy. It may be of
interest to inject such an SQM current into a single-molecule magnet
considered as a memory cell in order to temporarily switch off its easy-axis
anisotropy barrier in order to put it into ``writing'' mode. This also relates
to the recently studied tunnel-induced renormalization of the intrinsic
anisotropy of molecular magnets in contact with spin-polarized electrodes
{\cite{Misiorny11a,May11,Misiorny13}}. One may even envisage the utilization
of SQM as a resource, as an alternative to conventional spintronics, i.e.,
utilize the storage, transport, manipulation and readout of spin anisotropy
without transporting or affecting spin polarization. The possibility of pure
SQM currents pointed out in this paper indicates that this is principle
conceivable and warrants further study. Altogether the above indicates that
the theory of a generalized ``spin-multipoletronics'', is a real possibility,
if not a necessity when spintronics moves to the nanoscale.

{{We acknowledge G. E. W. Bauer, M. B{\"u}ttiker, D. P. DiVincenzo, J.
K{\"o}nig, M. Misiorny, R. Saptsov and J. Splettst{\"o}sser for useful
discussions.}}

\appendix

\section{Storage of Spin-quadrupole Moment}\label{sec:appMultipole}

In this appendix we give the calculation of the local average SQM stored in a
ferromagnet, cf. {\Sec{sec:sqmAvEx}}. We present three derivations, each of
which unveils different physical and technical aspects used in the main part.
The first, most straightforward approach is given in Sec. {\sec{sec:Pauli}}.
It shows how the Pauli exclusion hole arises in {\Eq{eq:Qstandardav2}}, which
provides the key to the physical interpretation of exchange SQM. Secondly, we
give a technically more sophisticated derivation in Sec.
{\sec{sec:spinTraceTechnique}}, which will be helpful to understand all steps
of our transport calculations. It expresses the Pauli exclusion hole in a
coordinate-free form in {\Eq{eq:Qrrfinal}}. Thirdly, we present in Sec.
{\sec{app:tripletCorr}} a derivation that makes explicit the spin-triplet
content of the correlations by vector coupling of the spins of electron pairs.
Finally, we discuss the spin-anisotropy function and derive a closed
expression for the exchange SQM in the flat band approximation (cf.
{\sec{sec:wide-flat-band}}). Throughout the appendix we focus on understanding
the exchange contributions to the SQM, which we showed in the absence of
tunneling to appear only locally for $( r = r')$, cf. {\sec{sec:sqmAvEx}}. We
therefore only consider one electrode $r$ with one band $n$ and subsequently
drop these indices in all expressions below, e. g. $c_{r n k \sigma}
\rightarrow c_{k \sigma}$, when convenient.

\subsection{Exchange SQM and the Pauli Principle}\label{sec:Pauli}

We furthermore take a coordinate system for which $\vec{e}_z = \op{\vec{J}}$
and quantize the spin along this vector. The calculation of the average local
SQM starts from the operator {\Eq{eq:Qalphaalphaprime}} in the main text.
{{We insert the second-quantized form {\eq{eq:chargespinOp}} of the spin
operator and anticommute $c^{\dag}_{k_1 \sigma_1'}$ twice to the right:}}
\begin{eqnarray}
  \tens{Q}^{} & = & \sum_{\{ k_i \sigma_i' \sigma_i \}}  \vec{s}_{\sigma_2'
  \sigma_2} \odot \vec{s}_{\sigma_1' \sigma_1} c^{\dag}_{k_2 \sigma_2'} c_{k_2
  \sigma_2} c^{\dag}_{k_1 \sigma_1'} c_{k_1 \sigma_1} \\
  & = & \sum_{\{ k_i \sigma_i' \sigma_i \}}  \vec{s}_{\sigma_2' \sigma_2}
  \odot \vec{s}_{\sigma_1' \sigma_1} c^{\dag}_{k_1 \sigma_1'} c^{\dag}_{k_2
  \sigma_2'} c_{k_2 \sigma_2} c_{k_1 \sigma_1}  \label{eq:Qstandard}
\end{eqnarray}
This generates a term $\delta_{k_2 k_1} \delta_{\sigma_2 \sigma'_1}
c^{\dag}_{k_2 \sigma_2'} c_{k_1 \sigma_1}$, which we omitted because it
vanishes after performing the sum over the spin indices by virtue of $\vec{s}
\odot \vec{s} = 0$: for all $\sigma_2', \sigma_1$
\begin{eqnarray}
  \sum_{\sigma_2 \sigma_1'} \vec{s}_{\sigma_2' \sigma_2} \odot
  \vec{s}_{\sigma_1' \sigma_1} \delta_{\sigma_2 \sigma_1'} & = &
  \bra{\sigma_2'} \vec{s} \odot \vec{s} \ket{\sigma_1} = 0. 
\end{eqnarray}
Computing the average of {\Eq{eq:Qstandard}} using Wick's theorem $\langle
c^{\dag}_{k' \sigma'} c_{k \sigma} \rangle = \delta_{k k'} \delta_{\sigma
\sigma'} f_+ ( \varepsilon_{k \sigma})$ in the standard way with $f_+ (
\omega)$ denoting the Fermi function,
\begin{eqnarray}
  &  & \brkt{c^{\dag}_{k_1 \sigma_1'} c^{\dag}_{k_2 \sigma_2'} c_{k_2
  \sigma_2} c_{k_1 \sigma_1}}  \label{eq:appWick}\\
  & = & \brkt{c^{\dag}_{k_1 \sigma_1'} c_{k_1 \sigma_1}} \brkt{c^{\dag}_{k_2
  \sigma_2'} c_{k_2 \sigma_2}} - \brkt{c^{\dag}_{k_1 \sigma_1'} c_{k_2
  \sigma_2}} \brkt{c^{\dag}_{k_2 \sigma_2'} c_{k_1 \sigma_1}} \nonumber
\end{eqnarray}
yields a direct and an exchange part:
\begin{eqnarray}
  \brkt{\tens{Q}} & = &  \sum_{k_2 k_1 \sigma_2 \sigma_1} f_+ (
  \varepsilon_{k_2 \sigma_2}) f_+ ( \varepsilon_{k_1 \sigma_1}) \times
  \nonumber\\
  &  & \left( \vec{s}_{\sigma_1 \sigma_2} \odot \vec{s}_{\sigma_2 \sigma_1} -
  \delta_{k_2 k_1} \vec{s}_{\sigma_2 \sigma_2} \odot \vec{s}_{\sigma_1
  \sigma_1} \right)  \label{eq:Qstandardav}\\
  & = & \tfrac{1}{4}  \sum_{k_2 k_1} ( 1 - \delta_{k_2 k_1}) \nonumber\\
  &  & \sum_{\sigma_2 \sigma_1} \sigma_1 \sigma_2 f_+ ( \varepsilon_{k_2
  \sigma_2}) f_+ ( \varepsilon_{k_1 \sigma_1}) \vec{e}_z \odot \vec{e}_z . 
  \label{eq:Qstandardav2}
\end{eqnarray}
Here we used the result
\begin{equation}
  \begin{array}{lllll}
    \vec{s}_{\sigma_1 \sigma_1} \odot \vec{s}_{\sigma_2 \sigma_2} & = &
    \vec{s}_{\sigma_1 \sigma_2} \odot \vec{s}_{\sigma_2 \sigma_1} & = &
    \tfrac{\sigma_1 \sigma_2}{4}  \vec{e}_z \odot \vec{e}_z
  \end{array} .
\end{equation}
Clearly, $\vec{s}_{\sigma_1 \sigma_1} \odot \vec{s}_{\sigma_2 \sigma_2} =
\sigma_1 \sigma_2  \tfrac{1}{4}  \vec{e}_z \odot \vec{e}_z$, whereas for
$\sigma_1 = - \sigma_2$ we have $\vec{s}_{\sigma_1 \sigma_2} \odot
\vec{s}_{\sigma_2 \sigma_1} = \tfrac{1}{2} \left( \vec{e}_x + i \sigma_1 
\vec{e}_y \right) \odot \tfrac{1}{2} \left( \vec{e}_x - i \sigma_1  \vec{e}_y
\right) = \tfrac{1}{4} \left( \vec{e}_x \odot \vec{e}_x + \vec{e}_y \odot
\vec{e}_y \right) = - \tfrac{1}{4} \vec{e}_z \odot \vec{e}_z$. The last step
follows from $\sum_i \vec{e}_i \odot \vec{e}_i = 0$ which is just the
traceless, symmetric part of the unit tensor by the coordinate-space
completeness relation $\sum_i \vec{e}_i \vec{e}_i = \tens{I}$.

With the two-particle operator {\eq{eq:Qstandard}} in the standard
second-quantized form (see also {\Eq{eq:secondquantize}} below) the
contributions to the average $\brkt{\tens{Q}}$ can be discussed as scattering
processes, treating $\tens{Q}$ as if it were an interaction (although it is
tensor-valued). In {\Fig{fig:Feynman}} we represent the contributions to the
average SQM {\eq{eq:Qstandardav}} by Feynman diagrams for scattering processes
between an initial pair of states $( 1, 2)$ to a final to pair of states $(
1', 2')$.
{
  \begin{figure}[h]
    \includegraphics[width=0.9\linewidth]{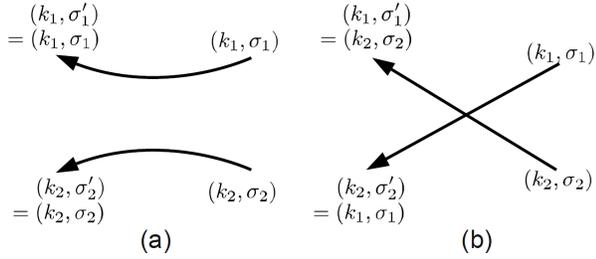}
    \caption{Feynman diagrams for calculating the
(a) direct and (b) the exchange contribution the the average
SQM.\label{fig:Feynman}}
\end{figure}}

The momenta in the final states are the same as for the initial states, $k_i =
k_{i'}$, since the SQM operator does not act on the orbital part of the wave
function (cf. {\Eq{eq:SQMmatrixElement}}). Two different scattering processes
are permitted: the first one is a direct scattering, for which the electron in
initial state $i$ ends up in state $i'$, restricting the spin indices to
$\sigma_i = \sigma_i'$ (while already $k_i = k_{i'}$). These direct scattering
contributions, multiplied with their tensor-valued amplitudes
$\vec{s}_{\sigma_1 \sigma_2} \odot \vec{s}_{\sigma_2 \sigma_1}$ in
{\eq{eq:Qstandardav}} add up to the direct SQM. This contribution to the spin
anisotropy is thus generated by two electrons (labelled by their states $1$
and 2) as if they were {\emph{distinguishable}}, i.e., by treating the
two-particle scattering classically.

The second type of scattering process, in which the particles are
{\emph{exchanged}}, accounts for the fact that electrons are
{\emph{in}}distinguishable. This is only possible if the momenta are the
{\emph{same}}, $k_1 = k_2$, and furthermore, the spins are exchanged,
$\sigma_1' = \sigma_2$ and $\sigma_2' = \sigma_1$. This exchange
contribution to the average SQM entirely cancels the direct contribution for
equal $k_1 = k_2$, correcting for the treatment of electrons as
distinguishable particles. In other words, the exchange SQM accounts for Pauli
``holes'' in the triplet spin-spin correlation tensor. The Pauli principle
thus counters the direct classical contribution to the spin anisotropy.

Indistinguishability becomes important if we consider pairs of electrons from
the same $k$-mode. Due to Pauli's principle, their wave function must have a
symmetric orbital part and an antisymmetric spin part, that is, they form a
spin singlet with zero SQM, i. e., triplet correlations are forbidden. This is
analogous to the direct and exchange contributions to the average Coulomb
interaction with respect to Slater determinants (e.g., in Hartree-Fock
theory). In that case, nearby electrons with parallel spins repel each other
due to the exchange potential. We mention that as expected from this analogy,
thermal fluctuations suppress the effect of the Pauli principle on SQM as
well, c.f. {\Eq{eq:qexApp}} below.

\subsection{Spin-Trace Technique}\label{sec:spinTraceTechnique}

We now reformulate the above calculation in a way that will be used in the
transport calculations (cf. Sec. {\sec{sec:energySpinSep}}): the result
{\eq{eq:Qrrfinal}} then assumes the coordinate-free form presented in Sec.
{\sec{sec:sqmAvEx}}.
\begin{eqnarray}
  \brkt{\tens{Q}^{r r}}_{\mathrm{ex}} & = & \sum_{k \sigma \sigma' \tau \tau'}
  \left( \vec{s}^r_{\sigma \sigma'} \odot \vec{s}^r_{\tau \tau'} \right)
  \delta_{\sigma \tau'} \delta_{\sigma' \tau} \nonumber\\
  &  & \times f_+^r ( \varepsilon^r_{k \sigma}) f^r_- ( \varepsilon^r_{k
  \sigma'}) .  \label{eq:appExContr}
\end{eqnarray}
Here we reintroduced the electrode index. To recast this into
{\emph{covariant}} expression, we first introduce the two-particle density of
states {\eq{eq:twoDOS}},
\begin{eqnarray}
  \nu^r_{\sigma \sigma'} ( \omega, \omega') & = & \sum_{n,  k} \delta
  ( \omega - \varepsilon^r_{k \sigma}) \delta ( \omega' - \varepsilon^r_{k
  \sigma'}),  \label{eq:nussprime}
\end{eqnarray}
and rewrite Eq. (\ref{eq:appExContr}) in terms of frequency integrals
\begin{eqnarray}
  &  & \brkt{\tens{Q}^{r r}}_{\mathrm{ex}} = \sum_{k \sigma \sigma' \tau \tau'}
  \int d \omega d \omega' \left( \vec{s}^r_{\sigma \sigma'} \odot
  \vec{s}^{r'}_{\tau \tau'} \right) \nonumber\\
  &  & \delta_{\sigma \tau'} \delta_{\sigma' \tau} \nu_{\sigma \sigma'} (
  \omega, \omega') f_+^r ( \omega) f^r_- ( \omega') .  \label{eq:appExContr2}
\end{eqnarray}
The 2DOS $\nu^r_{\sigma \sigma'} ( \omega)$ can be expressed as a matrix
element in spin space by
\begin{eqnarray}
  &  & \nu^r_{\sigma \sigma'} ( \omega, \omega') \delta_{\sigma \tau'}
  \delta_{\sigma' \tau} \nonumber\\
  & = & 2 \sum_{\mu_1 , \mu_2}  \text{}^r \langle \tau' |
  \check{r}_{\mu_1} | \sigma \rangle^r  \tens{A}^r_{\mu_1 \mu_2} ( \omega,
  \omega') ^r \langle \sigma' | \check{r}_{\mu_2} | \tau \rangle^r .
  \hspace{1em}  \label{eq:twoparticle}
\end{eqnarray}
Here we used the four component operator $\vec{\check{r}}$ with $\check{r}_0 =
\mathbbm{1} \text{/} \sqrt{2}$ and $\check{r}_i = \sqrt{2} s_i \text{}$ for $i
= x, y, z$. The four dimensional matrix $\tens{A}^r_{\mu_1 \mu_2}$
incorporates all relevant 2DOS information. It decomposes into a scalar, two
vectors and a tensor in coordinate space:
\begin{eqnarray}
  \tens{A}_{0 0}^r & = & \frac{1}{4} \overset{}{\sum_{\sigma \sigma'}}
  \nu^r_{\sigma \sigma'} ( \omega, \omega'),  \label{eq:A00}\\
  \tens{A}_{i 0}^r & = & \frac{1}{4} \overset{}{\sum_{\sigma \sigma'}} \sigma
  \nu^r_{\sigma \sigma'} ( \omega, \omega') \hat{J}^r_i,  \label{eq:Ai0}\\
  \tens{A}_{0 j}^r & = & \frac{1}{4} \overset{}{\sum_{\sigma \sigma'}} \sigma'
  \nu^r_{\sigma \sigma'} ( \omega, \omega') \hat{J}^r_j,  \label{eq:A0j}\\
  \tens{A}_{i j}^r & = & \frac{1}{4} \overset{}{\sum_{\sigma \sigma'}} \sigma
  \sigma' \nu^r_{\sigma \sigma'} ( \omega, \omega') \hat{J}^r_i \hat{J}^r_j . 
  \label{eq:Aij}
\end{eqnarray}
Inserting {\Eq{eq:twoparticle}} into {\Eq{eq:appExContr2}} and recasting the
sum over the spin indices as a trace in spin space yields
\begin{eqnarray}
  \brkt{\tens{Q}^{r r}}_{\mathrm{ex}} & = & \int d \omega d \omega' f^r_+ (
  \omega) f^r_- ( \omega') \nonumber\\
  &  & 2 \sum_{\mu_1  \mu_2} \tens{A}^r_{\mu_1 \mu_2} ( \omega,
  \omega')  \Tr \left[ \check{r}_{\mu_1} \vec{s} \odot \check{r}_{\mu_2}
  \vec{s} \right] \\
  & = & \int d \omega d \omega'  \overset{}{\tfrac{1}{4} \sum_{\sigma
  \sigma'}} \sigma \sigma' \nu^r_{\sigma \sigma'} ( \omega, \omega')
  \nonumber\\
  &  & \times f^r_+ ( \omega) f^r_- ( \omega')  \widehat{\vec{J}}^r \odot
  \widehat{\vec{J}}^r .  \label{eq:Qrrfinal}
\end{eqnarray}
This recovers the results {\eq{eq:Qtriplet}}-{\eq{eq:asigma}} obtained in the
main text from {\Eq{eq:Qstandardav2}}. The explicit calculation in the last
step is now reduced to using spin-1/2 operator algebra $s_i s_j = \tfrac{1}{4}
\delta_{i j} \mathbbm{1} + \tfrac{1}{2} \sum_k i \varepsilon_{i j k} s_k$ and
$\vec{s} \odot \vec{s} = 0$, i.e., without using matrix elements. These steps
are analogous to the evaluation of the diagrammatic expressions for the SQM
current in App. {\sec{app:FirstOrder}}.

\subsection{SQM and Triplet Spin Correlations}\label{app:tripletCorr}

Finally, we express the SQM tensor operator $\tens{Q}$ in the second-quantized
form. This allows one to perform vector-coupling of the pairs of involved
spin, thereby making explicit that only triplet correlations are ``measured''
by $\brkt{\tens{Q}}$, something that did not become clear in the above
calculations. This is merely important for the physical understanding, but
seems to bring no advantage for calculations. The many-body quadrupole
operator is a sum over quadrupole operators of pairs of particles, the latter
labeled by $a, b = 1, 2, 3, \ldots$
\begin{eqnarray}
  \tens{Q} & = & \sum_{a < b} \tens{Q}^{a b} 
\end{eqnarray}
with \ Cartesian components $i, j = x, y, z$:
\begin{eqnarray}
  Q^{a b}_{i j} & = & 2 \left( \tfrac{1}{2} ( s^a_i s^b_j + s^a_j s^b_i) -
  \tfrac{1}{3} \mathcal{\delta}_{i j} \sum_k s^a_k s_k^b \right) . 
  \label{eq:qab2}
\end{eqnarray}
Here we inserted the total spin operator $\vec{S} = \sum_a \vec{s}^a_{}$ into
{\Eq{eq:Qdef}} and used the result $\tens{Q}^{a a} = 0$ (`a `single electron
has no anisotropy'', cf. Sec. {\sec{sec:atomic}}). The SQM from pair $\langle
a b \rangle$ can also be expressed by introducing coupling the two spins to
$\vec{S}^{a b} = \vec{s}^a + \vec{s}^b$,
\begin{eqnarray}
  Q^{a b}_{i j} & = & \tfrac{1}{2} ( S^{a b}_i S^{a b}_j + S^{a b}_j S^{a
  b}_i) - \tfrac{1}{3} \delta_{i j} \left( \vec{S}^{a b} \right)^2, 
  \label{eq:qab1}
\end{eqnarray}
using $s^a_i s_j^a = \tfrac{1}{4} \delta_{i j} + i \tfrac{1}{2} \sum_k
\varepsilon_{i j k} s_k^a$. Note the factor 2 in {\Eq{eq:qab2}}. The general
second quantization prescription immediately gives
\begin{eqnarray}
  \tens{Q} & = & \sum_{\{ k_i \sigma_i \}} \tfrac{1}{2} \bra{k_2' \sigma_2'
  k_1' \sigma_1'} \mathcal{\tens{Q}}^{12} \ket{k_2 \sigma_2 k_1 \sigma_1}
  \nonumber\\
  &  & c^{\dag}_{k_1' \sigma_1'} c^{\dag}_{k_2' \sigma_2'} c_{k_2 \sigma_2}
  c_{k_1 \sigma_1} .  \label{eq:secondquantize}
\end{eqnarray}
We now make explicit use the particular property of the matrix elements of the
pair SQM $\tens{Q}^{12}$. First, we note that $\tens{Q}^{12}$ acts only on the
spin of the electrons,
\begin{eqnarray}
  &  & \bra{k_2' \sigma_2' k_1' \sigma_1'} \mathcal{\tens{Q}}^{12} \ket{k_2
  \sigma_2 k_1 \sigma_1} \nonumber\\
  & = & \delta_{k_2' k_2} \delta_{k_1' k_1} \bra{\sigma_2' \sigma_1'}
  \mathcal{\tens{Q}}^{12} \ket{\sigma_2 \sigma_1} . 
  \label{eq:SQMmatrixElement}
\end{eqnarray}
If we inserted this into {\Eq{eq:secondquantize}} we would recover
{\Eq{eq:Qstandard}}. Instead of this, we now introduce a singlet-triplet basis
for each pair of \ considered spins $\sigma_1$ and $\sigma_2$ above:
\begin{eqnarray}
  \ket{S} & = & \tfrac{1}{\sqrt{2}} \sum_{\sigma} \sigma \ket{\sigma
  \bar{\sigma}}, \\
  \ket{T 0} & = & \tfrac{1}{\sqrt{2}} \sum_{\sigma} \ket{\sigma \bar{\sigma}},
  \\
  \ket{T m} & = & \ket{m m}, \text{ \ \ \ \ } m = \pm 
\end{eqnarray}
where $\bar{\sigma} = - \sigma$. The crucial point is that $\tens{Q}^{12}$
only has matrix elements in the triplet sector. This follows from the fact
that $\tens{Q}^{12}$ is symmetric under exchange of the spins, i.e., $\left[
P, \tens{Q}^{12} \right] = 0$ where $P$ is the exchange operator. Therefore
$\tens{Q}^{12}$ is block-diagonal with respect to the eigen spaces of $P$,
which are here the singlet and triplet states satisfying $P \ket{S} = -
\ket{S}$ and $P \ket{T m} = + \ket{T m}$. Thus, $\langle S | \tens{Q}^{12} | T
m \rangle = \langle T m | \tens{Q}^{12} | S \rangle = 0$. Moreover, the
diagonal singlet block is zero, $\bra{S} \tens{Q}^{12} \ket{S} = 0$ by
{\Eq{eq:qab1}} with $a = 1, b = 2$ and $\vec{S}^{1 2} \ket{S} = 0$, completing
the proof. As a result
\begin{eqnarray}
  &  & \bra{\sigma_2' \sigma_1'} \mathcal{\tens{Q}}^{12} \ket{\sigma_2
  \sigma_1} = \nonumber\\
  &  & \sum_{m m'} \braket{\sigma_2' \sigma_1'}{T m} \bra{T m} \tens{Q}
  \mathcal{}^{12} \ket{T m'} \braket{T m'}{\sigma_2 \sigma_1} . \hspace{1em} 
  \label{eq:QwithTriplet}
\end{eqnarray}
Inserting {\Eq{eq:QwithTriplet}} into {\Eq{eq:secondquantize}}, we obtain the
central result of the appendix,
\begin{eqnarray}
  \tens{Q} & = & \sum_{m m'} \tfrac{1}{2} \bra{T m'} \mathcal{\tens{Q}}^{12}
  \ket{T m} \sum_{\{ k_i \}} E^{m' \dag}_{k_2 k_1} E^m_{k_2 k_1}, 
  \label{eq:QwithTripletOp}
\end{eqnarray}
with two-particle operators that explicitly generate \emph{only triplet
pairs}:
\begin{eqnarray}
  E^m_{k_2 k_1} & = & \sum_{\sigma_2 \sigma_1} \braket{T m}{\sigma_2 \sigma_1}
  c_{k_2 \sigma_2} c_{k_1 \sigma_1} \\
  & = & \left\{\begin{array}{ll}
    c_{k_2 m} c_{k_1 m} & m = \pm 1\\
    \frac{1}{\sqrt{2}} \sum_{\sigma} c_{k_2 \sigma} c_{k_1 \bar{\sigma}} & m =
    0
  \end{array}\right. . 
\end{eqnarray}
Considered as an interaction, $\tens{Q}$ thus only scatters triplet correlated
pairs of electrons. Due to the restrictions on the spins in these operators
$E^m_{k_2 k_1}$, the averages are
\begin{eqnarray}
  &  & \brkt{E^{m' \dag}_{k_2 k_1} E^m_{k_2 k_1}} = \delta_{m m'} \times
  \nonumber\\
  &  & \left\{\begin{array}{ll}
    f ( \varepsilon_{k_2 m}) f ( \varepsilon_{k_1 m}) ( 1 - \delta_{k_2 k_1}),
    & m = \pm 1\\
    \tfrac{1}{2} \sum_{\sigma} f ( \varepsilon_{k_2 \sigma}) f (
    \varepsilon_{k_1 \bar{\sigma}}) ( 1 - \delta_{k_2 k_1}), & m = 0
  \end{array}\right. 
\end{eqnarray}
with the tensor-valued matrix elements $\bra{T +} \mathcal{\tens{Q}}^{12}
\ket{T +} = \bra{T -} \mathcal{\tens{Q}}^{12} \ket{T -} = - \bra{T 0}
\mathcal{\tens{Q}}^{12} \ket{T 0} / 2 = \tfrac{1}{2}  \vec{e}_z \odot
\vec{e}_z$ given by Eqs. {\eq{eq:Qtripmatel1}}-{\eq{eq:Qtripmatel0}} in the
main text. These relations follow from the fact that $\tens{Q}^{12}$ is
traceless in the Hilbert space, $\sum_{m = 0, \pm 1} \bra{T m}
\mathcal{\tens{Q}}^{12} \ket{T m} = 0$ and that the $m = \pm$ states have the
identical spin-anisotropy. We recover {\Eq{eq:Qstandardav2}}:
\begin{eqnarray}
  &  & \brkt{\tens{Q}} = \tfrac{1}{4} \vec{e}_z \odot \vec{e}_z \sum_{k_2
  k_1} ( 1 - \delta_{k_2 k_1}) \nonumber\\
  &  & \times \left( \sum_{m = \pm} f_+ ( \varepsilon_{k_2 m}) f_+ (
  \varepsilon_{k_1 m}) \right. \nonumber\\
  &  & - \left. \sum_{\sigma = \pm} f_+ ( \varepsilon_{k_2 \sigma}) f_+ (
  \varepsilon_{k_1 \bar{\sigma}}) \right) 
\end{eqnarray}
This derivation, however, shows explicitly that the $m = \pm 1$ terms
contribute the same, uniaxial anisotropy tensor $\bra{T \pm}
\mathcal{\tens{Q}}^{12} \ket{T \pm}$, whereas the $m = 0$ term contributes an
easy-plane anisotropy tensor $\bra{T 0} \mathcal{\tens{Q}}^{12} \ket{T 0} = -
2 \bra{T +} \mathcal{\tens{Q}}^{12} \ket{T +}$. Moreover, the Pauli-exclusion
hole factor $1 - \delta_{k_2 k_1}$ is immediately explicit. Thus,
$\brkt{\tens{Q}}$ can be calculated by first accounting for triplet
correlations between spins of pairs of electrons in all possible orbitals,
including the same orbital,
\begin{eqnarray}
  \brkt{\tens{Q}}_{\mathrm{dir}} & = & \tfrac{1}{4}  \sum_{k_2 k_1 \sigma_2
  \sigma_1} \sigma_2 f ( \varepsilon_{k_2 \sigma_2}) \sigma_1 f (
  \varepsilon_{k_1 \sigma_1}) \vec{e}_z \odot \vec{e}_z \nonumber\\
  & = &  \brkt{\vec{S}} \odot \brkt{\vec{S}}, 
\end{eqnarray}
giving {\Eq{eq:redcontr}}, and then subsequently cancelling the latter
violation of the Pauli principle by the exchange term,
\begin{eqnarray}
  \brkt{\tens{Q}}_{\mathrm{ex}} & = & - q_{\mathrm{ex}} \vec{e}_z \odot \vec{e}_z 
\end{eqnarray}
with the positive exchange magnitude
\begin{eqnarray}
  q_{\mathrm{ex}} & = & \tfrac{1}{4} \sum_k ( f_+ ( \varepsilon_{k \uparrow}) -
  f_+ ( \varepsilon_{k \downarrow}))^2 .  \label{eq:qexApp}
\end{eqnarray}
We obtain {\Eq{eq:qexksum}} from the main text. Finally, we show that the
two-particle DOS $\nu_{\sigma \sigma'}$ can be decomposed explicitly into
{\emph{triplet}} DOS components: converting the sums in {\Eq{eq:qexApp}} to
integrals we obtain
\begin{eqnarray}
  q_{\mathrm{ex}} & = & \int d \omega d \omega' f_+ ( \omega) f_+ ( \omega')
  \sum_{\sigma \sigma'} \sigma \sigma' \nu^r_{\sigma \sigma'} ( \omega,
  \omega') .  \label{eq:qexss}
\end{eqnarray}
The only relevant combination of the 2DOS in the above expression can be
recast as
\begin{eqnarray}
  \sum_{\sigma \sigma'} \sigma \sigma' \nu^r_{\sigma \sigma'} & = & \nu^r_{T
  +} + \nu_{T -}^r - \sqrt{2} \nu^r_{T 0} 
\end{eqnarray}
with the triplet exchange 2DOS function ($m = \pm$)
\begin{eqnarray}
  \nu^r_{T m} ( \omega, \omega') & \assign & \nu^r_{m m} ( \omega, \omega')
  , \\
  \nu^r_{T 0} ( \omega, \omega') & \assign & \frac{1}{\sqrt{2}} \sum_{\sigma}
  \nu^r_{\sigma \bar{\sigma}} ( \omega, \omega') . 
\end{eqnarray}
This gives a precise decomposition into triplet spin correlations that
contribute to $\brkt{\tens{Q}^{r r}}_{\mathrm{ex}}$.

\subsection{Spin-anisotropy
Function}\label{app:closedFormula}\label{app:spinAnisotropy}

Finally, we further substantiate the physical interpretation of the anisotropy
function, which plays a key role in the main text. The basic idea of
``quadrupolarization'' of two triplet-correlated electrons is simply to
``count'' {{whether}} the spins are parallel ($\uparrow \uparrow$ or
$\downarrow \downarrow$, counted as $+$), or antiparallel ($\uparrow
\downarrow$ or $\downarrow \uparrow$, counted as $-$). In both cases their
individual orientations, i.e., their \emph{dipolarization} $\uparrow$ or
$\downarrow$, is ignored. {\Eq{eq:qexss}} precisely expresses this notion for
the exchange SQM. {{It is instructive to start from the $k$-sum
representation {\eq{eq:qexApp}} and to write it as}}
\begin{eqnarray}
  q_{\mathrm{ex}} & = & \sum_{k \sigma} f_+ ( \varepsilon_{k \sigma}) a_{k
  \sigma} . 
\end{eqnarray}
Here, \emph{given} that an electron with spin $\sigma$ occupies orbital
$k$, we ``count'' by
\begin{eqnarray}
  a_{k \sigma} & = & \sum_{\sigma'} \tfrac{\sigma \sigma'}{4} f (
  \varepsilon_{k \sigma'})  \label{eq:ak}
\end{eqnarray}
the average quadrupolarization contribution from electrons in that same
orbital $k$: parallel spin $\sigma' = \sigma$ gives $+ f_+ ( \varepsilon_{k
\sigma})$, antiparallel $\sigma' = \bar{\sigma}$ gives $- f ( \varepsilon_{k
\bar{\sigma}})$.{\footnote{Note that these two electrons are treated here
{\emph{as if}} they could occupy the same $k$-mode, which is of course
forbidden by the Pauli principle. As explained in Sec. {\sec{sec:Pauli}}, this
only corrects for the mistake that is made when the direct SQM is calculated
by treating all electrons as distinguishable objects, that is, when ignoring
Pauli principle.}} Converting the sum to an integral, we obtain
{\Eq{eq:Qtriplet}} {{of the main text:}}
\begin{eqnarray}
  q_{\mathrm{ex}} & = & \int \mathd \omega f_+ ( \omega)  \bar{\nu} ( \omega) a
  ( \omega) .  \label{eq:qexFlatBand}
\end{eqnarray}
The anisotropy function $a ( \omega) = \sum_{\sigma} a_{\sigma} ( \omega)$ has
two contributions:
\begin{eqnarray}
  \bar{\nu} ( \omega) a_{\sigma} ( \omega) & \assign & \sum_k a_{k \sigma}
  \delta ( \omega - \varepsilon_{k \sigma}) .  \label{eq:asigmak}
\end{eqnarray}
The quantity $\bar{\nu} ( \omega) a_{\sigma} ( \omega)$ is the exchange
quadrupolarization of a spin $\sigma$ electron at energy $\omega$. One should
note that the function $a_{k \sigma}$, defined by {\Eq{eq:ak}}, does not only
depend on the energy $\varepsilon_{k \sigma}$, but also on the energy
$\varepsilon_{k \bar{\sigma}}$. Since $\varepsilon_{k \bar{\sigma}}$ is not
necessarily an implicit function of $\varepsilon_{k \sigma}$ for arbitrary
band structures, one can in general reformulate {\Eq{eq:asigmak}} only in
terms of the 2DOS {\eq{eq:twoDOS}}:
\begin{eqnarray}
  \bar{\nu} ( \omega) a_{\sigma} ( \omega) & = & \int \mathd \omega' f_+ (
  \omega') \sum_{\sigma'} \tfrac{\sigma \sigma'}{4} \nu_{\sigma \sigma'} (
  \omega, \omega'),  \label{eq:2dosa}
\end{eqnarray}
resulting in {\Eq{eq:asigma}} of the main text.

However, for the Stoner model, which we discuss from hereon, the simple
relation $\varepsilon_{k \bar{\sigma}} = \varepsilon_{k \sigma} - \sigma J
\text{/} 2$ can be exploited to express $a_{k \sigma}$ as a function of
$\varepsilon_{k \sigma}$ only. We therefore obtain the simpler result
\begin{eqnarray}
  \bar{\nu} ( \omega) a_{\sigma} ( \omega) & = & \nu_{\sigma} ( \omega) ( f_+
  ( \omega) - f_+ ( \omega + \sigma J)),  \label{eq:tripletCorrSimple}
\end{eqnarray}
which only depends on the 1DOS $\nu_{\sigma}$. This unfortunately hides the
underlying two-particle nature of the exchange SQM, but aids the
interpretation of the total spin-anisotropy function:
{\Eq{eq:tripletCorrSimple}} shows that the contribution $a_{\uparrow} (
\omega)$ from up-spins is positive and comes from the range of energies $\mu -
J < \omega < \mu$, whereas the contribution $a_{\downarrow} ( \omega)$ from
down-spins is negative and comes from energies \ $\mu < \omega < \mu + J$
(both up to thermal smearing). Adding both contributions yields for the full
spin-anisotropy function after some manipulations
\begin{eqnarray}
  a ( \omega) & = & \tfrac{1}{4} [ 2 f_+ ( \omega) - f_+ ( \omega + J) - f_+ (
  \omega - J)] \nonumber\\
  &  & + \tfrac{1}{4} n ( \omega) [ f_+ ( \omega + J) - f_+ ( \omega - J)] . 
  \label{eq:aApp}
\end{eqnarray}
Here it should be noted that $n ( \omega)$ is not independent of $J$ but is a
function of it through {\Eq{eq:DOS}}, \ {\eq{eq:spinpol}} and
{\eq{eq:Stoner}}.{\footnote{In particular, one cannot set $n ( \omega) = 0$
for all $\omega$ without setting $J = 0$, since these are equivalent for the
Stoner dispersion considered.}} The combinations of Fermi-functions are
non-zero only for energies $| \omega - J | \lesssim T$.

It was pointed out in {\Sec{sec:spinFree}} for a purely thermally biased
tunnel junction that the finite SQM current that remains whenever the spin
current vanishes arises entirely from the exchange SQM, i.e., from the
\emph{antisymmetric part} of the spin-anisotropy function $a ( \omega)$
relative to $\mu$. Generally, when assuming a weakly energy dependent average
DOS $\bar{\nu} ( \omega)$ in the range $| \omega - J | \lesssim T$, the first
term {\Eq{eq:aApp}} always gives rise to such a term. The second term, in
which the spin-polarization $n ( \omega)$ is multiplied by a symmetric
function relative to $\mu$, can only cancel this function if $n ( \omega)$ is
strongly antisymmetric (i.e., $n ( \omega) = \pm 1$ for $\omega \gtrless 0$,
up to thermal smearing). If we further specialize to the approximation of a
flat band symmetric about $\mu$ (cf. {\Sec{sec:wide-flat-band}}), this second
term is exactly zero because the spin polarization vanishes in the window $[
\omega - J, \omega + J]$ up to thermal smearing. Only the first line of
{\Eq{eq:aApp}} remains and gives a thermally induced pure SQM current as
discussed in Sec. {\sec{sec:spinFree}}: substituting $x = ( \omega - \mu)
\text{/} T$, {\Eq{eq:qexFlatBand}} can be rewritten as
\begin{eqnarray}
  q_{\mathrm{ex}} & = & \frac{\bar{\nu} T}{4} \int \mathd x f ( x) [ 2 f ( x) -
  f ( x - j) - f ( x + j)], \quad  
\end{eqnarray}
where $f ( x) = [ e^x + 1]^{- 1}$ and $j = J \text{/} T$. As $a ( \omega)$ is
nonzero in an energy window $2 J$ centered at $\mu$, which is far away from
the band edges, we can replace the 1DOS by its constant value $\bar{\nu}$ and
extend the limits integration to $\pm \infty$. Using the identity $f ( x) = 1
- f ( - x)$ and the result
\begin{eqnarray}
  \int d x f ( x) f ( - ( x - y)) & = & \left\{ \begin{array}{ll}
    1 & y = 0\\
    y b ( y) & \text{else}
  \end{array}, \right.  \label{eq:intFF}
\end{eqnarray}
where $b ( x) = [ e^x - 1]^{- 1}$ is the Bose function, and taking the limit
$y \rightarrow 0$ we obtain {\Eq{eq:qrex}} of the main text:
\begin{eqnarray}
  q_{\mathrm{ex}} & = & \frac{\bar{\nu}}{4} [ - 2 + j b ( j) + ( - j) b ( - j)]
  \\
  & = & \frac{\bar{\nu}}{2} \left[ \frac{j}{2} \coth \left( \frac{j}{2}
  \right) - 1 \right] . 
\end{eqnarray}

\section{Symmetric and Traceless Tensors}\label{sec:appDiagonal}

In this appendix, we collect some relevant results on symmetric, traceless
tensors. We first show how a particular type of such tensors, constructed from
two real vectors $\vec{a}$ and $\vec{b}$,
\begin{eqnarray}
  & \begin{array}{lllll}
    \mathcal{A} & = & \vec{a} \odot \vec{b} & = & \frac{1}{2} \left( \vec{a}
    \vec{b} + \vec{b} \vec{a} \right) - \frac{1}{3} \left( \vec{a} \cdot
    \vec{b} \right) \mathcal{I},
  \end{array} &  \label{eq:symmtrace}
\end{eqnarray}
can be diagonalized using dyadic calculus {\cite{Lindellbook}}, i. e., without
introducing a coordinate system. The tensor $\mathcal{A}$ can be expressed in
terms of its principal values $\lambda_{\mu}$ and normalized vectors
$\op{\vec{v}}_{\mu}$ that define the principal axes as
\begin{eqnarray}
  \mathcal{A} & = & \overset{}{\sum_{\mu = 0, \pm}} \lambda_{\mu}
  \op{\vec{v}}_{\mu} \op{\vec{v}}_{\mu}^T . 
\end{eqnarray}
The results {{(\ref{eq:Izero})-(\ref{eq:eigen0pi}) of the main text}} can
then be obtained from Eq. (\ref{eq:SQMCurrentSimple2}) by setting $\vec{a} =
\op{\vec{J}}^L$ and $\vec{b} = E^L \op{\vec{J}}^L + A^L  \op{\vec{J}}^R + T^L 
\left( \op{\vec{J}}^L \times \op{\vec{J}}^R \right)$. To diagonalize
{\eq{eq:symmtrace}}, we have to distinguish two cases:

{{Case}} (i) $\mathbf{\vec{a} \nparallel \vec{b}}$: {{eigenvalues}}
\begin{eqnarray}
  \lambda_{\mu} & = & - \tfrac{1}{3} \left( \vec{a} \cdot \vec{b} \right) +
  \tfrac{1}{2} \delta_{\mu, \pm} \left[ \left( \vec{a} \cdot \vec{b} \right) +
  \mu a b \right],  \label{eq:lambdaNPar}
\end{eqnarray}
{{where $\mu = 0, \pm$ and normalized eigenvectors}}
\begin{eqnarray}
  \op{\vec{v}}_0 & = & \frac{1}{\left| \vec{a} \times \vec{b} \right|} \left(
  \vec{a} \times \vec{b} \right), \\
  \op{\vec{v}}_{\pm} & = & \frac{1}{2 a b \left( a b \pm \left( \vec{a} \cdot
  \vec{b} \right) \right)} \left[ a \vec{b} \pm b \vec{a} \right] 
  \label{eq:v}
\end{eqnarray}
with $a = \left| \vec{a} \right| , b = \left| \vec{b} \right|$.

{{Case}} (ii) $\vec{a} \parallel \vec{b}$, i. e., $\vec{b} = \alpha
\vec{a}$: {{eigenvalues for $\mu = 0, \pm$}}
\begin{eqnarray}
  \lambda_{\mu} & = & \left( \delta_{\mu, +} - \tfrac{1}{3} \right) \alpha
  a^2,  \label{eq:lambdaPar}
\end{eqnarray}
and $\op{\vec{v}}_+ = \vec{a} \text{/} a$ and $\op{\vec{v}}_0$, \
$\op{\vec{v}}_-$ \ {{are any two orthonormal vectors that span the plane
perpendicular to $\op{\vec{v}}_+$.}}

To prove case (i), we first note that one principal axis is obviously
$\vec{v}_0 = \vec{a} \times \vec{b} \neq 0$:
\begin{eqnarray}
  \mathcal{A} \cdot \vec{v}_0 & = & \left[ \tfrac{1}{2} \left( \vec{a} \vec{b}
  + \vec{b} \vec{a} \right) - \tfrac{1}{3} \left( \vec{a} \cdot \vec{b}
  \right) \mathcal{I} \right] \cdot \left( \vec{a} \times \vec{b} \right) \\
  & = & - \tfrac{1}{3}  \left( \vec{a} \cdot \vec{v} \right)  \vec{v}_0 =
  \lambda_0 \vec{v}_0 . 
\end{eqnarray}
In order to derive the the remaining principal values, we need to set up the
characteristic equation:
\begin{eqnarray}
  0 & = & \det \left( \mathcal{\tens{A}} - \lambda \tens{I} \right) \\
  & = & \det \left( \mathcal{\tens{A}} \right) - \lambda \mathrm{spm} \left(
  \tens{A} \right) + \lambda^2 \mathrm{tr} \left( \mathcal{\tens{A}} \right) -
  \lambda^3,  \label{eq:charEq}
\end{eqnarray}
where the coefficients are given by the trace, the sum of principal minors and
the determinant of $\mathcal{\tens{A}}$, respectively:
\begin{eqnarray}
  \mathrm{tr} \left( \tens{A} \right) & = & \sum_i \mathcal{\tens{A}}_{i i}, 
  \label{eq:tr}\\
  \mathrm{spm} \left( \tens{A} \right) & = & \tfrac{1}{2} \mathrm{tr} \left(
  \mathcal{\tens{A}}^{\times}_{\times} \mathcal{\tens{A}} \right), 
  \label{eq:spm}\\
  \det \left( \tens{A} \right) & = & \tfrac{1}{6} \left(
  \mathcal{\tens{A}}^{\times}_{\times} \tens{\mathcal{A}} \right) :
  \tens{\mathcal{A}} .  \label{eq:det}
\end{eqnarray}
Here we used the shorthand notations \ $\begin{array}{lll}
  \left( \mathcal{\tens{A}}^{\times}_{\times} \mathcal{\tens{A}} \right)_{i_1
  i_2} & = & \varepsilon_{i_1 j_1 k_1} \mathcal{\tens{A}}_{j_1 j_2}
  \mathcal{\tens{A}}_{k_1 k_2} \varepsilon_{j_2 k_2 i_2}
\end{array}$ and \ $\mathcal{\tens{B}} : \mathcal{\tens{A}} = \sum_{i, j}
\tens{\mathcal{B}}_{i j} \mathcal{\tens{A}}_{i j}$. Eqs.
(\ref{eq:tr})-(\ref{eq:det}) are the (only) three rotational invariants, i. e.
they are invariant under transformations $\mathcal{\tens{A}} \rightarrow
\mathcal{\tens{R}} \cdot \mathcal{\tens{A}} \cdot \mathcal{\tens{R}}^T$ where
$\mathcal{\tens{R}}$ is a rotation matrix: $\mathcal{\tens{R}} \cdot
\tens{\mathcal{R}}^T = \tens{\mathcal{R}}^T \cdot \mathcal{\tens{R}} =
\mathcal{\tens{I}}$. {{Inserting {\Eq{eq:symmtrace}} we obtain}}
\begin{eqnarray}
  \mathrm{tr} \left( \mathcal{\tens{A}} \right) & = & \left( \vec{a} \cdot
  \vec{b} \right),  \label{eq:trA}\\
  \mathrm{spm} \left( \mathcal{\tens{A}} \right) & = & - \tfrac{1}{3} \left(
  \vec{a} \cdot \vec{b} \right)^2 - \tfrac{1}{4}  \left( \vec{a} \times
  \vec{b} \right)^2, \\
  \det \left( \mathcal{\tens{A}} \right) & = & \tfrac{2}{27} \left( \vec{a}
  \cdot \vec{b} \right)^3 + \tfrac{1}{12} \left( \vec{a} \cdot \vec{b} \right)
  \left( \vec{a} \times \vec{b} \right)^2 .  \label{eq:detA}
\end{eqnarray}
Inserting {{these}} into {\Eq{eq:charEq}}, {{one finds}} that
$\lambda_0$, given by {\eq{eq:lambdaNPar}} {{with $\mu = 0$}}, is indeed a
principal value of $\mathcal{\tens{A}}$. By polynomial division we
obtain{{ a quadratic equation for the remaining principal values
$\lambda_{\pm}$, which is solved by {\Eq{eq:lambdaNPar}} with $\mu = \pm$}}.
The general solution of $\begin{array}{l}
  \left( \tens{\mathcal{A}} - \lambda_{\pm} \mathcal{\tens{I}} \right) \cdot
  \vec{v}_{\pm} = 0
\end{array}$is given by {\cite{Lindellbook}}:
\begin{eqnarray}
  \vec{v}_{\pm} & = & \vec{c} \cdot \left[ \left( \mathcal{\tens{A}} -
  \lambda_{\pm} \mathcal{\tens{I}} \right)^{\times}_{\times} \left(
  \mathcal{\tens{A}} - \lambda_{\pm} \mathcal{\tens{I}} \right) \right] \\
  & = & - \tfrac{1}{2} \vec{c} \cdot \left[ {\color{white} ]}
  \left. \left( \vec{a} \times \vec{b} \right) \right. \left( \vec{a} \times
  \vec{b} \right) \right. \nonumber\\
  &  & \left. + \left( 2 q_{\pm} \left( \vec{a} \cdot \vec{b} \right) -
  q_{\pm}^2 \right) \tens{\mathcal{I}} - q_{\pm} \left( \vec{a} \vec{b} +
  \vec{b} \vec{a} \right)  \right] . 
\end{eqnarray}
{{where \ $q_{\pm} = \left( \vec{a} \cdot \vec{b} \right) \pm \left|
\vec{a} \right| \left| \vec{b} \right|$. Here $\vec{c}$ is a vector such that
$\vec{v}_{\pm} \neq 0$ and either $\vec{c} = \vec{a}$ or $\vec{c} = \vec{b}$
fulfills this condition, yielding the result (\ref{eq:v}) after
normalization.}}

{{For case (ii) we have $\tens{\mathcal{A}} = \vec{a} \odot \alpha \vec{a}
= \alpha \left( \vec{a} \vec{a} - a^2 \mathcal{\tens{I}} / 3 \right)$. Since
$\vec{a} \times \vec{b} = 0$, the above results cannot be applied. The vector
$\vec{a}$ obviously defines a principal axes since $\mathcal{\tens{A}} \cdot
\vec{a} = + \tfrac{2}{3} \alpha a^2 \vec{a}$, and for any vector
$\vec{a}_{\perp}$ in the plane perpendicular to $\vec{a}$ we have
$\tens{\mathcal{A}} \cdot \vec{a}_{\perp} = - \tfrac{1}{3} \alpha a^2$. This
confirms the principal values {\eq{eq:lambdaPar}}, completing the proof.}}

\section{Rotation of the SQM by SQM Currents}\label{sec:approtation}

In this appendix we show that a SQM tensor $\tens{Q}$ is rotated if it does
not commute with the SQM current tensor $\tens{I}_{\tens{Q}}$ as pointed out
in {\sec{sec:angleDep}}. More formally, \ $\left[ \tens{Q},
\tens{I}_{\tens{Q}} \right]_- = 0$ implies the both tensors have collinear
principal axes. We outline the analogous situation for the spin vector: it is
geometrically clear that the spin $\vec{S}$ does not rotate, that is, $\vec{S}
= S ( t) \vec{e}_3$ in some time-independent basis if and only if the spin
current vector $\vec{I}_{\vec{S}} = \dot{\vec{S}}$ is collinear to $\vec{S}$
at all times.

We now outline a proof of this statement, which can be extended to the SQM.
Assume $\vec{S}$ does not rotate, then $\vec{S} = S ( t) \vec{e}_3$ \ in some
time independent basis $\vec{e}_i$ and. Then $\dot{\vec{S}} = \dot{S} ( t)_{}
\vec{e}_3 = \vec{I}_{\vec{S}}$. Thus, $\vec{I}_{\vec{S}} \cdot \vec{S} = \pm
\left| \vec{S} \right| \left| \vec{I}_{\vec{S}} \right| \vec{e}_3 \cdot
\vec{e}_3 = \pm \left| \vec{S} \right| \left| \vec{I}_{\vec{S}} \right|$.
Conversely, assume $\vec{S} \cdot \vec{I}_{\vec{S}} = \pm \left| \vec{S}
\right| \left| \vec{I}_{\vec{S}} \right|$, i. e. $\vec{S} = S ( t) \vec{e}_3 (
t)$ and $\vec{I}_{\vec{S}} = I_{\vec{S}} ( t)  \vec{e}_3 ( t)$, but for some
time-{\emph{dependent}} $\vec{e}_3 ( t)$. One can derive a contradiction from
the latter assumption: first, we have \ $\dot{\vec{S}} = S ( t)  \vec{e}_3 (
t) + S ( t)_{} \dot{\vec{e}}_3 ( t) = \vec{I}_{\vec{S}} = I_{\vec{S}} ( t) 
\vec{e}_3 ( t)$. By assumption, the normalization is preserved, that is,
$\frac{d}{\mathrm{dt}} \left( \vec{e}_3 \cdot \vec{e}_3 \right) = 2 \vec{e}_3
\cdot \dot{\vec{e}}_3 = 0$. Thus, if $\dot{\vec{e}}_3 \perp \vec{e}_3$, we
have $\dot{S}_{} ( t) = I_{\vec{S}} ( t)$ and either \ $_{} S ( t) = 0$ (which
implies a trivial spin) or $\dot{\vec{e}}_3 = 0$, contradicting our
assumption. This completes the proof.

We now prove a similar statement for the SQM tensor $\tens{Q}$ and its current
$\tens{I}_{\tens{Q}} = \dot{\tens{Q}}$,
\begin{eqnarray}
  \tens{Q}  \text{ has fixed principal axes} & \Leftrightarrow & \left[
  \tens{Q}, \tens{I}_{\tens{Q}} \right]_- = 0. 
\end{eqnarray}
Assume $\tens{Q} = \sum_i Q_i ( t) \vec{e}_i \vec{e}^T_i$ has a
time-independent basis, i.e., only the principal values $Q_i ( t)$ change but
the principal axes $\vec{e}_i$ do not rotate. Thus, $\dot{\tens{Q}} = \sum_i
\dot{Q}_i  ( t) \vec{e}_i \vec{e}^T_i = \tens{I}_{\tens{Q}} = \sum_{i j} I_{i
j} ( t) \vec{e}_i \vec{e}^T_j$, implying $I_{i j} ( t) = \delta_{i j} 
\dot{Q}_i ( t)$ and consequently $\left[ \tens{Q}, \tens{I}_{\tens{Q}}
\right]_- = 0$. Now assume the converse, i. e., $\tens{I}_{\tens{Q}}$ and
$\tens{Q}$ commute. Since both are real, symmetric tensors, they can be
diagonalized and since they commute they have common principal axes: $\tens{Q}
= \sum_i Q_i ( t)  \vec{e}_i ( t) \vec{e}^T_i ( t)$ and $\tens{I}_{\tens{Q}} =
\sum_i I_i ( t)  \vec{e}_i ( t) \vec{e}^T_i ( t)$. \ We next a derive a
contradiction from the assumption $\dot{\vec{e}}_i ( t) \neq 0$. We first find
$\dot{\tens{Q}} = \sum_i \left[ \dot{Q}_i  \vec{e}_i \vec{e}^T_i + Q_i  \left(
\dot{\vec{e}}_i \vec{e}^T_i + \vec{e}_i \dot{\vec{e}}^T_i \right) \right] =
\tens{I}_{\tens{Q}}$ and note that $\dot{\vec{e}}_i \vec{e}^T_i$ is
independent of $\vec{e}_k \vec{e}^T_k$ for all $k$ because $\dot{\vec{e}}_i
\perp \vec{e}_k$. Thus, we have $I_i = \dot{Q}_i$ and either $Q_i = 0$ for all
$i$ or \ $\dot{\vec{e}}_i \vec{e}^T_i + \vec{e}_i \dot{\vec{e}}^T_i = 0$,
which would implies $\dot{\vec{e}}_i = 0$. Thus, $\vec{e}_i$ is time
independent and the proof is complete.

\section{Charge, Spin and Spin-quadrupole Current in
O($\Gamma$)}\label{app:FirstOrder}

In this section, we apply the general diagrammatic technique developed in
{\App{app:covrealtime}} accounting only for contributions up to 1st order in
the tunneling rate $\Gamma$ assuming spin-conserving tunneling. For the charge
and spin current, the construction as described in {\sec{sec:diagramRules}}
gives two contributing diagrams, depicted in Fig. \ref{fig:diagrams1storder}
(a).

{
  \begin{figure}[h]
    \includegraphics[width=0.9\linewidth]{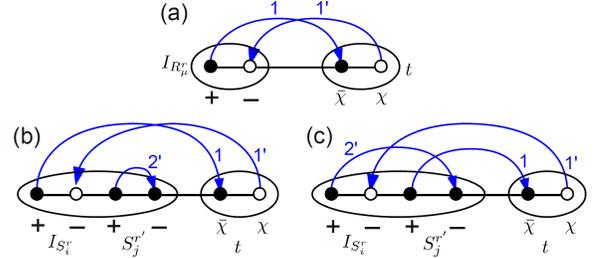}
    \caption{\label{fig:diagrams1storder}Diagrams
representing $O ( \Gamma)$-contributions to (a) charge and spin current, (b)
direct SQM current and (c) exchange SQM current.}
\end{figure}}

In first order, there is only one irreducible contraction possible.
Furthermore, the h. c. indices $\eta$ are fixed by the observable vertex.
However, there are two possibilities to chose the charge indices $\chi$ for
the tunneling double vertex. The total sign of the diagram equals $\chi$ since
we have (i) a factor $- 1$ due to one crossing (ii) a factor $\bar{\chi}$ due
to the early and late vertices and (iii) a factor + since there are no
intermediate vertices. This results in
\begin{eqnarray}
  \brkt{I_{R_{\mu}}^L} & = & 2 \mathrm{Im} \left[ \int d \omega_1 d \omega_{1'}
  \frac{1}{i 0 - \omega_{1'} + \omega_1}_{_{}} \sum_{\rho_1 \rho_{1'} \tau_1
  \tau_{1'}} \right. \nonumber\\
  &  & \sum_{\chi} \chi ( F^L_{\chi})_{\rho_1} ( \omega_1) (
  F_{\bar{\chi}}^R)_{\rho_{1'}} ( \omega_{1'})  ( \check{t}_{\tau_1}
  \check{t}_{\tau_2}) \nonumber\\
  &  &  \mathrm{tr} ( \check{r}_{\rho_1'} \check{r}_{\tau_2}
  \check{r}_{\rho_1} r_{\mu} \check{r}_{\tau_1}) ], 
\end{eqnarray}
where $( F_i)_{\rho_i} = \bar{\nu}^{r_i} ( \omega_i) f^{r_i
}_{\bar{\chi}_i} ( \omega_i) \sqrt{2} [\delta_{\rho_i, 0} + ( 1 -
\delta_{\rho_i, 0}) n^r \op{J}^r_{\rho_i}]$. Note that $r_{\mu}$ is associated
with the current vertex and therefore has no ``check''. For our case, the
tunneling is spin-independent so that $\check{t}_{\tau} = \sqrt{2} t
\delta_{\tau, 0}$. We will use furthermore $\check{r}_0 = r_0 \text{/}
\sqrt{2} = \mathbbm{1} \text{/} \sqrt{2}$, $\check{r}_i = \sqrt{2} r_i =
\sqrt{2} s_i$ for $i = 1, 2, 3$ and
\begin{eqnarray}
  &  & \mathrm{tr} ( \check{r}_{\rho_{1'}} \check{r}_{\rho_1} \check{r}_{\mu})
  = \frac{1}{\sqrt{2}} \delta_{\mu 0} \delta_{\rho_1 \rho_{1'}} \nonumber\\
  &  & + \frac{1}{\sqrt{2}} \bar{\delta}_{\mu, 0} ( \delta_{\rho_1 0}
  \delta_{\rho_{1'} \mu} + \delta_{\rho_{1'} 0} \delta_{\rho_1 \mu} + i
  \varepsilon_{\rho_{1'} \rho_1 \mu}) . 
\end{eqnarray}
We obtain the charge current {\eq{eq:chargeCurrent}} given in the main text,
\begin{eqnarray}
  \brkt{I_N^L} & = & 2 \mathrm{Im} \left[ \int_{1 1'}  \frac{1}{\pi} \Gamma^{L
  R}_{1 1'} ( - \Delta_{1 1'}) \left( 1 + \vec{n}^L_1 \cdot \vec{n}^R_{1'}
  \right) \right. \nonumber\\
  &  & \left. \left( P \frac{1}{\omega_1 - \omega_{1'}} - i \pi \delta (
  \omega_1 - \omega_{1'}) \right) \right] \nonumber\\
  & = & 2 \int_{1 1'} \Gamma_1 \Delta_1 \left( 1 + \vec{n}^L_1 \cdot
  \vec{n}^R_1 \right), 
\end{eqnarray}
and for the spin current we obtain {\Eq{eq:spinCurrent}} of the main text,
\begin{eqnarray}
  &  & \brkt{I_{S_i}^L} = \text{ } 2 \mathrm{Im} \left[ \int_{1 1'} \Gamma^{L
  R}_{1 1'} ( - \Delta_{1 1'}) \right. \nonumber\\
  &  & \text{ \ \ \ \ \ \ \ } \times \left( P \frac{1}{\omega_1 -
  \omega_{1'}} - i \pi \delta ( \omega_1 - \omega_{1'}) \right) \nonumber\\
  &  & \left. \text{ \ \ \ \ \ \ \ } \times \frac{1}{2} \left( \vec{n}^L_1 +
  \vec{n}^R_{1'} + i \left( - \vec{n}^L_1 \times \vec{n}^R_{1'} \right)
  \right) \right]  \label{eq:1storderSpinCurr}\\
  &  & = \int_1 \Gamma_1 \left[ \Delta_1 \left( \vec{n}^L_1 + \vec{n}^R_1
  \right) + \tfrac{\vec{n}^L_1}{\bar{\nu}^R_1} \times \vec{\beta}^R_1 +
  \vec{\beta}^L_1 \times \tfrac{\vec{n}^R_1}{\bar{\nu}^L_1} \right] . \, 
  \label{eq:appspincurr}
\end{eqnarray}
Here we introduced the short-hand notations $\Gamma_1 = \Gamma^{L R}_{1 1}$,
$n_1^r = n^r ( \omega_1)$ and furthermore
\begin{eqnarray}
  \Gamma^{L R}_{1 1'} = \Gamma^{R L}_{1 1'} & = & 2 \pi | t |^2 \bar{\nu}^L (
  \omega_1) \bar{\nu}^R ( \omega_{1'}) \\
  \Delta_{1 1'} & = & f^R_+ ( \omega_{1'}) - f^L_+ ( \omega_1) \\
  \beta^r ( \varepsilon) & = & - \int d \omega P \frac{f_+^r ( \omega)
  \bar{\nu}^r ( \omega) n^r ( \omega)}{\omega - \varepsilon} 
  \label{eq:exfield}
\end{eqnarray}
Here $P \left( \tfrac{1}{z} \right) = \mathrm{Re} \left( \frac{1}{z + i 0}
\right)$ denotes the principal value.

The calculation of the SQM current in O($\Gamma$) for spin-independent
tunneling proceeds in a similar way. However, due to its two-particle nature,
there are two pairs of diagrams with different contraction topologies, which
make up the direct (Fig. \ref{fig:diagrams1storder} (b)) and the exchange
contribution (Fig. \ref{fig:diagrams1storder} (c)) to the SQM current,
respectively. Each pair differs with respect to charge indices.

On the level of diagrams, one immediately sees that the expressions involving
the spin operator and the spin current operator factorize for the direct
contribution since the contraction labelled with 2' in Fig.
\ref{fig:diagrams1storder} (b) does not cross any other lines. This
corresponds to the product of the expectation value of two operators.
Therefore, without only further calculation, we obtain
\begin{eqnarray}
  \brkt{\tens{I}^{L L}_{\tens{Q}}}_{\text{dir}} & = & 2
  \brkt{\vec{I}_{\vec{S}}^L} \odot \brkt{\vec{S}^L} . 
\end{eqnarray}
where $\brkt{\vec{I}_{\vec{S}}^L}$ is the previously calculated spin current
{\eq{eq:appspincurr}}. The evaluation of the exchange contribution is more
complicated because the $2'$-contraction does cross other lines. Applying the
diagram rules from {\sec{sec:diagramRulesSQM}}, we obtain
\begin{eqnarray}
  \brkt{\tens{I}^{L L}_{\tens{Q}}}_{\text{ex}} & = & 2 \cdot 2 \sum_{\chi}
  \bar{\chi} \mathrm{Im} \int d \omega_1 d \omega_{1'} d \omega_{2'} \nonumber\\
  &  & \frac{2 t^2 f^L_- ( \omega_{2'}) f^L_{\chi} ( \omega_1) (
  F^R_{\bar{\chi}})_{\rho_{1'}} ( \omega_{1'})}{i 0 + \omega_1 - \omega_{1'}}
  \nonumber\\
  &  & \times \mathcal{A}^L_{\mu \nu} \mathrm{tr} ( \check{r}_{\mu}
  \mathbf{s} \odot \check{r}_{\nu} \mathbf{s}  \check{r}_{\rho_{1'}}) 
\end{eqnarray}
where the 2DOS {\eq{eq:twoDOS}} enters through the matrix $\tens{A}^L_{\mu
\nu}$ given by {\eq{eq:A00}}-{\eq{eq:Aij}}. The first factor $2$ comes from
Hermitian conjugation symmetry, the second factor of 2 is due to the product
rule when applying the derivative to the SQM operator, which is quadratic in
spin (cf. {\Eq{eq:ISQM}}) and the third one in the second line is associated
with the SQM current vertex (see Sec. {\sec{sec:diagramRulesSQM}}). The sign
factor is obtained as follows: The number of crossings is even and the
intermediate spin vertex has $\eta_e = +$, giving no sign, but from the early
and late vertex, we obtain a sign factor $\bar{\chi}$. It remains to calculate
the spin trace by employing the anticommutation relations of spin-1/2 operator
algebra and the identity $\vec{s} \odot \vec{s} = 0$ (cf. {\Eq{eq:symdyad}}):

{\wideeq{\begin{eqnarray}
  \brkt{\tens{I}^{L L}_{\tens{Q}}}_{\text{ex}} & = & \text{ \ } 2 \int_{1, 2'}
  f^L_- ( \omega_{2'}) \Gamma_1^{L R} \Delta_1 \left[ \frac{( \mathcal{A}_{1
  2'}^L)_{0 i}}{\bar{\nu}^L_1} n^R_1 \vec{e}_i \odot \op{\vec{J}}^R + \frac{(
  \mathcal{A}_{1 2'}^L)_{i j}}{\bar{\nu}^L_1} \vec{e}_i \odot \vec{e}_j
  \right] \nonumber\\
  &  & + 2 \int_{1, 1', 2'} f^L_- ( \omega_{2'}) \Gamma^{L R}_{1 1'}
  \Delta_{1 1'} P \frac{1}{\omega_1 - \omega_{1'}} \left[ \frac{(
  \mathcal{A}_{12'}^L)_{i j}}{\bar{\nu}^L_1} n^R_{1'}  \vec{e}_i \odot \left(
  \vec{e}_j \times \op{\vec{J}}^R \right) \right]  \label{eq:intSQM}\\
  & = & - 2 \int_1 \Gamma_1^{L R} \Delta_1 \left[ \tilde{a}^L_1 n^R_1 
  \op{\vec{J}}^L \odot \op{\vec{J}}^R + a^L_1 \op{\vec{J}}^L \odot
  \op{\vec{J}}^L \right] - 2 \int_{1, 1'} \Gamma^{L R}_{1 1'} \Delta_{1 1'} P
  \frac{1}{\omega_1 - \omega_{1'}} \left[ a^L_1 n^R_{1'} \op{\vec{J}}^L \odot
  \left( \op{\vec{J}}^L \times \op{\vec{J}}^R \right) \right] . \, 
  \label{eq:IQLL}
\end{eqnarray}}}

In the last step, we inserted $f^L_- ( \omega_{2'}) = 1 - f^L_+ (
\omega_{2'})$ into {\Eq{eq:intSQM}} and used that the contribution from 1
vanishes when summing over all spin indices. Restoring all indices explicitly,
we obtain {\Eq{eq:SQMCurrent}}, the main result of the paper. Moreover, we
identified the spin-anisotropy functions {\eq{eq:qLR}} and
{\eq{eq:tripletCorr}}:
\begin{eqnarray}
  a^L ( \omega) & = & \int d \omega' f^L_+ ( \omega') \sum_{\sigma \sigma'}
  \frac{\sigma \sigma'}{4} \frac{\mathcal{\nu}_{\sigma \sigma'}^L  ( \omega,
  \omega')}{\bar{\nu}^L ( \omega)}, \\
  \tilde{a}^L ( \omega) & = & \int d \omega' f^L_+ ( \omega') \sum_{\sigma
  \sigma'} \frac{\sigma'}{4} \frac{\mathcal{\nu}_{\sigma \sigma'}^L  ( \omega,
  \omega')}{\bar{\nu}^L ( \omega)} . 
\end{eqnarray}

\section{Covariant Real Time Diagrammatics}\label{app:covrealtime}

In this appendix, we give a self-contained derivation of the technique used
for calculating the charge, spin-dipole and SQM currents. Using this technique
the actual calculation becomes very compact and is presented
{\Sec{sec:firstOrder}}. The interest in presenting the derivation here lies in
three factors: (i) The real-time technique is more general and can be applied
to more complex systems containing strongly interacting localized systems.
Therefore, it is not often applied to non-interacting systems since other
approaches are available in that case. However, for the calculation of
multi-particle averages, its practical rules of calculation prove to be very
convenient. Therefore it is of interest to point out how the technique
simplifies when applied to non-interacting problems. (ii) We reformulate the
real-time technique here such that one can deal more efficiently with any
non-trivial spin dependencies. In a forthcoming work we show that this
generalizes to the more complex cases {\cite{Hell13c}}. (iii) Finally, it is
also of great help to have these simpler calculations, formulated in the same
way, for comparison the more complex ones (for which other approaches do not
work anymore) {\cite{Hell13c}}.

After reviewing the compact Liouville space notation
{\cite{Schoeller09a,Leijnse08a,Saptsov12a}} in ({\sec{sec:notation}}), we
indicate how the general real-time approach simplifies and show that the
calculation of operator expectation values in the long-time limit and to any
order in the tunnel-coupling $\Gamma = 2 \pi \nu^L \nu^R | t |^2$ amounts to
evaluation of irreducible diagrams in a perturbation expansion
({\sec{sec:perturbation}}). The central technical achievement of this paper is
a {\emph{covariant}} formulation of these diagram rules for the charge,
spin-dipole and SQM current: the expressions they produce are manifestly
invariant under the change of the coordinate system and of the
spin-quantization axis (cf. {\Sec{sec:candsTfinite}} and {\sec{sec:Tfinite}}).
They are thus coordinate-free both in real space and in Hilbert / Liouville
space. This reformulation is crucial to keep the calculation of the
non-equilibrium steady state average of the spin-quadrupole moment tractable.
The required steps are:

1. Separation of spin and energy dependence in the diagrammatic expressions,
recasting the spin part as traces over Pauli operators
({\sec{sec:energySpinSep}}),

2. Collection of all sign factors ({\sec{sec:signs}}),

3. Halving the number of diagrams by exploiting \ their complex conjugation
symmetry ({\sec{sec:hermConj}}),

4. Identification of {{observable-specific}} diagram rules for charge,
spin-dipole ({\sec{sec:diagramRules}}), and finally

5. Spin-quadrupole current ({\sec{sec:diagramRulesSQM}}).

{{Steps 1, 4 and 5}} are presented for the first time in this paper. We
note that the technique formulated is more general than the problem of
interest in two ways: it applies to (i) arbitrary orders of the tunnel
coupling and (ii) models with spin-dependent tunnel amplitudes.

\subsection{Compact Notation in Liouville Space}\label{sec:notation}

The calculation is formulated entirely in Liouville space, the space of linear
operators acting on a Hilbert space of a quantum mechanical system. The linear
transformations of operators, the elements of Liouville space, are called
superoperators. Although our notation follows Ref. {\cite{Schoeller09a}} and
recent developments reported in Ref. {\cite{Saptsov12a}}, we have made some
further convenient modifications which warrant some discussion.

Similar to any usual operator, any superoperator can be expressed in terms of
\emph{field superoperators}, which we define following {\cite{Saptsov12a}}
by
\begin{eqnarray}
  J_1 \cdot & = & ( - \chi_1 \eta_1)^{\mathcal{N}} \left\{ \begin{array}{ll}
    c_{r_1 \eta_1 n_1 k_1 \sigma_1} \cdot & \chi_1 \eta_1 = -\\
    \cdot c_{r_1 \eta_1 n_1 k_1 \sigma_1}  & \chi_1 \eta_1 = +
  \end{array} \right. .  \label{eq:J}
\end{eqnarray}
Here the $\cdot$ denotes the operator argument of the superoperator. Here
$\mathcal{N} \cdot = [ N, \cdot]$ is the particle number superoperator. The
subscript ``1'' is an abbreviation for all indices
\begin{eqnarray}
  1 & = & ( \chi_1, \eta_1, r_1, n_1, k_1, \sigma_1) .  \label{eq:1}
\end{eqnarray}
Here $\eta_1$ is the Hermitian conjugation index, which determines whether the
field operator is a creation operator ($c_{\eta_1 = -} = c^{\dag}$) or an
annihilation operator ($c_{\eta_1 = +} = c$) of an electron in electrode $r_1$
in band $n_1$ in mode $k_1$ with spin $\sigma_1$. New in this notation is the
charge index $\chi$, which distinguishes whether physically the total
superparticle number is increased ($\chi = +$) or decreased $( \chi = -
$) by the action of $J_1$ (note $\eta_1$ does not have such physical
meaning: an annihilation (creation) operator acting from the right
{\emph{increases}} ({\emph{decreases}}) the superparticle number). We prefer
this physically more meaningful charge index $\chi$ instead of the commonly
combination of the Keldysh index $p = - \chi \eta$.

The time evolution of the density operator in {\Sec{sec:perturbation}} is
generated by the Liouvillian superoperator $L = L_0 + L_T$ describing the
internal evolutions, $L_0 \cdot = [ H_0, \cdot]$, and the one due to
tunneling, $L_T = [ H_T, \cdot]$. Using Eq. (\ref{eq:J}), we obtain
\begin{eqnarray}
  L_0 & = & \chi_1 \delta_{\eta_1, -} \varepsilon_1 J_1 J_{\bar{1}} 
  \label{eq:L0}
\end{eqnarray}
with $\varepsilon_1 = \varepsilon^{r_1}_{n_1 k_1 \sigma_1}$ and
\begin{eqnarray}
  L_T & = & T_{1 1'} J_1 J_{1'},  \label{eq:LT}
\end{eqnarray}
where
\begin{eqnarray}
  T_{1 1'} & = & \bar{\chi}_{1'} \delta_{\eta_1  \bar{\eta}_{1'}}
  \delta_{\chi_1  \bar{\chi}_{1'}} \delta_{r_1 L} \delta_{r_{1'} R}  [ T^{L
  R}_{\sigma_1 \sigma_{1'}}]^{\eta_{1'}}  \label{eq:tLT}
\end{eqnarray}
Here we extended the use the Hermitian-conjugation index $\eta$ as
{\emph{superscript}} to indicate complex conjugation, i. e., $[ T^{L
R}_{\sigma_1 \sigma_{1'}}]^+ \assign T^{L R}_{\sigma_1 \sigma_{1'}}$ and $[
T^{L R}_{\sigma_1 \sigma_{1'}}]^- \assign T^{L R^{\ast}}_{\sigma_1
\sigma_{1'}}$. In both Eqs. {\eq{eq:L0}} and {\eq{eq:LT}} we implicitly sum
over all indices contained {\Eq{eq:1}}.

Our main interest is to deal efficiently with the dependence of expressions on
the choice of the spin-quantization axis. However, for non-collinear
spin-polarized systems, there exists no specific choice for the
spin-quantization axis that simplifies the calculations considerably. The best
strategy is therefore to completely remove a reference to the spin
quantization axis in all expressions. We start with the tunneling Liouvillian,
for which we allow any type of symmetry breaking tunneling processes, for
example due to a magnetic impurity in the barrier. The most general tunneling
amplitudes reads
\begin{eqnarray}
  T^{L R}_{\sigma \sigma'} & = & _L \langle \sigma | \check{\vec{t}} \cdot
  \check{\vec{r}} | \sigma' \rangle_R,  \label{eq:tsigma0}
\end{eqnarray}
where $\check{\vec{r}} \cdot \check{\vec{t}} = \sum_{\mu = 0}^3
\check{r}_{\mu} \check{t}_{\mu}$. Here $\check{\vec{r}}$ is a four-component
vector of operators $\check{r}_0 = \mathbbm{1} \text{/} \sqrt{2}$ and
$\check{r}_i = \sqrt{2} s_i$ for $i = x , y, z$ forming an basis for
the Liouville space formed by operators on the spin-1/2 Hilbert space. The
basis is orthonormal with respect to the scalar product $( A, B) = \mathrm{tr} (
A^{\dag} B)$. The tunneling is completely specified by a four-component vector
$\check{\vec{t}}$: if $L_T$ is spin-conserving, as assumed in the main part of
the paper, the spatial components of $\check{t}_{\mu} = \delta_{\mu, 0}
\sqrt{2} t$ must be zero (and {\Eq{eq:tsigma0}} reduces to
{\Eq{eq:tsigmacons}} of {\Sec{sec:model}}). Spin non-conserving tunneling
processes are thus introduced by any further non-zero components $\check{t}_i
= t_i \text{/} \sqrt{2}$ for $i = x, y, z$. Note that the spin-dependence
through the bra and ket in {\Eq{eq:tsigma0}} merely reflects the choice of
(different, arbitrary) quantization axes for the field operators in the
reservoirs connected by the tunneling: clearly it should cancel out of the
final answer. Written in the form {\eq{eq:tsigma0}}, the tunneling Liouvillian
is explicitly covariant: changing either of these quantization axis merely
changes the meaning of the dummy indices $\sigma, \sigma'$. There is also no
explicit dependence on the coordinate system either, since $\check{\vec{t}}
\cdot \check{\vec{r}}$ is a coordinate-free expression.

\subsection{Wick's theorem for Super Operators}\label{sec:Wick}

For the perturbative calculation of expectation values in Sec.
{\sec{sec:perturbation}}, we will use Wick's theorem for the super field
operators as defined in {\Eq{eq:J}}, which reads as {\cite{Saptsov10}}:
\begin{eqnarray}
  \langle J_n \ldots J_1 \rangle & = & \mathrm{tr} ( J_n \ldots J_1 \rho_0)
  \nonumber\\
  & = & \sum_{\mathrm{contr} .}^{} ( - 1)^{N_P}  \prod_{i > j} \gamma_{i j} 
  \label{eq:Wick}
\end{eqnarray}
with the grand canonical distribution
\begin{eqnarray}
  \rho_0 & = & \prod_r \frac{1}{Z^r} e^{- ( H^r_0 - \mu N^r) / T^r}, 
\end{eqnarray}
the partition function $Z^r = \mathrm{tr}_r (   e^{- ( H^r_0
- \mu N^r) / T^r})$ and the contraction function
\begin{eqnarray}
  \gamma_{1 1'} & = &  ( - \chi_{1'} \eta_{1'}) \delta_{1 \bar{1}'}
  \left|_{\mathrm{excl} .  \chi} f_1 . \label{eq:contraction} \right.
\end{eqnarray}
Here $f_1 = f ( \bar{\chi}_1 ( \varepsilon^{r_1}_{k_1 \sigma_1} - \mu^{r_1}) /
T^{r_1})$ with the Fermi function $f ( x) = 1 \text{/} ( e^x + 1)$. In
{\Eq{eq:contraction}}, $\delta_{1 \bar{1}'} |_{\mathrm{excl} .  \chi} 
$ denotes a Kronecker symbol for all indices $1$ and $\bar{1}'$
except for the charge indices $\chi$. In agreement with physical intuition,
the charge index $\chi_1$, at the beginning of the process determines the type
of distribution function appearing: $\chi_{1'} = + ( -)$ corresponds to a
particle (hole). In Eq. (\ref{eq:Wick}), we sum as usual over all possible
pair contractions and $N_P$ is the signature of the permutation that is needed
to disentangle all pairs of contracted superoperators while keeping the order
of the contracted operators within a pair. The easy form of {\Eq{eq:Wick}}
relies on the fact that the field superoperators obey anti-commutation
relations:
\begin{eqnarray}
  {}[ J_1, J_{1'}]_+ & = & ( - \chi_1 \eta_1) \delta_{1 \bar{1}'} 
  \label{eq:anticomm}
\end{eqnarray}
with $\delta_{1 \bar{1}'} = \delta_{r_1 r_{1'}} \delta_{k_1 k_{1'}}
\delta_{n_1 n_{1'}} \delta_{\sigma_1 \sigma_{1'}} \delta_{\chi_1
\chi_{\bar{1}'}} \delta_{\eta_1 \eta_{\bar{1}'}}$ with the conjugate
multi-index:
\begin{eqnarray}
  \bar{1} & = & ( - \chi_1, - \eta_1, r_1, n_1, k_1, \sigma_1) . 
\end{eqnarray}
The inclusion of the \emph{fermion-parity} superoperator $( - \chi_1
\eta_1)^{\mathcal{N}}$ in Eq. (\ref{eq:J}) is crucial for the validity of Eq.
(\ref{eq:anticomm}) from which {\Eq{eq:Wick}} basically follows, as pointed
out by Saptsov et. al. {\cite{Saptsov12a}}.

\subsection{Perturbative Calculation of Expectation
Values}\label{sec:perturbation}

The expectation value of an observable $A$ is given by
\begin{eqnarray}
  \langle A \rangle ( t) & = & \mathrm{Tr} ( A \rho^{\mathrm{tot}} ( t)) 
  \label{eq:Aaverage}
\end{eqnarray}
where $\rho^{\mathrm{tot}} ( t) = e^{- i L ( t - t_0)} \rho_0$ is the
time-dependent density operator of the system and $L = L_0 + L_T$ is the
Liouvillian. We are interested in the long-time limit of {\Eq{eq:Aaverage}},
which by virtue of the final value theorem
\begin{eqnarray}
  & \begin{array}{lllll}
    A & : = & \lim_{t \rightarrow \infty} \mathrm{tr} ( A \rho ( t)) & = & ( -
    i) \lim_{z \rightarrow i 0} z \langle A \rangle ( z)
  \end{array} &  \label{eq:stationarylimit}
\end{eqnarray}
follows from the Laplace transform $\langle A \rangle ( z) =
\int_{t_0}^{\infty} d t e^{i z t} \langle A \rangle ( t)$ of $\langle A
\rangle ( t)$ with \ $\mathrm{Im} ( z) > 0$. We obtain
\begin{eqnarray}
  \langle A \rangle ( z) & = & \mathrm{Tr} \left( \! A \frac{i}{z - L_0 - L_T}
  \rho_0 \right) .  \label{eq:laplace}
\end{eqnarray}
The trace can be evaluated if we rewrite the resolvent $( z - L_0 - L_T)^{-
1}$ in terms of a power series in $L_T$, apply Wick's theorem, collect
diagrams irreducibly contracted to the $A$ operator into a self-energy kernel
$\Sigma_A^{\mathrm{irr}} ( z)$ and resum the series
\begin{eqnarray}
  \langle A \rangle ( z) & = & \Sigma_A^{\mathrm{irr}} ( z) . 
\end{eqnarray}
To compare this simple result to the usual situation considered in the
real-time approach, we assume for a moment that the leads are tunnel-coupled
to an {\emph{interacting}} system with a few degrees of freedom. Since only
the non-interacting leads can integrated out by applying Wick's theorem, the
objective is to derive an exact effective theory for the reduced density
operator of the system $\rho = \tr{\mathrm{res}} \rho^{\mathrm{tot}}$. By merely
replacing $\mathrm{Tr} ( A \ldots) \rightarrow \tr{\mathrm{res}} ( \ldots)$, one
may take the same steps as above to express the Laplace transform of the
reduced density matrix as
\begin{eqnarray}
  \rho ( z) & = & \tr{\mathrm{res}}  \left( \frac{i}{z - L_0 - L_T} \rho_0
  \right)  \label{eq:rhoz}\\
  & = & \frac{i}{z - L - \Sigma ( z)} \rho ( t_0) 
\end{eqnarray}
where $\Sigma ( z)$ is the (reducible) self-energy. {\Eq{eq:rhoz}} is used as
a starting point for a diagrammatic perturbation theory in the coupling $L_T$.
In our case, we have simply no ``system'', i.e., $L = \Sigma ( z) = 0$ and
$\rho ( z) = i / z$ for $\rho ( t_0) = \tr{\mathrm{res}} \rho^{\mathrm{tot}} (
t_0) = 1$ when we take the trace over leads.

Most of the steps that follow up can be generalized to the case where there
is a non-trivial system coupled to the reservoirs {\cite{Hell13c}}, making the
following analysis of interest.

The left action of $A$, considered as a superoperator, can be expressed in
terms of field superoperators, i. e.,
\begin{eqnarray}
  A \cdot & = & A_{a_1 \ldots a_m a_{1'} \ldots a_{m'}} J_{a_1} \ldots J_{a_m}
  J_{a_{1'}} \ldots J_{a_{m'}}, 
\end{eqnarray}
where we assume $r_{a_i} = L$ and $r_{a_{i'}} = R$, which can always be
achieved by a rearrangement of the field superoperators by virtue of the
anticommutation relation (\ref{eq:anticomm}). Using $L_T = T_{1 1'} J_1
J_{1'}$ and $L_0 J_1 = ( L_0 - x_1) J_1$ with $x_1 = \eta_1 \varepsilon_1$, we
can shift all field superoperators to the left. Since \ $\rho_0$ is an
eigenstate of the internal Liouvillian, that is, $L_0 \rho_0 = 0$, we can pull
$\rho_0$ also to the left, setting all $L_0$ to zero in the resolvents and
finally apply Wick's theorem Eq. (\ref{eq:Wick}). We obtain for the $n -
\mathrm{th}$ order contribution to the Laplace transform:
\begin{eqnarray}
  \Sigma_A^{\mathrm{irr}} ( z) |_n  & = & \sum_{\mathrm{contr} .
  , \{ k \}} A_{a_1 \ldots a_{m'}} T_{n n'} \ldots T_{1 1'}
  \nonumber\\
  &  & \times  \prod_{k = 1}^n \frac{1}{z + X_k}  ( - 1)^{N_p}  \prod_{i < j}
  \gamma_{i j} .  \label{eq:AnWick}
\end{eqnarray}
The sum includes all possible pair contractions and all indices $\{ k \} = \{
a_1, \ldots ., a_{m'} , n, n', \ldots ., 1, 1' \}$. The frequencies in
the propagators read
\begin{eqnarray}
  X_i & = & \sum_{j \leqslant i} ( \eta_j \varepsilon_j + \eta_{j'}
  \varepsilon_{j'}) .  \label{eq:Xi}
\end{eqnarray}
We represent the expressions contributing to {\Eq{eq:AnWick}} by diagrams as
follows: each tunneling amplitude $T_{i i'}$ is associated with a double
vertex (two dots) on a line. The line represents the free propagation of the
system, directed from right (earlier times) to left (later times), so that the
order of the vertices naturally corresponds to the order of the field
operators in the expression Eq. (\ref{eq:AnWick}). The left-most element is a
2m-vertex with 2m dots, which represents the m-particle observable $A$. As
usual {\cite{Schoeller09a}}, contractions are depicted by lines {\emph{above}}
the propagator line connecting two associated dots $i$ and $j$. Furthermore,
the electrode indices of the dots are fixed: all $i$'-indices belong to $r =
L$ (filled dots in {\Fig{fig:diagramExamples}}), whence all $i'$-indices
belong to $r = R$ (empty dots in {\Fig{fig:diagramExamples}}).

{
  \begin{figure}[h]
    \includegraphics[width=0.9\linewidth]{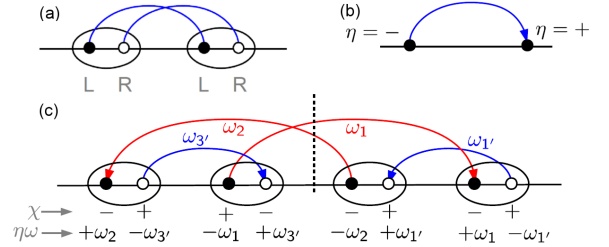}
    \caption{\label{fig:diagramExamples}Examples
for Liouville space diagrams: (a) two contracted tunneling Liouvillians, (b)
Hermitian conjugation indices are expressed by directions of arrows, (c)
reading off the propagator for a certain segment (for the segment indicated by
the vertical dashed line, we obtain $X_2 = \omega_1 - \omega_2$). Charge
indices $\chi$ are denoted by $+$ or $-$ below the propagator line, the
factors $\eta \omega$ appearing in {\Eq{eq:Xi}} are denoted are shown for the
sake of completeness.}
\end{figure}}

The factor $X_i$ of the $i$-th propagator segment can be readily read off from
the diagrams as illustrated in {\Fig{fig:diagramExamples}}: firstly, all
energies associated with contractions that connect dots that lie on the same
side of this segment do not contribute to $X_i$ (cf. Fig.
\ref{fig:diagramExamples} (c), grey contractions). For ``later'' dots this is
clear as they are always excluded in the sum of definition (\ref{eq:Xi}). For
``earlier'' dots the respective energy occurs twice in the sum with opposite
sign $\eta_i$ (enforced to the contraction {\eq{eq:contraction}}). Thus, only
the contraction lines that intersect a vertical line drawn at the $i$-th
propagator segment contribute to $X_i$ (red contractions in Fig.
\ref{fig:diagramExamples} (c)). To determine $X_i$ completely, we need to know
the h. c. indices $\eta_i$, which we will depict by arrows attached to the
contraction lines: if a vertex has $\eta_i = + ( -)$, the arrow points towards
(away from) this vertex (see {\Fig{fig:diagramExamples}} (b)).

Finally, we indicate the charge index $\chi = \pm$ of every dot in the diagram
by a sign below the propagator line (see {\Fig{fig:diagramExamples}} (c)).
Note that charge indices $\chi$ of a double vertex must be opposite by Eq.
(\ref{eq:tLT}). Our diagrams therefore represent the algebraic expression
associated with both a fixed contraction structure and {\emph{fixed}} $( \eta,
\chi)$-indices. The latter is a distinction to the Liouville space diagrams
used, e.g, in {\cite{Schoeller09a}}: distinct combinations of $( \eta, \chi)$
represent {\emph{distinct}} diagrams. Hence, the $n$-th order contribution
$\langle A \rangle_n$ to the expectation value of $A$ is represented by a set
of diagrams covering all allowed combinations contractions {\emph{and}} all
combinations for $( \eta, \chi)$'s.

For the ensuing discussion, we will first ignore any sign factors in $\langle
A \rangle_n ( z)$, for example due to the contraction functions. This
discussion and the question which diagrams are allowed is most conveniently
postponed to the very end of our derivations.

\subsection{Energy-Spin Separation}\label{sec:energySpinSep}

We now arrive at the crucial part of the derivation of {\emph{covariant}}
diagram rules: the separation of the spin-dependent and energy-dependent parts
in $\Sigma_A^{\mathrm{irr}} ( i 0)$ and recasting the former as a trace in spin
space similar to Sec. {\sec{sec:spinTraceTechnique}}. For the sake of
simplicity, we first replace the observable vertex by a tunneling vertex and
discuss modifications afterwards.

In Eq. (\ref{eq:AnWick}), we have three different spin-dependent factors: (i)
the energies $\varepsilon_i = \varepsilon_{n_i k_i \sigma_i}^{r_i}$, (ii) the
contractions $\gamma_{i j} \propto \delta_{\sigma_i \sigma_j}$ and (iii) the
tunneling amplitudes $t_{\sigma \sigma'}$.

(i)+(ii): Any pair of contracted indices, say $i$ and $j$ occur in \
{\Eq{eq:AnWick}} with a sum of the form
\begin{eqnarray}
  &  & \sum_{n_i, k_i} \delta_{\sigma_i \sigma_j} g ( \varepsilon_{n_i k_i
  \sigma_i}^{r_i}),  \label{eq:insertDOS}
\end{eqnarray}
where $g$ is some function of $\varepsilon^{r_i}_{n_i k_i \sigma_i} =
\varepsilon^{r_j}_{n_j k_j \sigma_j}$. We can get rid of the spin-dependence
of the energies by introducing the spin-dependent DOS, proceeding analogous to
the derivation of {\Eq{eq:covChargeSpin}}. {\Eq{eq:insertDOS}} then equals
\begin{eqnarray}
  &  & \int d \omega_i \bar{\nu}^{r_i} ( \omega_i) g ( \omega_i) \left[
  \text{}_{r_i} \langle \sigma_i | \check{\vec{r}} \cdot \check{\vec{n}}^{r_i}
  ( \omega_i) \ket{\sigma'}_{r_i} \right]^{\bar{\eta}_i} . 
\end{eqnarray}
The scalar product now involves a new 4-vector with $\check{\vec{n}} ( \omega)
= \sqrt{2} \left( 1, n^r ( \omega) \op{\vec{J}}^r \right)$. Furthermore, we
artificially introduced the complex conjugation by $\bar{\eta}$ even though
the matrix elements $\text{}_r \langle \sigma | \check{\vec{r}} \cdot
\check{\vec{n}}^r ( \omega) | \sigma' \rangle_r$ are real. This will become
advantageous for later manipulations. We therefore simply replace
$\varepsilon_i$ by $\omega_i$ in {\Eq{eq:AnWick}} and the sum over the index
$i$ now abbreviates also an integration over $\omega_i$ instead of a summation
over $n_i , k_i$.

(iii): The spin-dependence of the tunneling amplitudes can be rewritten as a
matrix element in spin space, too:
\begin{eqnarray}
  {}[ T^{L R}_{\sigma \sigma'}]^{\eta'} & = & \left[  \text{}_L \bra{\sigma} 
  \check{\vec{t}} \cdot \check{\vec{r}}  \ket{\sigma'}_R  \right]^{\eta'} . 
  \label{eq:t}
\end{eqnarray}
We next separate the frequency dependent parts (propagators, Fermi functions,
$\check{\vec{t}} $ and $\check{\vec{n}}$-vectors) and the
spin-dependent matrix elements, which leads to
\begin{eqnarray}
   A |_n & = & \sum_{\{ \kappa , \rho, \omega_i \}} (
  \mathrm{signs}) ( \mathrm{prop} .) \left[ \ldots \check{t}_{\kappa_i} \ldots
  \frac{( \check{F}_i)_{\rho_i}}{\pi} \ldots \right] \nonumber\\
  &  & \times \sum_{\{ \sigma \}} \left( \ldots (
  \check{r}_{\kappa_i})^{\eta_{i'}}_{\sigma_i \sigma_{i'}} \ldots (
  \check{r}_{\rho_i})^{\bar{\eta}_{\sigma_{k'}}}_{\sigma_k \sigma_{k'}} \ldots
  \right) .  \label{eq:Ansep}
\end{eqnarray}
The sum over the $\kappa_i$'s and $\rho_i$'s indicates the scalar products of
the respective 4-vectors with the matrix elements $(
\check{r}_{\mu})_{\sigma_i \sigma_{i'}}$ of the charge-spin operator
$\check{r}_{\mu}$. For simplicity, we do not indicate here the quantization
axis for the basis states used to express these matrix elements since we will
see below that this choice drops our. We moreover introduced the abbreviation
\begin{eqnarray}
  ( F_i)_{\rho_i} & = & \bar{\nu}^{r_i} ( \omega_i) \check{n}_{\rho_i}^{r_i} (
  \omega_i) f^{r_i }_{\bar{\chi}_i} ( \omega_i) . 
\end{eqnarray}
Interestingly, the sum over the spin indices factorizes into sums over the
spin indices of vertices that are connected in the diagrams in a {\emph{loop}}
(when formally considering contraction lines to be connected at each double
vertex, cf. {\Fig{fig:diagramExamples}} (c), which consists of a single loop).
The contribution of each loop can be evaluated independently as follows: if we
start at an arbitrary dot and follow the directed contraction line. For every
loop element we encounter (vertex, contraction), we put charge-spin operators
in a sequence from the right to the left into a trace. For example, starting
at the dot labelled 1' in {\Fig{fig:diagramExamples}} (c), we obtain the
expression $\mathrm{tr} ( \check{r}_{\kappa_{1'}} \check{r}_{\rho_1}
\check{r}_{\kappa_{3'}} \check{r}_{\rho_{3'}} \check{r}_{\kappa_{4'}}
\check{r}_{\rho_2} \check{r}_{\kappa_{2'}} \check{r}_{\rho_{1'}})$. Thus, Eq.
(\ref{eq:Ansep}) becomes
\begin{eqnarray}
   A |_n & \sim & ( \mathrm{signs}) \int_{\{ \omega_i \}} (
  \mathrm{propagators}) \nonumber\\
  &  & \times \prod_{\mathrm{loops}} \mathrm{tr} \left[ \ldots \left(
  \check{\vec{t}} \cdot \check{\vec{r}} \right) \ldots \left( \vec{F}_i \cdot
  \check{\vec{r}} \right) \ldots \right],  \label{eq:Antrace}
\end{eqnarray}
where we moved the frequency-dependent 4-vectors $\vec{t}$ and $\vec{F}_i$
into the spin traces and restored the scalar products.

The rest of this section is dedicated to prove this simple rule. We therefore
start from the sequence of matrix elements $\check{r}_{\sigma \sigma'}$,
obtained by writing them down from the right to the left in the order in which
we encounter down when following a directed loop. We have to prove that pairs
of the same spin indices are next to each other when we change all Hermitian
conjugation indices of the matrix elements in to in {\Eq{eq:Ansep}} to +
(except for the left- and right-most one), so that the expression can be
recast as a trace. We will suppress the $\mu$-indices for convenience and we
let $\sigma_{\mathrm{prec}}$ denote a spin index of the preceding element we
have already run through and $\sigma_{\mathrm{succ}}$ will refer to a spin index
of the succeeding loop element. We have to insert the following factors:

(i) For each contraction, put a factor
$\check{r}^{\bar{\eta}_{\mathrm{early}}}_{\sigma_{\mathrm{late}}
\sigma_{\mathrm{early}}}$ where $\eta_{\mathrm{early}}$ is the h. c. index of the
earlier vertex involved. We first note that the h. c. index $\eta_i$ of the
vertex $i$ at which we start to follow a contraction line has always $\eta_i =
-$ (the arrow points away from the vertex). We have to distinguish two cases:
(ia) if we start at the early vertex (e. g. contraction $1'$ in Fig.
\ref{fig:chargeSpinCurr}), we have $\eta_{\mathrm{early}} = -$ and thus
$\begin{array}{lll}
  \check{r}^{\bar{\eta}_{\mathrm{early}}}_{\sigma_{\mathrm{late}}
  \sigma_{\mathrm{early}}} & = & \check{r}^+_{\sigma_{\mathrm{succ}}
  \sigma_{\mathrm{prec}}}
\end{array}$whence in case (ib) we start at the late vertex (e. g. contraction
$1$ in Fig. \ref{fig:diagramExamples} (c)), so $\eta_{\mathrm{early}} = +$ and
thus $\begin{array}{lll}
  \check{r}^{\bar{\eta}_{\mathrm{early}}}_{\sigma_{\mathrm{late}}
  \sigma_{\mathrm{early}}} & = & \check{r}^-_{\sigma_{\mathrm{prec}}
  \sigma_{\mathrm{succ}}} = \check{r}^+_{\sigma_{\mathrm{succ}}
  \sigma_{\mathrm{prec}}}
\end{array}$, using the Hermiticity of the Pauli matrices.

(ii) For each double vertex, we put a factor
$\check{r}^{\eta_{\mathrm{early}}}_{\sigma_{\mathrm{late}}
\sigma_{\mathrm{early}}}$. Again, there two cases: (a) if $\eta_{\mathrm{early}} =
+$, we arrive at the earlier vertex, so we have $\begin{array}{lll}
  \check{r}^{\eta_{\mathrm{early}}}_{\sigma_{\mathrm{late}} \sigma_{\mathrm{early}}}
  & = & \check{r}^+_{\sigma_{\mathrm{succ}} \sigma_{\mathrm{prec}}}
\end{array}$ (e. g., later double vertex in Fig. \ref{fig:diagramExamples}
(c)) whence (b) if $\eta_{\mathrm{early}} = -$ (e. g. earlier vertex in Fig.
\ref{fig:diagramExamples} (c)), we arrive at the later vertex, so we have
$\begin{array}{lll}
  \check{r}^{\eta_{\mathrm{early}}}_{\sigma_{\mathrm{left}} \sigma_{\mathrm{right}}}
  & = & \check{r}^-_{\sigma_{\mathrm{prec}} \sigma_{\mathrm{succ}}} =
  \check{r}^+_{\sigma_{\mathrm{succ}} \sigma_{\mathrm{prec}}}
\end{array}$. These considerations prove the simple rule:
\begin{eqnarray}
  \sum_{\{ \sigma_i, \sigma_{i'} \}} \check{r}^+_{\sigma_n \sigma_{n'}} \ldots
  \check{r}^+_{\sigma_1 \sigma_{1'}} \check{r}^+_{\sigma_{1'} \sigma_n} & = &
  \mathrm{tr} ( \check{r} \ldots \check{r}) . 
\end{eqnarray}
Here we used that the matrix elements of adjacent Pauli operators are taken
for spin states with respect to the {\emph{same}} quantization axis, i. e., we
can combine $\sum_{\sigma_{1'}} \check{r}^+_{\sigma_1 \sigma_{1'}}
\check{r}^+_{\sigma_{1'} \sigma_n} = \text{}_{r_1} \langle \sigma_1 | \op{r} |
\sigma_{1'} \rangle_{r_1'}  \text{}_{r_1'} \langle \sigma_{1'} | \op{r} |
\sigma_n \rangle_{r_n} = \text{}_{r_1} \langle \sigma_1 | \op{r}  \op{r} |
\sigma_n \rangle_{r_n}$ and so on.

\subsection{Sign Factors}\label{sec:signs}

Having tackled the most tedious part, the spin and energy dependence, it
remains to collect all sign factors of an expression from its representing
diagram, yielding
\begin{eqnarray}
  \mathrm{signs} & = & ( - 1)^{\text{\#cr.}} \prod_{\substacktwo{\mathrm{early} +
  \mathrm{late}}{\mathrm{vertices}}} \chi_e 
  \prod_{\substacktwo{\mathrm{intermediate}}{\mathrm{vertices}}} \eta_e . 
  \label{eq:signs}
\end{eqnarray}
Here \#cr. is the number of crossing contractions lines and ``$e$'' always
refers to the earlier vertex of each double vertex. The meaning of early /
intermediate / late vertices is depicted in {\Fig{fig:signfactors}} and
explained in the proof of {\Eq{eq:signs}}.

To prove {\Eq{eq:signs}}, we first note that there are three origins for
signs: (i) an overall permutation factor $( - 1)^{N_p}$from Wick's theorem
{\eq{eq:Wick}}, (ii) a factor $( - \chi_{\mathrm{early}} \eta_{\mathrm{early}})$
for every contraction (see {\Eq{eq:contraction}}) and (iii) a factor $( -
\chi_{\mathrm{early}})$ for every double vertex \ (see {\Eq{eq:tLT}}). (i) is
readily obtained from the \ number of crossing contraction lines in the
diagrams, {{giving the first factor of {\Eq{eq:signs}}}}. The signs due to
(ii) and (iii) can be determined together: first of all, the minus signs can
be omitted since the number of contractions and double vertices is always
even. For the further procedure, it is helpful to distinguish three types of
vertices {{sketched in {\Fig{fig:signfactors}}}}: (a) ``late double
vertices'' where both vertices are the later ones in their contractions. Then
no signs due to contractions have to be considered and only the charge index
of the earlier vertex occurs. We furthermore have (b) ``early double
vertices'' where both vertices are the earlier ones in their contractions. The
signs due to {\emph{both}} contractions cancel each other since charge and h.
c. indices of both vertices in any double vertex are opposite. Thus, the sign
for this type of vertex is again given by the earlier charge index. Finally,
there are (c) ``intermediate double vertices'' where one of the vertices is
earlier and the other one is later in its contraction. The sign factor is in
this case the h. c. index $\eta_{\mathrm{early}}$ of the earlier vertex of the
double vertex.

{
  \begin{figure}[h]
    \includegraphics[width=0.9\linewidth]{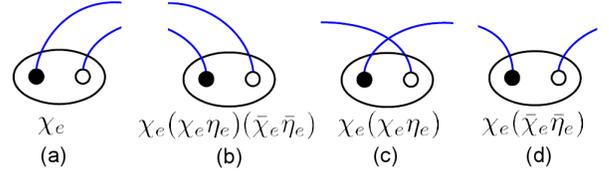}
    \caption{\label{fig:signfactors}Different types of
vertices: late double vertex (a), early double vertex (b) and intermediate
double vertices (c), (d). The corresponding sign factors are denoted below the
vertex. Note that $\chi_e, \eta_e$ refer to the earlier vertex of the double
vertex and {\emph{not}} to being earlier in the contractions.}
\end{figure}}

\subsection{Hermitian Conjugated Partner Diagrams}\label{sec:hermConj}

Since any observable must have a real expectation value, the imaginary parts
of all diagrams have to cancel each other. In fact diagrams come in complex
conjugates pairs, which are obtained from each other by inverting the h. c.
indices, or, diagrammatically speaking, by inverting all arrow directions.
Therefore, it is sufficient to take only {\emph{one}} diagram as a
{\emph{representative}} of both. We therefore obtain schematically
{{instead of {\Eq{eq:Antrace}}}}
\begin{eqnarray}
   A |_n & \sim & 2 \mathrm{Re} ( \mathrm{signs}) \int_{\{ \omega_i \}}
  ( \mathrm{propagators}) \nonumber\\
  &  & \times \prod_{\substacktwo{\mathrm{oriented}}{\mathrm{loops}}} \mathrm{tr}
  \left[ \ldots \left( \check{\vec{t}} \cdot \check{\vec{r}} \right) \ldots
  \left( \vec{F}_i \cdot \check{\vec{r}} \right) \ldots \right] . 
  \label{eq:AntraceHC}
\end{eqnarray}
We will next explicitly {{show how the complex-conjugation symmetry
arises}} in our formulation of real-time diagrammatics. Inverting all h. c.
{{indices}}, $\eta_i \rightarrow \bar{\eta}_i$, involves the following
modifications:

(i) The sign factor {{is negated}}. All charge indices may be kept fixed
except for {{those}} of the current vertex (see Eq.
(\ref{eq:replSpinFactor})). This involves an additional minus sign from the
early and late double vertices. The sign due to the intermediate vertices does
not change since the number $\# \mathrm{IV}$ of this type of vertices is always
even. Reverting all $\eta_e \rightarrow \bar{\eta}_e$ therefore results in an
additional factor $( - 1)^{\# \mathrm{IV}} = 1$ compared to the original sign
factor.

(ii) The order of the Pauli operators in the spin trace is inverted, yielding
the complex conjugate of the original trace expression:
\begin{eqnarray}
  \mathrm{tr} ( \check{r}_1 \ldots \check{r}_n) & \rightarrow & \mathrm{tr} (
  \check{r}_n \ldots \check{r}_1) = \mathrm{tr} ( \check{r}_1 \ldots
  \check{r}_n)^{\ast}, 
\end{eqnarray}
where we used $\check{r}_1^{\dag} = \check{r}_1$ and $( \check{r}_n^{\dag}
\ldots \check{r}_1^{\dag}) = ( \check{r}_1 \ldots \check{r}_n)^{\dag}$.

(iii) The propagators are mapped onto their negative Hermitian conjugate
since \ $X_i \rightarrow - X_i$
\begin{eqnarray}
  \frac{1}{i 0 + X_i} & \rightarrow & \frac{1}{i 0 + ( - X_i)} = - \left(
  \frac{1}{i 0 + X_i} \right)^{\ast} . 
\end{eqnarray}
As the number of propagators is odd, we obtain an additional minus sign.

Since all remaining factors are real (cf. Eq. (\ref{eq:Antrace})) and have no
further $\eta$-dependence, we conclude that the mapping $\eta_i \rightarrow
\bar{\eta}_i$ results in the {\emph{positive}} complex conjugate contribution
of the initial expression. Adding both these partner diagrams gives two times
the real part of one of the two diagrams, proving {\Eq{eq:AntraceHC}}

\subsection{Covariant Diagram Rules for Charge and Spin
Current}\label{sec:diagramRules}

{{We first describe the modifications of the covariant diagram rules
required for the charge and spin current, which can be treated simultaneously
as outlined in {\sec{sec:NSnetwork}}.}} The expressions for the current
operator read (see {\Eq{eq:TCurrent}}):
\begin{eqnarray}
  I_{R^r_{\mu}} & = & \sum_{n n' k k' \sigma \sigma'} ( - i r_{\mu} T)^{r
  \bar{r}}_{\sigma \sigma'} c^{\dag}_{r n k \sigma} c_{\bar{r} n' k' \sigma'}
  - \text{h.c.}  \label{eq:TCurrentApp}
\end{eqnarray}
with $( r T)_{\sigma \sigma'} = \sum_{\tau} r_{\sigma \tau} T^{r
\bar{r}}_{\tau \sigma'}$. The structure of the current is very similar \ to
the tunneling Hamiltonian with the simple replacement $T_{\sigma \sigma'}^{r
\bar{r}} \rightarrow ( - i r_{\mu} T)^{r \bar{r}}_{\sigma \sigma'}$.
Introducing a superoperator for the left action $\mathcal{I}_{R^r_{\mu}} \cdot
= I_{R^r_{\mu}} \cdot$, we may therefore treat the corresponding current
vertices similar to the tunneling Liouvillian by making the replacement
\begin{eqnarray}
  &  & \bar{\chi}_{1'} \left[ \text{}_L \langle \sigma_1 | \check{\vec{t}}
  \cdot \check{\vec{r}} | \sigma_{1'} \rangle_R \right]^{\eta_{1'}}
  \nonumber\\
  & \rightarrow & \delta_{\bar{\chi}_{1'} \eta_{1'}, +}  [ ( - i) \text{}_L
  \langle \sigma_1 | i_{R^r_{\mu}} | \sigma_{1'} \rangle_R]^{\eta_{1'}} 
  \label{eq:replSpinFactor}
\end{eqnarray}
with $i_{R^r_{\mu}}$ given by {\Eq{eq:iRmu}} below. The restriction
$\delta_{\bar{\chi}_{1'} \eta_{1'}, +}$ is due to the fact that all field
operators in $\mathcal{I}_{R^r_{\mu}}$ act from the left whereas the tunneling
Liouvillian as a commutator also possesses a part that acts from the right.
Consequently, we have account for the following modifications:

(i) Insert $i_{R^r_{\mu}}$ instead of $\left( \check{\vec{t}} \cdot
\check{\vec{r}} \right)$ into the spin traces.

(ii) Set $\eta_{1'} = \bar{\chi}_{1'} = +$. The first equality follows from
the factor $\delta_{\bar{\chi}_{1'} \eta_{1'}, +}$ in {\Eq{eq:replSpinFactor}}
and the second from the freedom to chose one representative of the complex
conjugated diagrams, which we fix by an explicit choice for $\eta_{1'}$. Note
that the ``missing'' factor $\bar{\chi}_{1'} = +$ in the second line of
{\Eq{eq:replSpinFactor}} does give any further modifications for this choice.

(iii) Replace $\mathrm{Re} ( \ldots) \rightarrow \mathrm{Im} ( \ldots)$ in
{\Eq{eq:AntraceHC}}, which is due to the additional factor $( - i)^{\eta_{1'}
= +} = - i$ in {\Eq{eq:replSpinFactor}}.

{{We now present}} a systematic way to draw all diagrams and read off the
respective $n$-th order contributions to the expectation value $
I_{R^r_{\mu}} |_n$, schematically given by

{\wideeq{\begin{eqnarray}
   I_{R^r_{\mu}} |_n & \sim & \underset{\substacktwo{\mathrm{irr} .
  \mathrm{contr} .}{\{ \chi, \eta \}}}{\sum} 2 \mathrm{Im} \left\{ ( \mathrm{signs})
  \int_{\{ \omega_i \}} \left( \text{...} \frac{1}{i 0 + X_i} \ldots \right) 
  \prod_{\mathrm{loops}} \mathrm{tr} \left[ \ldots \left( \check{\vec{t}} \cdot
  \check{\vec{r}} \right) \ldots \left( \vec{F}_i \cdot \check{\vec{r}}
  \right) \ldots \right] \right\} .  \label{eq:Anreal}
\end{eqnarray}}}

The first sum in Eq. (\ref{eq:Anreal}) adds up all possible diagrams, whereas
the residual term corresponds to an {{individual}} diagram. All allowed
diagrams contributing to $ I_{R^r_{\mu}} |_n$ are obtained by the
following drawing instructions (an easy example is given in Fig.
\ref{fig:diagramExamples} (c)).

1. \textbf{Propagator and vertices.} For the $n$-th order contribution, put
$n$ double vertices on a propagator line. The later (earlier) vertex of each
double vertex refers to the left (right) electrode. Mark the latest vertex
with the symbol $I_{R^r_{\mu}}$ designating this one to be the observable
vertex. {\emph{Symbol:}} a horizontal line represents the propagator line;
double dots, encircled by a line depict the double vertices; full (empty) dots
refer to the left (right) electrode.

2. \textbf{Contractions.} Construct all possible irreducible contractions.
Full (empty) vertices are only permitted to be contracted with full (empty)
vertices. {\emph{Symbol: }}contractions depicted full lines above the
propagator; attach an ``$i$'', denoting the Liouville index of the
{\emph{earlier}} dot in each contraction.

3. \textbf{Hermitian Conjugation Indices.} Construct all possibilities for
the choice of the h. c. indices, obeying the following rules: (i) the h. c. of
one double vertex have to be opposite, (ii) the earlier h. c. index of the
observable vertex is +, (iii) the h. c. indices have to alternate in each
loop. {\emph{Symbol}}: Arrow pointing to (away from) a dot is associated with
$\eta = +$ ($\eta = -$).

4. \textbf{Charge indices.} Construct all possible charge index
arrangements, restricted by (i) the charge indices of a double vertex have to
be opposite, (ii) the charge indices of \ the observable vertex have to be
opposite to the h. c. indices. {\emph{Symbol}}: +,-- below the vertex

Translating diagrams into algebraic expressions, schematically indicated in
{\Eq{eq:Anreal}}, proceeds as follows:

1. \textbf{Propagators}: For each propagator segment following a
{\emph{tunneling}} vertex write down a factor
\begin{eqnarray}
  \frac{1}{i 0 + X_{}} & \text{with} & X = \sum ( \omega_{\mathrm{early}} -
  \omega_{\mathrm{late}}) . 
\end{eqnarray}
The sum in $X$ involves all frequencies associated with contractions that
cross over that segment from the left to the right (earlier frequency) or from
the right to the left (later frequency), respectively.

2. \textbf{Loopwise evaluation of spin traces}: Start at any point and
follow the loop in direction of the arrow. Insert from the right to the left
into a trace in spin space the following terms when encountering

- a tunneling vertex: $\check{\vec{t}} \cdot \check{\vec{r}}$,

- an observable vertex:
\begin{eqnarray}
  i_{R^r_{\mu}} & = & \left\{ \begin{array}{ll}
     - \left( \check{\vec{t}} \cdot \check{\vec{r}} \right) r_{\mu}, &
    r = R\\
    + r_{\mu} \left( \check{\vec{t}} \cdot \check{\vec{r}} \right), & r = L
  \end{array} .  \label{eq:iRmu} \right.
\end{eqnarray}

- a contraction: $\vec{F}_i \cdot \check{\vec{r}}$ where $i$ refers to the
earlier dot, $( F_i)_{\rho_i} = \bar{\nu}^{r_i} ( \omega_i) f^{r_i
}_{\bar{\chi}_i} ( \omega_i) \check{n}^{r_i}_{\rho_i} ( \omega_i)$ and
\begin{equation}
  \check{n}^r_{\rho} = \sqrt{2} \left\{ \begin{array}{ll}
    1 & \rho = 0\\
    n^r \op{J}^r_{\rho} & \rho = 1, 2, 3
  \end{array} . \right.
\end{equation}
Repeat this for all other loops in the diagram.

3. \textbf{Signs}: Put a factor
\begin{eqnarray*}
  & ( - 1)^{\text{\#cr.}} \underset{\substacktwo{\mathrm{early} +
  \mathrm{late}}{\mathrm{vertices}}}{\prod} \chi_e 
  \underset{\substacktwo{\mathrm{intermediate}}{\mathrm{vertices}}}{\prod} \eta_e,
  & 
\end{eqnarray*}
where $\# \mathrm{cr}$. is the number of crossings of contractions, the
subscript ``$e$'' refers to the earlier dot of a vertex. For the meaning of
late, early and intermediate double vertices, see Fig. \ref{fig:signfactors}.

4. \textbf{Complete expectation value}: Multiply all expressions obtained
from steps 1, 2 and 3 and integrate over all frequencies $\{ \omega_i \}$,
where $i$ refers to the indices associated with the contractions. Take $2
\mathrm{Im} ( \ldots)$ of this expression and sum up the contributions of all
valid diagrams.

\subsection{Covariant Diagram Rules for Spin-quadrupole
Current}\label{sec:diagramRulesSQM}

The calculation of the expectation value of the SQM current requires some
modifications compared to that of the charge and spin-dipole current. The
reason is that the spin-quadrupole current operator for node $\langle r r'
\rangle$, i. e. $\tens{I}^{r r'}_{\tens{Q}} = i \left[ H_T, \tens{Q}^{r r'}
\right]$, is a dyadic of two vector operators:
\begin{eqnarray}
  & \tens{I}^{r r'}_{\tens{Q}} = \frac{1}{2} \left[ g^{r r'} 
  \vec{I}_{\vec{S}^r} \odot \vec{S}^{r'} + ( r \leftrightarrow r') + h.c.
  \right] . &  \label{eq:QCurrent}
\end{eqnarray}
We remind the reader of the factor $g^{r r'} = 2$ if $r \neq r'$ and $g^{r r}
= 1$. Crudely speaking, the diagrams for the spin-quadrupole current have the
same structure as for the spin current, but include an additional spin vertex.
However, the spin operator has to be treated differently from the tunneling or
current vertices {\eq{eq:TCurrentApp}} as is contains only {\emph{one}} sum
over the $k$-modes (see {\Eq{eq:chargespinOp}}) instead of two. We will
discuss this point later in detail.

We will restrict ourselves in the following to spin-{\emph{independent}}
tunneling. If and only if this is the case, we may replace
$\vec{I}_{\vec{S}}^r$ by $2 \vec{I}^{r \bar{r}}_{\vec{S}^r}$ where
\begin{eqnarray}
  \vec{I}_{\vec{S}^r}^{r \bar{r}} & = & \sum_{k, k', \sigma, \sigma'} \left[ -
  i \text{}_r \langle \sigma | \vec{i}_{\vec{S}^r} | \sigma' \rangle_{\bar{r}}
  \right] c_{r k \sigma}^{\dag} c_{\bar{r} k' \sigma'} 
\end{eqnarray}
with $\bar{r} = R, L$ for $r = L, R$ and $\vec{i}_{\vec{S}^r}$ defined in
{\Eq{eq:iRmu}} for $\mu = x, y, z$. Note that $\vec{I}_{\vec{S}^r}^{r
\bar{r}}$ only accounts for the contribution to the spin current for electrode
$r$ due to tunneling from subsystem $\bar{r}$ to $r$, but not for tunneling
from $r$ to $\bar{r}$. Using Eq. (\ref{eq:QCurrent}), the expectation value of
the spin-quadrupole current is then given by
\begin{eqnarray}
  \brkt{\tens{I}^{r r'}_{\tens{Q}}} & = & 2 g^{r r'} \mathrm{Re} \, \mathrm{tr}
  \left( \vec{I}_{\vec{S}^r}^{r \bar{r}} \odot \vec{S}^{r'} \right) \rho ( t)
  + ( r \leftrightarrow r') . \,  \label{eq:IQExp}
\end{eqnarray}
In the following, we will discuss the modifications that are necessary to
calculate Eq. (\ref{eq:IQExp}) starting from the spin current. First of all,
$\tens{I}_{\tens{Q}^{r r'}}$ contains four field operators so that the
associated observable vertex is a quadruple vertex instead of a double vertex.
This is sketched in Fig. \ref{fig:SQMvertex}: the later two vertices refer to
the spin current operator (whose dots refer to {\emph{two}} electrodes) and
the earlier vertices refer to the spin operator (whose dots refer to only
{\emph{one}} electrode). As all field operators act from the left, all charge
indices are opposite to the h. c. indices. The action of the spin operator can
be expressed by
\begin{eqnarray}
  S_j & \sim & \delta_{\eta_{1'}, +} \delta_{- \eta_{1'} \chi_{1'}, +}
  \delta_{\chi_1, \bar{\chi}_{1'}} \delta_{\eta_1, \bar{\eta}_{1'}}
  \delta_{k_1 k_{1'}} ( s^r_j)_{\sigma_1 \sigma_{1'}} J_1 J_{1'} . \nonumber\\
  &  &  \label{eq:spinVertex}
\end{eqnarray}
This fixes $\eta_{\mathrm{early}} = \bar{\chi}_{\mathrm{early}} = +$.

{
  \begin{figure}[h]
    \includegraphics[width=0.9\linewidth]{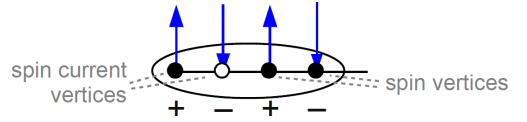}
    \caption{\label{fig:SQMvertex}SQM current vertex}
\end{figure}}

The technical challenge for implementing the spin-quadrupole current vertex is
the single $k$-summation in the spin operator. If we contract the two dots
within the spin operator this is no problem since the contraction also sets
the $k$-indices to be the same anyway. However, if the two dots are contracted
with other vertices, their $k$'s are not independent, so we cannot introduce
the one-particle DOS for both contractions individually (denoted by $1
$ and $1'$ in the following). Instead, we obtain an expression of the
form:
\begin{equation}
  \sum_{\tau_1 \tau_{1'}} \sum_{k_1 } \delta_{\sigma_1 \tau_{1'}}
  \delta_{\sigma_{1'} \tau_1} g ( \varepsilon^{r'}_{n_1 k_1 \tau_{1'}},
  \varepsilon^{r'}_{n_1 k_1 \tau_1})  \text{}^{r'} \bra{\sigma_1}_{} \vec{s}
  \ket{\sigma_{1'}}^{r'}, \label{eq:beforetwoDOS}
\end{equation}
where $g$ is some function and the Cartesian index $j$ refers to the spin
operator. We may treat this expression similar to the calculation of the
exchange contribution to the spin-quadrupole moment in equilibrium (see Sec.
{\sec{sec:DOSModel}}) by introducing the two-particle density of states
$\nu^r_{\sigma_1 \sigma_{1'}} ( \omega_1, \omega_{1'})$ {\eq{eq:twoDOS}} and
expressing the latter by {\Eq{eq:twoparticle}}. Then the term in
(\ref{eq:beforetwoDOS}) equals
\begin{eqnarray}
  &  & \int_{\{ \omega_1, \omega_{1'} \}} g ( \omega_1, \omega_{1'}) 2
  \tens{A}^{r'}_{\mu \nu}  \text{}_{r'} \bra{\sigma_1}  \check{r}_{\mu}
  \vec{s} \check{r}_{\nu} \ket{\sigma_{1'}} \text{}_r, 
\end{eqnarray}
where the 2DOS {\eq{eq:twoDOS}} enters through the matrix $\tens{A}^r_{\mu
\nu}$ given by {\eq{eq:A00}}-{\eq{eq:Aij}}. Consequently, the frequency and
energy part of the diagrams can be treated without modification as explained
in Sec. {\sec{sec:energySpinSep}}. Furthermore, the signs can be treated
without modifications as well as we can simply omit the factor
$\bar{\chi}_{1'} = +$ in Eq. (\ref{eq:spinVertex}). In contrast, the spin-part
is altered: we now insert for the spin vertex and the two associated
contractions the factor $\sqrt{2} \mathcal{A}^r_{\mu \nu} \check{r}_{\mu} s_j
\check{r}_{\nu}$ into the spin trace.

Finally, we mention that the symmetrization in $r \leftrightarrow r'$ and
multiplying with $g^{r r'}$in {\Eq{eq:IQExp}} is equivalent to multiplying
with a factor of $2$ and symmetrizing in $r \leftrightarrow r'$ only for $r
\neq r'$.

We now summarize the modifications of the diagram rules for the SQM current
compared to the spin current. We first state the drawing rules:

1. \textbf{Propagator and vertices:} the observable vertex consists of four
vertices, the later two depicting $\vec{I}_{\vec{S}^r}$ and the earlier two
depicting $\vec{S}^{r'}$.

2. \textbf{Contractions.} No changes.

3. \textbf{Indices.} The h. c. indices of the observable vertex are fixed to
(--,+,--,+) and the charge indices are opposite to this.

The changes in the rules for translation into algebraic expressions are:

1. \textbf{Propagators.} No changes.

2. \textbf{Evaluation of loops.} If the vertices associated with
$\vec{S}^{r'}$ in Eq. (\ref{eq:IQExp}) are

(a) contracted with eachoter, then insert $\vec{F}_+^{r'} ( \omega_{1'}) \cdot
\check{\vec{r}}  \vec{s}$.

(b) contracted with other vertices, then insert \ $2 t^2 f^{r_1
}_{\bar{\chi}_1} ( \omega_1) f^{r_{1'} }_{\bar{\chi}_1'} (
\omega_{1'}) \tens{A}^{r'}_{\mu \nu} \check{r}_{\mu} \vec{s} 
\check{r}_{\nu}$.

This accounts for both the vertices and their contractions. Here 1 (1') refers
to the indices of the contractions of the later (earlier) vertex.

3. \textbf{Signs.} For the evaluation of the signs, treat the SQM current
vertex as two double vertices.

4. \textbf{Complete expectation value.} Multiply with a factor of $2$ and
retain the symmetric and traceless part of the tensor-valued result $\sim
\vec{i}_{\vec{S}^r} \otimes \vec{s}$. If $r \neq r'$, symmetrize the result in
$r \leftrightarrow r'$.

\bibliographystyle{apsrev}

\begin{thebibliography}{28}
\expandafter\ifx\csname natexlab\endcsname\relax\def\natexlab#1{#1}\fi
\expandafter\ifx\csname bibnamefont\endcsname\relax
  \def\bibnamefont#1{#1}\fi
\expandafter\ifx\csname bibfnamefont\endcsname\relax
  \def\bibfnamefont#1{#1}\fi
\expandafter\ifx\csname citenamefont\endcsname\relax
  \def\citenamefont#1{#1}\fi
\expandafter\ifx\csname url\endcsname\relax
  \def\url#1{\texttt{#1}}\fi
\expandafter\ifx\csname urlprefix\endcsname\relax\def\urlprefix{URL }\fi
\providecommand{\bibinfo}[2]{#2}
\providecommand{\eprint}[2][]{\url{#2}}

\bibitem[{\citenamefont{{M. Julliere}}(1975)}]{Julliere75}
\bibinfo{author}{\bibnamefont{{M. Julliere}}}, \bibinfo{journal}{Phys. Lett. A}
  \textbf{\bibinfo{volume}{54}}, \bibinfo{pages}{225} (\bibinfo{year}{1975}).

\bibitem[{\citenamefont{{J. C. Slonczewski}}(1989)}]{Slonczewski89}
\bibinfo{author}{\bibnamefont{{J. C. Slonczewski}}}, \bibinfo{journal}{Phys.
  Rev. B} \textbf{\bibinfo{volume}{39}}, \bibinfo{pages}{6995}
  (\bibinfo{year}{1989}).

\bibitem[{\citenamefont{Potok et~al.}(2002)\citenamefont{Potok, Folk, Marcus,
  and Umansky}}]{Potok02}
\bibinfo{author}{\bibfnamefont{R.~M.} \bibnamefont{Potok}},
  \bibinfo{author}{\bibfnamefont{J.~A.} \bibnamefont{Folk}},
  \bibinfo{author}{\bibfnamefont{C.~M.} \bibnamefont{Marcus}},
  \bibnamefont{and} \bibinfo{author}{\bibfnamefont{V.}~\bibnamefont{Umansky}},
  \bibinfo{journal}{Phys. Rev. Lett.} \textbf{\bibinfo{volume}{89}},
  \bibinfo{pages}{266602} (\bibinfo{year}{2002}).

\bibitem[{\citenamefont{Braun et~al.}(2005)\citenamefont{Braun, K{\"o}nig, and
  Martinek}}]{Braun05}
\bibinfo{author}{\bibfnamefont{M.}~\bibnamefont{Braun}},
  \bibinfo{author}{\bibfnamefont{J.}~\bibnamefont{K{\"o}nig}},
  \bibnamefont{and} \bibinfo{author}{\bibfnamefont{J.}~\bibnamefont{Martinek}},
  \bibinfo{journal}{Superlatt. and Microstruct.} \textbf{\bibinfo{volume}{37}},
  \bibinfo{pages}{333} (\bibinfo{year}{2005}).

\bibitem[{\citenamefont{Schmidt}(2005)}]{Schmidt05}
\bibinfo{author}{\bibfnamefont{G.}~\bibnamefont{Schmidt}},
  \bibinfo{journal}{Journal of Physics D: Applied Physics}
  \textbf{\bibinfo{volume}{38}} (\bibinfo{year}{2005}).

\bibitem[{\citenamefont{Glazman and Raikh}(1988)}]{Glazman88}
\bibinfo{author}{\bibfnamefont{L.~I.} \bibnamefont{Glazman}} \bibnamefont{and}
  \bibinfo{author}{\bibfnamefont{M.~E.} \bibnamefont{Raikh}},
  \bibinfo{journal}{JETP\ Lett.} \textbf{\bibinfo{volume}{47}},
  \bibinfo{pages}{452} (\bibinfo{year}{1988}), \bibinfo{note}{[Pis'ma v ZhETF
  \textbf{47}, 378 (1988)]}.

\bibitem[{\citenamefont{Goldhaber-Gordon
  et~al.}(1998)\citenamefont{Goldhaber-Gordon, Shtrikman, Mahalu,
  Abusch-Magder, Meirav, and Kastner}}]{Goldhaber98}
\bibinfo{author}{\bibfnamefont{D.}~\bibnamefont{Goldhaber-Gordon}},
  \bibinfo{author}{\bibfnamefont{H.}~\bibnamefont{Shtrikman}},
  \bibinfo{author}{\bibfnamefont{D.}~\bibnamefont{Mahalu}},
  \bibinfo{author}{\bibfnamefont{D.}~\bibnamefont{Abusch-Magder}},
  \bibinfo{author}{\bibfnamefont{U.}~\bibnamefont{Meirav}}, \bibnamefont{and}
  \bibinfo{author}{\bibfnamefont{M.~A.} \bibnamefont{Kastner}},
  \bibinfo{journal}{Nature} \textbf{\bibinfo{volume}{391}},
  \bibinfo{pages}{156} (\bibinfo{year}{1998}).

\bibitem[{\citenamefont{{K. Ono} et~al.}(2002)\citenamefont{{K. Ono}, {D. G.
  Austing}, {Y. Tokura}, and {S. Tarucha}}}]{Ono02}
\bibinfo{author}{\bibnamefont{{K. Ono}}}, \bibinfo{author}{\bibnamefont{{D. G.
  Austing}}}, \bibinfo{author}{\bibnamefont{{Y. Tokura}}}, \bibnamefont{and}
  \bibinfo{author}{\bibnamefont{{S. Tarucha}}}, \bibinfo{journal}{Science}
  \textbf{\bibinfo{volume}{297}}, \bibinfo{pages}{1313} (\bibinfo{year}{2002}).

\bibitem[{\citenamefont{Weinmann et~al.}(1994)\citenamefont{Weinmann,
  H\"ausler, Pfaff, Kramer, and Weiss}}]{Weinmann94}
\bibinfo{author}{\bibfnamefont{D.}~\bibnamefont{Weinmann}},
  \bibinfo{author}{\bibfnamefont{W.}~\bibnamefont{H\"ausler}},
  \bibinfo{author}{\bibfnamefont{W.}~\bibnamefont{Pfaff}},
  \bibinfo{author}{\bibfnamefont{B.}~\bibnamefont{Kramer}}, \bibnamefont{and}
  \bibinfo{author}{\bibfnamefont{U.}~\bibnamefont{Weiss}},
  \bibinfo{journal}{Eur. Phys. Lett.} \textbf{\bibinfo{volume}{26}},
  \bibinfo{pages}{467} (\bibinfo{year}{1994}).

\bibitem[{\citenamefont{Chye et~al.}(2002)\citenamefont{Chye, White,
  Johnston-Halperin, Gerardot, Awschalom, and Petroff}}]{Chye02}
\bibinfo{author}{\bibfnamefont{Y.}~\bibnamefont{Chye}},
  \bibinfo{author}{\bibfnamefont{M.~E.} \bibnamefont{White}},
  \bibinfo{author}{\bibfnamefont{E.}~\bibnamefont{Johnston-Halperin}},
  \bibinfo{author}{\bibfnamefont{B.~D.} \bibnamefont{Gerardot}},
  \bibinfo{author}{\bibfnamefont{D.~D.} \bibnamefont{Awschalom}},
  \bibnamefont{and} \bibinfo{author}{\bibfnamefont{P.~M.}
  \bibnamefont{Petroff}}, \bibinfo{journal}{Phys. Rev. B}
  \textbf{\bibinfo{volume}{66}}, \bibinfo{pages}{201301}
  (\bibinfo{year}{2002}).

\bibitem[{\citenamefont{Braun et~al.}(2004)\citenamefont{Braun, K{\"o}nig, and
  Martinek}}]{Braun04set}
\bibinfo{author}{\bibfnamefont{M.}~\bibnamefont{Braun}},
  \bibinfo{author}{\bibfnamefont{J.}~\bibnamefont{K{\"o}nig}},
  \bibnamefont{and} \bibinfo{author}{\bibfnamefont{J.}~\bibnamefont{Martinek}},
  \bibinfo{journal}{Phys. Rev. B} \textbf{\bibinfo{volume}{70}},
  \bibinfo{pages}{195345} (\bibinfo{year}{2004}).

\bibitem[{\citenamefont{D.~Gatteschi and Villain}(2006)}]{Gatteschi}
\bibinfo{author}{\bibfnamefont{R.~S.} \bibnamefont{D.~Gatteschi}}
  \bibnamefont{and} \bibinfo{author}{\bibfnamefont{J.}~\bibnamefont{Villain}},
  \emph{\bibinfo{title}{Molecular Nanomagnets}} (\bibinfo{publisher}{OUP
  Oxford}, \bibinfo{year}{2006}).

\bibitem[{\citenamefont{Brune and Gambardella}(2009)}]{Brune09}
\bibinfo{author}{\bibfnamefont{H.}~\bibnamefont{Brune}} \bibnamefont{and}
  \bibinfo{author}{\bibfnamefont{P.}~\bibnamefont{Gambardella}},
  \bibinfo{journal}{Surface Science} \textbf{\bibinfo{volume}{603}},
  \bibinfo{pages}{1812} (\bibinfo{year}{2009}).

\bibitem[{\citenamefont{Sothmann and K\"onig}(2010)}]{Sothmann10}
\bibinfo{author}{\bibfnamefont{B.}~\bibnamefont{Sothmann}} \bibnamefont{and}
  \bibinfo{author}{\bibfnamefont{J.}~\bibnamefont{K\"onig}},
  \bibinfo{journal}{Phys.\ Rev.\ B} \textbf{\bibinfo{volume}{82}},
  \bibinfo{pages}{245319} (\bibinfo{year}{2010}).

\bibitem[{\citenamefont{Baumg\"artel et~al.}(2011)\citenamefont{Baumg\"artel,
  Hell, Das, and Wegewijs}}]{Baumgaertel11}
\bibinfo{author}{\bibfnamefont{M. M. E.}~\bibnamefont{Baumg\"artel}},
  \bibinfo{author}{\bibfnamefont{M.}~\bibnamefont{Hell}},
  \bibinfo{author}{\bibfnamefont{S.}~\bibnamefont{Das}}, \bibnamefont{and}
  \bibinfo{author}{\bibfnamefont{M.~R.} \bibnamefont{Wegewijs}},
  \bibinfo{journal}{Phys.\ Rev.\ Lett.} \textbf{\bibinfo{volume}{107}},
  \bibinfo{pages}{087202} (\bibinfo{year}{2011}).

\bibitem[{\citenamefont{Misiorny
  et~al.}(2012{\natexlab{a}})\citenamefont{Misiorny, Hell, and
  Wegewijs}}]{Misiorny13}
\bibinfo{author}{\bibfnamefont{M.}~\bibnamefont{Misiorny}},
  \bibinfo{author}{\bibfnamefont{M.}~\bibnamefont{Hell}}, \bibnamefont{and}
  \bibinfo{author}{\bibfnamefont{M.~R.} \bibnamefont{Wegewijs}},
  \bibinfo{journal}{submitted}  (\bibinfo{year}{2012}{\natexlab{a}}).

\bibitem[{\citenamefont{Misiorny
  et~al.}(2012{\natexlab{b}})\citenamefont{Misiorny, Weymann, and
  Barna{\'s}}}]{Misiorny12a}
\bibinfo{author}{\bibfnamefont{M.}~\bibnamefont{Misiorny}},
  \bibinfo{author}{\bibfnamefont{I.}~\bibnamefont{Weymann}}, \bibnamefont{and}
  \bibinfo{author}{\bibfnamefont{J.}~\bibnamefont{Barna{\'s}}},
  \bibinfo{journal}{Phys.\ Rev.\ B} \textbf{\bibinfo{volume}{86}},
  \bibinfo{pages}{245415} (\bibinfo{year}{2012}{\natexlab{b}}).

\bibitem[{\citenamefont{Cottingham and Greenwood}(2001)}]{Cottingham}
\bibinfo{author}{\bibfnamefont{W.~N.} \bibnamefont{Cottingham}}
  \bibnamefont{and} \bibinfo{author}{\bibfnamefont{D.~A.}
  \bibnamefont{Greenwood}}, \emph{\bibinfo{title}{An Introduction to Nuclear
  Physics}}, vol.~\bibinfo{volume}{1} (\bibinfo{publisher}{Cambridge University
  Press}, \bibinfo{year}{2001}).

\bibitem[{\citenamefont{Hell and Wegewijs}(2013)}]{Hell13c}
\bibinfo{author}{\bibfnamefont{M.}~\bibnamefont{Hell}} \bibnamefont{and}
  \bibinfo{author}{\bibfnamefont{M.~R.} \bibnamefont{Wegewijs}},
  \bibinfo{journal}{in preparation}  (\bibinfo{year}{2013}).

\bibitem[{\citenamefont{Gatteschi and Sessoli}(2003)}]{Gatteschi03rev}
\bibinfo{author}{\bibfnamefont{D.}~\bibnamefont{Gatteschi}} \bibnamefont{and}
  \bibinfo{author}{\bibfnamefont{R.}~\bibnamefont{Sessoli}},
  \bibinfo{journal}{Angew. Chem. Int. Ed.} \textbf{\bibinfo{volume}{42}},
  \bibinfo{pages}{268} (\bibinfo{year}{2003}).

\bibitem[{\citenamefont{{J. K\"onig} and {J. Martinek}}(2003)}]{Koenig03}
\bibinfo{author}{\bibnamefont{{J. K\"onig}}} \bibnamefont{and}
  \bibinfo{author}{\bibnamefont{{J. Martinek}}}, \bibinfo{journal}{Phys. Rev.
  Lett.} \textbf{\bibinfo{volume}{90}}, \bibinfo{pages}{166602}
  (\bibinfo{year}{2003}).

\bibitem[{\citenamefont{Misiorny et~al.}(2011)\citenamefont{Misiorny, Weymann,
  and Barna{\'s}}}]{Misiorny11a}
\bibinfo{author}{\bibfnamefont{M.}~\bibnamefont{Misiorny}},
  \bibinfo{author}{\bibfnamefont{I.}~\bibnamefont{Weymann}}, \bibnamefont{and}
  \bibinfo{author}{\bibfnamefont{J.}~\bibnamefont{Barna{\'s}}},
  \bibinfo{journal}{Phys. Rev. Lett.} \textbf{\bibinfo{volume}{106}},
  \bibinfo{pages}{126602} (\bibinfo{year}{2011}).

\bibitem[{\citenamefont{May et~al.}(2011)\citenamefont{May, Wegewijs, and
  Hofstetter}}]{May11}
\bibinfo{author}{\bibfnamefont{F.}~\bibnamefont{May}},
  \bibinfo{author}{\bibfnamefont{M.~R.} \bibnamefont{Wegewijs}},
  \bibnamefont{and}
  \bibinfo{author}{\bibfnamefont{W.}~\bibnamefont{Hofstetter}},
  \bibinfo{journal}{Beilstein J. Nanotechnol.} \textbf{\bibinfo{volume}{2}},
  \bibinfo{pages}{693} (\bibinfo{year}{2011}).

\bibitem[{\citenamefont{Lindell}(1992)}]{Lindellbook}
\bibinfo{author}{\bibfnamefont{I.~V.} \bibnamefont{Lindell}},
  \emph{\bibinfo{title}{Methods for electromagnetic field analyisis}}, IEEE
  Series on Electromagnetic Wave Theory (\bibinfo{publisher}{IEEE Press},
  \bibinfo{year}{1992}).

\bibitem[{\citenamefont{Schoeller}(2009)}]{Schoeller09a}
\bibinfo{author}{\bibfnamefont{H.}~\bibnamefont{Schoeller}},
  \bibinfo{journal}{Eur. Phys. Journ. B} \textbf{\bibinfo{volume}{168}},
  \bibinfo{pages}{179} (\bibinfo{year}{2009}).

\bibitem[{\citenamefont{Leijnse and Wegewijs}(2008)}]{Leijnse08a}
\bibinfo{author}{\bibfnamefont{M.}~\bibnamefont{Leijnse}} \bibnamefont{and}
  \bibinfo{author}{\bibfnamefont{M.~R.} \bibnamefont{Wegewijs}},
  \bibinfo{journal}{Phys.\ Rev.\ B} \textbf{\bibinfo{volume}{78}},
  \bibinfo{pages}{235424} (\bibinfo{year}{2008}).

\bibitem[{\citenamefont{Saptsov and Wegewijs}(2012)}]{Saptsov12a}
\bibinfo{author}{\bibfnamefont{R.~B.} \bibnamefont{Saptsov}} \bibnamefont{and}
  \bibinfo{author}{\bibfnamefont{M.~R.} \bibnamefont{Wegewijs}},
  \bibinfo{journal}{Phys. Rev. B} \textbf{\bibinfo{volume}{86}},
  \bibinfo{pages}{235432} (\bibinfo{year}{2012}).

\bibitem[{\citenamefont{Saptsov}(2010)}]{Saptsov10}
\bibinfo{author}{\bibfnamefont{R.~B.} \bibnamefont{Saptsov}},
  \bibinfo{journal}{private notes}  (\bibinfo{year}{2010}).

\end{thebibliography}
\end{document}